\documentclass[11pt,a4paper]{article}
\usepackage{jheppub}
\pdfoutput=1

\usepackage{siunitx}
\usepackage{pdfpages}
\usepackage{float}
\usepackage{textcomp}               
\usepackage{braket,slashed,bm}
\usepackage{array,multirow}
\usepackage[normalem]{ulem}
\usepackage[usenames,dvipsnames]{xcolor}
\usepackage{cancel,youngtab}
\usepackage{multirow}
\usepackage{graphicx}
\usepackage{subcaption}
\usepackage{anyfontsize}

\usepackage[T1]{fontenc} 
\usepackage{mathrsfs}
\usepackage{booktabs}
\usepackage{adjustbox}
\usepackage{mathtools}
\usepackage{hhline}
\usepackage{soul}
\usepackage{verbatim}
\usepackage[section]{placeins}
\usepackage{pythonhighlight}

\usepackage{tikz}
\usetikzlibrary{arrows,decorations.pathmorphing,backgrounds,positioning,fit,petri,automata,shadows,calendar,mindmap,
decorations.markings,calc}
\usepackage{ctable}

\definecolor{labelkey}{rgb}{0,0.5,0.0}

\usepackage{xspace}
\usepackage{slashed}
\usepackage{array,multirow}
\usepackage{graphicx}
\usepackage{graphics}
\usepackage{epsfig}
\usepackage{amsfonts}
\usepackage{amsmath}
\usepackage{amssymb}
\usepackage{dsfont}
\usepackage{color}
\usepackage{pdflscape}
\usepackage{pifont}
\usepackage{pgfplots}
\usepackage{tikz} 
\usepackage{pgfplotstable}
\usepackage{bbold}
\usepackage{stmaryrd}
\usepackage[capitalise, english]{cleveref}
\usepackage{subcaption}
\usepackage{diagbox}
\DeclareSIUnit\barn{b}

\newcommand{\lam}{\lambda}

\newcommand{\kap}{\kappa}

\newcommand{\CMSewkfourlep}{\texttt{CMS-ewk-4$\ell$}\xspace}
\newcommand{\CMSewkthreelep}{\texttt{CMS-ewk-3$\ell$}\xspace}
\newcommand{\CMSewktwotautwolep}{\texttt{CMS-ewk-2$\tau$2$\ell$}\xspace}
\newcommand{\CMSewktwotauonelep}{\texttt{CMS-ewk-2$\tau$1$\ell$}\xspace}
\newcommand{\CMSewkonetauthreelep}{\texttt{CMS-ewk-1$\tau$3$\ell$}\xspace}
\newcommand{\ATLASstrongSS}{\texttt{ATLAS-gluino-SS/3$\ell$}\xspace}
\newcommand{\ATLASstrongSSb}{\texttt{ATLAS-gluino-SS/3$\ell$-1b}\xspace}
\newcommand{\ATLASrpv}{\texttt{ATLAS-RPV-1$\ell$/SS}\xspace}

\makeatletter
\DeclareRobustCommand{\shortto}{%
  \mathrel{\mathpalette\short@to\relax}%
}

\newcommand{\short@to}[2]{%
  \mkern2mu
  \clipbox{{.5\width} 0 0 0}{$\m@th#1\vphantom{+}{\shortrightarrow}$}%
  }
\makeatother

\usepackage{bm}                                  

\newcommand{\beq}{\begin{equation}}
\newcommand{\eeq}{\end{equation}}
\newcommand{\be}{\begin{equation}}
\newcommand{\ee}{\end{equation}}
\newcommand{\bea}{\begin{eqnarray}}
\newcommand{\eea}{\end{eqnarray}}
\newcommand{\ben}{\begin{eqnarray*}}
\newcommand{\een}{\end{eqnarray*}}

\newcommand{\bma}{\begin{pmatrix}}
\newcommand{\ema}{\end{pmatrix}}

\def\lixo#1{}

\def\slashchar#1{\setbox0=\hbox{$#1$}           
  \dimen0=\wd0                                    
  \setbox1=\hbox{/} \dimen1=\wd1                  
  \ifdim\dimen0>\dimen1                           
    \rlap{\hbox to \dimen0{\hfil/\hfil}}            
    #1                                             
  \else                                          
    \rlap{\hbox to \dimen1{\hfil$#1$\hfil}}        
    /                                           
 \fi}                                           %

\newcommand\lsim{\lesssim}
\newcommand\gsim{\gtrsim}

\newcommand{\dslash}[1]{#1 \llap{/\kern-0.5pt}}
\newcommand{\Dslash}[1]{#1 \llap{/\kern+1.5pt}}
\newcommand{\DDslash}[1]{#1 \llap{/\kern+2.3pt}}
\newcommand{\dslashh}[1]{#1 \llap{/\kern+1pt}}

\definecolor{ao}{rgb}{0.0, 0.5, 0.0}
\definecolor{ForestGreen}{rgb}{0.0, 0.27, 0.13}
\definecolor{charcoal}{rgb}{0.21, 0.27, 0.31}
\definecolor{coolgrey}{rgb}{0.55, 0.57, 0.67}
\definecolor{carrotorange}{rgb}{0.93, 0.57, 0.13}
\definecolor{buff}{rgb}{0.94, 0.86, 0.51}
\definecolor{junglegreen}{rgb}{0.16, 0.67, 0.53}
\definecolor{darkpastelblue}{rgb}{0.47, 0.62, 0.8}
\definecolor{taupegray}{rgb}{0.55, 0.52, 0.54}
\definecolor{sunset}{rgb}{0.98, 0.84, 0.65}
\definecolor{vegasgold}{rgb}{0.77, 0.7, 0.35}
\definecolor{darkolivegreen}{rgb}{0.33, 0.42, 0.18}
\definecolor{atomictangerine}{rgb}{1.0, 0.6, 0.4}
\definecolor{cardinal}{rgb}{0.77, 0.12, 0.23}

\definecolor{cadmiumred}{RGB}{199, 22, 22}
\definecolor{chromeyellow}{RGB}{30, 136, 229}
\definecolor{coral}{RGB}{220, 166, 6}
\definecolor{cadetblue}{RGB}{0, 100, 77}
\definecolor{naplesyellow}{RGB}{161, 18, 159}
\definecolor{cambridgeblue}{RGB}{133, 132, 140}

\definecolor{buff}{RGB}{199, 22, 22}
\definecolor{carrotorange}{RGB}{30, 136, 229}
\definecolor{coolgrey}{RGB}{220, 166, 6}
\definecolor{junglegreen}{RGB}{0, 77, 64}
\definecolor{charcoal}{RGB}{161, 18, 159}

\definecolor{darkolivegreen}{RGB}{199, 22, 22}
\definecolor{cardinal}{RGB}{30, 136, 229}
\definecolor{darkpastelblue}{RGB}{220, 166, 6}
\definecolor{taupegray}{RGB}{0, 77, 64}
\definecolor{atomictangerine}{RGB}{161, 18, 159}
\definecolor{vegasgold}{RGB}{133, 132, 140}

\newcommand{\ETmiss}{\ensuremath{{E_T^\text{miss}}}\xspace}

	\preprint{\begin{flushright} BONN-TH-2023-04 \\
 IFIC/23-21
	\end{flushright}}	

	\title{The ABC of RPV: Classification of R-Parity Violating Signatures at the LHC for Small Couplings
		}
	
		\author[a]{Herbi~K.~Dreiner,}
		\emailAdd{dreiner@uni-bonn.de}

            \author[b]{Yong Sheng Koay,}
		\emailAdd{yongsheng.koay@physics.uu.se}

            \author[a]{Dominik K\"ohler,}
		\emailAdd{koehler@physik.uni-bonn.de}

		\author[c,d]{V\'ictor Mart\'in Lozano,}
		\emailAdd{victor.lozano@ific.uv.es}

    	\author[e,f]{Javier Montejo Berlingen,}
		\emailAdd{jmontejo@cern.ch}
		
		\author[a]{Saurabh Nangia,}
		\emailAdd{nangia@physik.uni-bonn.de}

		\author[g]{and Nadja Strobbe.}
		\emailAdd{nstrobbe@umn.edu}

		\affiliation[a]{Bethe Center for Theoretical Physics \& Physikalisches Institut der Universit\"at Bonn,\\ Nu{\ss}allee 12, 53115 Bonn, Germany}
            \affiliation[b]{Department of Physics and Astronomy, Uppsala University, Box 516, SE-751 20 Uppsala, Sweden}
		\affiliation[c]{Departament de Física Teòrica, Universitat de València-CSIC,
E-46100, Burjassot, Spain}
        \affiliation[d]{Instituto de F\'isica Corpuscular, CSIC-Universitat de València, 46980 Paterna, Spain}
		\affiliation[e]{Instituto de F\'isica de Altas Energ\'ias, Campus UAB, 08193, Bellaterra (Barcelona), Spain}
		\affiliation[f]{International Center for Quantum-field Measurement Systems for Studies of the Universe and Particles (QUP), 1-1 Oho, Tsukuba, Ibaraki 305-0801, Japan}
		\affiliation[g]{School of Physics \& Astronomy, University of Minnesota, Minneapolis, MN 55455, USA}

\abstract{We perform a classification of all potential supersymmetric $R$-parity violating signatures at the LHC to address the question: \emph{are existing bounds on supersymmetric models robust, or are there still signatures not covered by existing searches, allowing LHC-scale supersymmetry to be hiding?} We analyze all possible scenarios with one dominant RPV trilinear coupling at a time, allowing for arbitrary LSPs and mass spectra. We consider direct production of the LSP, as well as production via gauge-cascades, and find 6 different experimental signatures for the $LL\Bar E$-case, 6 for the $LQ\Bar D$-case, and 5 for the $\Bar U\Bar D\Bar D$-case; together these provide complete coverage of the RPV-MSSM landscape. This set of signatures is confronted with the existing searches by \texttt{ATLAS} and \texttt{CMS}. We find all signatures have been covered at the LHC, although not at the sensitivity level needed to probe the direct production of all LSP types. For the case of a dominant $LL\bar E$-operator, we use \texttt{CheckMATE} to quantify the current lower bounds on the supersymmetric masses and find the limits to be comparable to or better than the $R$-parity conserving case. Our treatment can be  easily extended to scenarios with more than one non-zero RPV coupling.}

\pgfplotsset{compat=1.8}
\begin{document}
\maketitle


\section{Introduction}\label{sec:introduction}
Supersymmetry (SUSY)~\cite{Golfand:1971iw, Volkov:1973ix, Wess:1973kz, Wess:1974tw} is a well-motivated extension of the 
Standard Model (SM). It uniquely extends the SM 
algebra~\cite{Coleman:1967ad, Haag:1974qh}, addresses the `naturalness problem' of the Higgs boson~\cite{Gildener:1976ai, Veltman:1980mj}, and has many further appealing features, 
as reviewed in Refs.~\cite{Nilles:1983ge, Martin:1997ns, Drees:2004jm}. Extensive experimental effort has been devoted 
in its search, particularly at the Large Hadron Collider (LHC) by the \texttt{ATLAS} and \texttt{CMS} 
Collaborations. However, no evidence for SUSY has been found so far with lower mass bounds reaching $\mathcal{O}\left(1-2\right)\SI{} 
{\tera\electronvolt}$ for the colored sector~\cite{ATLAS:2022rcw,ATLAS:2022ihe,ATLAS:2022ckd,ATLAS:2021jyv,ATLAS:2021hza,ATLAS:2021yij,ATLAS:2021twp,ATLAS:2020xzu,ATLAS:2020syg,ATLAS:2020xgt,ATLAS:2020dsf,ATLAS:2019fag,ATLAS:2019gdh,CMS:2022vpy,CMS:2021eha,CMS:2021edw,CMS:2020bfa,CMS:2020cpy,CMS:2020fia,CMS:2020cur,CMS:2019lrh,CMS:2019ybf,CMS:2019zmd,CMS:2022idi}, and $\mathcal{O}\left(100-1000\right)\SI{} 
{\giga\electronvolt}$ for the electroweak sector~\cite{ATLAS:2022rcw,ATLAS:2022hbt,ATLAS:2021yqv,ATLAS:2021moa,ATLAS:2019wgx,ATLAS:2019lng,ATLAS:2019gti,ATLAS:2020pgy,CMS:2022vpy,CMS:2021cox,CMS:2017moi,CMS:2021few,CMS:2018eqb,CMS:2022rqk,CMS:2021edw,CMS:2020bfa,CMS:2019zmn,CMS:2019pov,CMS:2022sfi,CMS:2019eln}, with some dependence on the model details.

As we prepare for more data through Run 3 at the LHC, and
especially in the high-luminosity era, it is an excellent 
opportunity to assess the current status of supersymmetric searches and gain insight into how we should proceed. An interesting question is: 
\textit{Are the above bounds robust, or are there gaps/loopholes 
that could still allow LHC-scale SUSY to be hiding?} Typically, \texttt{ATLAS} and \texttt{CMS} derive these limits within the framework of various simplified models or a limited number of 
complete models such as the Constrained Minimal Supersymmetric Standard Model (CMSSM); it is not clear whether these results can be used to conclude that low-scale SUSY has been definitively excluded.

The above question was first addressed in Ref.~\cite{Bechtle:2011dm}, and -- after LHC Run 1 -- in Ref.~\cite{Evans:2013jna}, in more detail. In the latter, it was argued that any `natural' SUSY model\footnote{The `naturalness' criterion in Ref.~\cite{Evans:2013jna} requires the Higgsinos and stops to be light.} with kinematically accessible gluinos -- independent of model details -- results in final states containing at least one of the following ingredients: large missing transverse momentum $\left(E^{
\text{miss}}_\text{T}\right)$, high multiplicity of objects $\left(\geq 8\right)$, or a significant number of top quarks. Using this, the authors showed that combining just five existing \texttt{ATLAS} and \texttt{CMS} searches, and one newly proposed search~\cite{Lisanti:2011tm} excludes almost any `natural' SUSY model containing gluinos lighter than $\SI{1}{\tera\electronvolt}$. Of course, using current data and a similar 
strategy should yield a higher mass bound. Nevertheless, the 
demonstration that a minimal set of searches can target almost any SUSY setup, independent of details concerning the model, mass 
spectrum, UV-completion, etc., is noteworthy. Such an approach is desirable, especially since it informs us about potential gaps 
that may exist in our SUSY coverage. For instance, the search proposed by  Ref.~\cite{Lisanti:2011tm} represented a real gap 
that has since been filled by \texttt{ATLAS} and \texttt{CMS} in Refs.~\cite{ATLAS:2021fbt,CMS:2021knz}.

In this work, we wish to consider the same question but with two important differences. First, beyond assuming the MSSM particle 
content, we remain completely blind to the particle-spectrum 
details. In particular, we do not require that the gluinos are 
kinematically accessible. With the LHC transitioning from an era 
of energy upgrades to one of increasing luminosity, we should 
seriously entertain the possibility that the colored sector may be 
heavy, while a focus on rarer production channels may yield fruit. 
We also do not make any `naturalness' requirements in the sense of Ref.~\cite{Evans:2013jna}. 

Second, our focus will be on the R-parity Violating MSSM (RPV-MSSM). The most general, renormalizable superpotential with the 
MSSM particle content includes lepton- and baryon-number violating 
operators, together referred to as RPV terms~\cite{Weinberg:1981wj, Martin:1997ns, Dreiner:1997uz, Barbier:2004ez}. These are usually set to zero by imposing a 
discrete $\mathbb{Z}_2$-symmetry called R-parity as they can lead 
to proton decay~\cite{Farrar:1978xj, Chamoun:2020aft} at rates in 
excess of the strict experimental bound~\cite{Workman:2022ynf}. 
However, the proton-decay problem can be averted without removing 
all RPV terms~\cite{Ibanez:1991pr,Ibanez:1991hv,Dreiner:2012ae}; in general, there is no theoretical or phenomenological reason 
to consider the MSSM without RPV terms~\cite{Dreiner:1997uz}. On the other hand, as we demonstrate in~\cref{sec:model}, the 
different configurations of couplings and types of the lightest SUSY particle (LSP) in the RPV-MSSM lead to a bewildering number 
of possible signatures. In particular, with the requirements on 
gluino and higgsino masses absent, a large number of these 
signatures do not possess any of the characteristics listed in Ref.~\cite{Evans:2013jna}. In comparison, the `vanilla' MSSM is 
less interesting as it tends to retain its characteristic significant $\ETmiss$ signal, irrespective of spectrum 
details.\footnote{We note that the $\ETmiss$ signature can be 
diluted even in the case of the MSSM through scenarios with a compressed spectrum or a `Hidden Valley'; see, for instance, Refs.~\cite{Dreiner:2012gx, Evans:2013jna} for details. Despite 
the varied phenomenology offered by these models, we believe that it is more efficient to thoroughly explore the minimal setup 
provided by the RPV-MSSM before adding further complexities.} This 
makes a systematic treatment and classification particularly crucial in the case of the RPV-MSSM.

To summarize, we study the coverage of the most general RPV-MSSM setup at the LHC, without making any assumptions about the particle-spectrum details. We seek a minimal set of searches that would provide complete coverage; this will allow us to identify any potential gaps in our current searches. We will restrict ourselves to the case of small RPV couplings in this work, leaving the large-coupling case for a 
dedicated study in the future. Thus, the production of sparticles 
is unchanged from the MSSM case and we only need to consider pair-production channels. The final state signatures will be altered, however, due to the RPV couplings affecting decays. 

The paper is organized as follows. In~\cref{sec:model}, we set notation and state the assumptions of our framework. Further, we explicitly describe the vast phenomenology of the RPV-MSSM, in order to demonstrate our point about the need for a systematic method of classification. In~\cref{sec:classification}, we provide such a systematic classification by grouping signatures in a meaningful way, according to the coupling and nature of the LSP. Our approach allows us to identify a minimal set of searches that would provide complete RPV-MSSM coverage at the LHC, and discuss the current status of such a program. Then, in~\cref{sec:results}, we demonstrate 
applications of our framework -- as a first study -- for the case of a dominant $LL\bar E$ RPV-operator. We consider 
several benchmark scenarios with such lepton-number 
violating operators, involving the full range of LSP types, and derive
exclusion limits. Our results demonstrate that, irrespective of model details, the minimal set of searches 
proposed in this work can be used to derive strong limits. Finally, we conclude 
and discuss the implications and limitations of our work, and provide an outlook in~\cref{sec:conclusions}. Additionally, we provide a set of 
appendices containing supplementary details about our simulation procedure (~\cref{sec:A}), information that can be used to optimize future searches (~\cref{sec:B}), and an introduction to \texttt{abc-rpv} (~\cref{sec:C}), an accompanying RPV Python library\footnote{Available at: \href{https://github.com/kys-sheng/abctestrun.git}{\texttt{https://github.com/kys-sheng/abc-rpv.git}}} that can be used to generate all the signature tables in this paper.


\section{Framework}
\label{sec:model}
\subsection{Conventions and Assumptions}
With the MSSM particle content and the $N=1$ supersymmetry algebra, the most general $\mathrm{SU}(3)_{C}\times\mathrm{SU}(2)_{L}\times\mathrm{U}(1)_Y$-invariant, renormalizable superpotential is,
\begin{equation}
	W = W_{\mathrm{MSSM}} + W_{\mathrm{LNV}} + W_{\mathrm{BNV}}\,,
	\label{eq:TFeq1}
\end{equation}
where $W_{\mathrm{MSSM}}$ is the usual MSSM superpotential -- see, for instance,
Ref.~\cite{Allanach:2003eb} -- while, 
\begin{align}
	W_{\mathrm{LNV}} = \frac{1}{2}\lam^{ijk}L_iL_j\Bar{E}_k + 
	\lam'^{ijk}L_iQ_j\Bar{D}_k + \kappa^{i}H_uL_i\,, \;\qquad 
	W_{\mathrm{BNV}} = 
	\frac{1}{2}\lam''^{ijk}\Bar{U}_i\Bar{D}_j\Bar{D}_k\,,
	\label{eq:TFeq1a}
\end{align}
violate lepton- and baryon-number, respectively. Together, $W_{\mathrm{RPV}} 
\equiv W_{\mathrm{LNV}} + W_{\mathrm{BNV}}$, are 
called the RPV superpotential terms. In our notation, $L\;(\Bar{E})$ and $Q\;(\Bar{U},\;\Bar{D})$ are the MSSM lepton- and quark-doublet (-singlet) 
chiral superfields, respectively, while $H_u$ labels the (up-type) $\mathrm{SU}(2)_{L}$-doublet Higgs chiral superfield. We do not write gauge indices explicitly but retain the generational ones: $i,j,k \in \{1,2,3\}$ with a summation implied over repeated indices. The $\lam$'s and
the $\kap$'s are the trilinear and bilinear couplings, respectively. 

We shall employ the particle content of the MSSM and the superpotential of~\cref{eq:TFeq1} as the basis for this study. As mentioned in 
the Introduction, some terms in the superpotential can lead to rapid proton decay. In general, this 
requires combinations of certain $LQ\Bar{D}$ and $\Bar{U}\Bar{D}\Bar{D}$ operators.\footnote{One 
exception is if the lightest neutralino is lighter than the proton in which case the decay can occur 
via $\Bar{U}\Bar{D}\Bar{D}$ operators alone~\cite{Chamoun:2020aft}.} As long as these combinations 
are kept small, the proton's lifetime remains consistent with the bounds. Indeed, there are 
symmetries that can achieve this -- see, for instance, Refs.~\cite{Ibanez:1991hv, Ibanez:1991pr, 
Dreiner:2012ae}. In this study, we will not bother with the details of how this is done; our focus will be on classifying all possible collider signatures coming from the various couplings. We will, however, ignore the bilinear couplings. These are severely 
constrained by neutrino mass data~\cite{Allanach:2003eb} and are expected to be relevant for colliders 
only in limited contexts~\cite{Barbier:2004ez, deCampos:2007bn}. Furthermore, at a fixed energy scale 
they can be rotated away \cite{Hall:1983id, Dreiner:2003hw}.

The optimal search strategy for RPV-MSSM scenarios at colliders depends on the magnitude of the RPV couplings. We will restrict ourselves to the case where these couplings are small enough such that the production of sparticles and their cascade decays down to the LSP remain unchanged from the MSSM case, but large enough so that the LSP decays promptly in the detector (we also require the cascade decays of the other sparticles to be prompt). While the exact magnitudes depend on the spectrum details, we can estimate it to roughly mean the range,
\begin{equation}
\sqrt{\frac{\left(\beta\gamma\right)\SI{e-12}{\giga\electronvolt}}{m_{\text{LSP}}}} \lsim \lam \ll g\,,
\label{eq:TFeq2}
\end{equation}
where $\lam$ is the relevant RPV coupling, $g$ is a gauge coupling, $m_{\text{LSP}}$ is 
the mass of some LSP that has a two body-decay via the 
RPV coupling, and $\beta$ and $\gamma$ are its velocity 
and Lorentz factor, respectively. The left condition is 
derived from the requirement that the LSP has a decay 
length of about $\SI{1}{\centi\metre}$ in the lab frame.\footnote{We have 
considered a two-body decay here. For comparison, a similar estimate for an LSP with mass $\SI{500}{\giga\electronvolt}$ undergoing a three-body decay via a virtual sfermion of mass $\SI{1}{\tera\electronvolt}$ (this is how a neutralino decays, for instance) gives the range $\mathcal{O}\left(10^{-5}\right) \lsim \lam \ll 
\mathcal{O}\left(10^{-1}\right)$. We note that, in some cases, four-body decays are also possible, \textit{e.g.}, a slepton LSP decaying via $\lambda''$ couplings.} For an LSP mass of $\SI{1}{\tera\electronvolt}$,~\cref{eq:TFeq2} implies the range $\mathcal{O}\left(10^{-7}\right) \lsim \lam \ll 
\mathcal{O}\left(10^{-1}\right)$. Considering 
$\lam$ values smaller or 
larger than the above range leads to unique features that require separate studies. The former can lead to 
new kinds of signals such as displaced vertices or long-lived particles, and both topics have received some attention in recent times~\cite{Lee:2018pag,Alimena:2019zri,Alimena:2021mdu,ATLAS:2023oti,ATLAS:2020wjh,ATLAS:2020xyo,CMS:2020iwv,CMS:2021kdm,CMS:2021sch,CMS:2022qej,CMS:2022wjc,CMS:2021tkn}. The latter also leads to interesting features; in particular, single production of sparticles
\cite{Dimopoulos:1988fr, Dreiner:1991pe, Dreiner:2000vf,Dreiner:2012np,Monteux:2016gag}, and RPV effects in cascade chains can lead to phenomenological changes requiring a dedicated study that we shall pursue in the future, as a continuation of this work.

One assumption, related to the above point, that we will need to make in this work is that the LSP is not too light, \textit{i.e.}, $m_{\text{LSP}} > \mathcal{O}\left(\SI{200}{\giga\electronvolt}\right)$. While current mass bounds on most SUSY particles place them well above this limit, a bino-like neutralino is still allowed to be massless~\cite{Dreiner:2003wh, Dreiner:2009ic}. Requiring the above condition ensures that the decay of the LSP can be prompt without requiring the RPV couplings to be too large. Further, it allows the LSP to decay into all SM fermions (except for, perhaps, the top quark).\footnote{Note that, throughout this work, we will neglect all SM Yukawas, except for that of the top quark.} Dedicated LHC studies for a very light neutralino can be found in, for instance, Refs.~\cite{deVries:2015mfw, Dreiner:2020qbi, Dreiner:2022swd}. 

\begin{table}[h]
\caption{Summary of notation for labeling the RPV-MSSM particle content used in this work. For the particles not mentioned in the table, we use standard notation.}
\begin{center}
\begin{tabular}{cc}
\hline \hline
{\bf Symbol} & {\bf Particles} \\
\hline \hline
\vspace{-1em}&\\
$\ell$ & $e/\mu$ \\
$L$ & $\ell/\tau$\\
$j_l$ & $u/d/c/s$ jets\\
$j_3$ & $t/b$ jets \\
$j$ & $j_l/j_3$ jets \\
$V$ & $W/Z/h$\\
$\tilde{\ell} (\tilde{\nu})$ & $\tilde{e}_L(\tilde{\nu}_{e})/\tilde{\mu}_L (\tilde{\nu}_{\mu})$\\
$\tilde{e}$ & $\tilde{e}_R/\tilde{\mu}_R$\\
$\tilde{q}$ & $\tilde{u}_L/\tilde{d}_L/\tilde{c}_L/\tilde{s}_L$\\
$\tilde{u}$ & $\tilde{u}_R/\tilde{c}_R$\\
$\tilde{d}$ & $\tilde{d}_R/\tilde{s}_R$\\
$\tilde{q}_3$ & $\tilde{t}_L/\tilde{b}_L$\\
$\tilde{t}$ & $\tilde{t}_R$\\
$\tilde{b}$ & $\tilde{b}_R$\\
$\tilde{B}$ & Bino\\
$\tilde{W}$ & Winos (charged/neutral)\\
$\tilde{H}$ & Higgsinos (charged/neutral)\\
\hline \hline
\end{tabular}
\end{center}

\label{tab:notation}
\end{table}

Finally, before concluding this subsection, we introduce our notation for labeling the particle content in~\cref{tab:notation}. We will find the groupings we define useful in presenting our results later. For simplicity, we will also assume all SUSY particles belonging to a particular grouping are mass degenerate -- \textit{i.e.}, we treat mass splittings between components of the same doublet (for instance, $\tilde{H}^{\pm}$ and $\tilde{H}^0$), as well as between first and second generation sparticles as negligible. The former assumption holds true to a very good approximation~\cite{Martin:1997ns}. The latter is not essential for our framework but allows us to be concise; generalization is straightforward.

\subsection{The RPV Landscape}
The presence of even small RPV couplings can drastically change collider
phenomenology compared to the MSSM. In the latter case, SUSY particles are 
pair-produced at colliders and undergo gauge-cascade decays into the LSP -- 
typically the neutralino\footnote{The nature of the LSP in the MSSM follows 
from the strict constraints on charged or colored stable 
particles~\cite{Ellis:1990nb, Barbier:2004ez}.} -- which then escapes the 
detector unobserved, giving the characteristic $\ETmiss$ signature. The 
presence of RPV couplings changes this simple 
picture in two main ways. First, the LSP is no longer constrained to be 
the neutralino but can be any SUSY particle \cite{Dreiner:2008ca, Dercks:2017lfq}. 
Second, the RPV couplings make the LSP unstable; the $\ETmiss$ signature is now replaced (diluted, or even completely absent) by the objects 
arising in this decay, which are determined by the dominant RPV coupling. The total number of possible signatures for the RPV-MSSM at a hadron 
collider can be summarized as (adapted from Ref.~\cite{Dercks:2017lfq}, 
see also Ref.~\cite{Dreiner:2012wm}):

\begin{equation}
\text { RPV signature }=\left(\begin{array}{c}
\tilde{g} \tilde{g}\\
\tilde{g} \tilde{q}, \tilde{g} \tilde{u} \ldots\\
\tilde{q} \tilde{q}, \tilde{q}_3 \tilde{q}_3, \tilde{q}\tilde{u} \ldots\\
\tilde{\ell} \tilde{\ell}, \tilde{\tau}_L \tilde{\tau}_L, \tilde{\ell}\tilde{\nu} \ldots\\
\tilde{H} \tilde{H}\\
\tilde{W} \tilde{W}\\
\tilde{B} \tilde{B}
\end{array}\right)_{\substack{\text { Production } \\
\text{ Channels }}}
\otimes
\left(\begin{array}{c}
\tilde{B} \\
\tilde{H} \\
\tilde{W} \\
\tilde{\ell}(\tilde{\nu})\\
\tilde{\tau}_L(\tilde{\nu}_{\tau})\\
 \tilde{e}\\
 \tilde{\tau}_R\\
\tilde{q}\\
\tilde{u}\\
\tilde{d} \\
\tilde{q}_3 \\
\tilde{t} \\
\tilde{b} \\
\tilde{g}
\end{array}\right)_{\substack{\text { Possible } \\
\text{ LSPs }}} \otimes\left(\begin{array}{c}
L_1 L_2 \bar{E}_1 \\
\ldots \\
L_1 Q_1 \bar{D}_1 \\
\ldots \\
\bar{U}_3 \bar{D}_2 \bar{D}_3
\end{array}\right)_{\substack{\text {LSP }\\ 
\text{Decay}}}
\label{eq:totalsigs}
\end{equation}

There are 45 different RPV trilinear couplings to consider above. Further, the final state will depend on the details of the cascade decays which, in turn, are determined by the mass orderings in the SUSY spectrum: the total number of possibilities is immense! The first systematic analysis of these signatures was performed in Ref.~\cite{Dreiner:1991pe}, for the particular case of a neutralino LSP. A 
more general classification, allowing for all possible LSPs, has been presented
in Ref.~\cite{Dreiner:2012wm} (see also Ref.~\cite{Konar:2010bi}). However, the
study assumes that the lightest colored particle is kinematically accessible at
the collider. In this work, we extend this by also including the possibility 
that the colored sector lies beyond LHC energies. More importantly, the 
emphasis in Ref.~\cite{Dreiner:2012wm} was on finding signatures arising most 
frequently from~\cref{eq:totalsigs}, when one considers the space of all 
possible mass orderings of the SUSY spectrum. Our approach here is different: 
we wish to create a minimal set of signatures that provides complete coverage 
for the space of RPV-MSSM models, irrespective of how frequently an individual 
signature may arise. Furthermore, we will concretely tie this to the LHC 
search program, discussing the current experimental coverage and 
identifying possible gaps; this aspect is absent in Ref.~\cite{Dreiner:2012wm}. 
Ref.~\cite{Dercks:2017lfq} has studied it for the case of the RPV-CMSSM, but 
a more general model-independent treatment is missing in the literature.

\section{Classification of Signatures: The RPV Dictionary}
\label{sec:classification}
We now describe our approach for classifying the most general RPV-MSSM signatures. Since we assume the RPV couplings are small, sparticles are pair-produced at the LHC via gauge interactions, as in the MSSM. The production channels that we consider are listed in~\cref{eq:totalsigs} on the left; the 
mass spectrum determines which of these are kinematically 
accessible. The produced sparticles -- if not the LSP -- will then 
cascade-decay via gauge interactions until the LSP is 
reached with the details of the cascade also depending on the model (\textit{i.e.}, 
the spectrum). The LSP, once produced, decays promptly via the relevant RPV 
coupling. 

\begin{figure*}
\centering
\includegraphics[width=\textwidth]{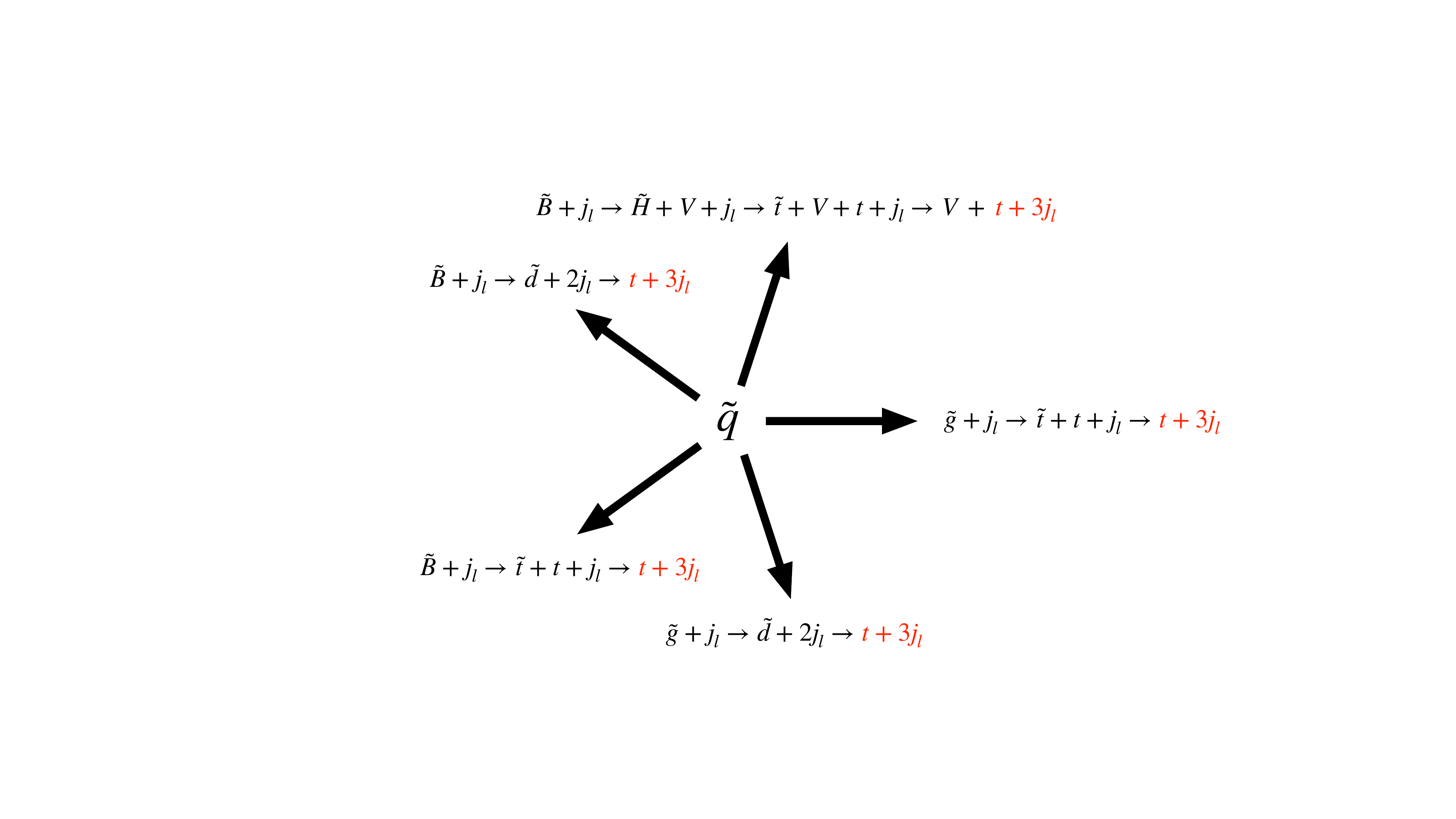}
\caption{Some possible paths a $\tilde{q}$ LSP can take while decaying through $\lam''_{312}$. Since 
$\tilde{q}$ is the LSP here, all the intermediate sparticles are 
virtual. See~\cref{tab:notation} for the notation employed.}
\label{fig:squark_decays}
\end{figure*}

In our model-independent approach, we target the last step above: the LSP 
decay. The essential features of the signatures can be 
characterized by specifying the nature of the LSP and the RPV coupling, 
independent of any spectrum-specific details such as the exact chain 
leading to the LSP production, the mass hierarchies, etc. This is 
obviously true when the LSP couples directly to the relevant RPV decay 
operator, leading to a two-body decay. However, it is also true 
more generally. To illustrate this point, we consider a scenario with a 
$\tilde{q}$ LSP (first or second generation squark doublet, 
\textit{cf.}~\cref{tab:notation}), with $\lam''_{312}$ the only non-zero 
RPV coupling. In this case, there is no direct two-body decay available 
for $\tilde{q}$. Instead, it must decay via a virtual $\tilde{t}$ or 
$\tilde{d}$; some of the paths it can take are depicted in~\cref{fig:squark_decays}. 
Without specifying the model spectrum, it is impossible to state which path will be 
favored. However, note that in each case we end up with the final state $t+3j_l+X$.\footnote{There is one subtlety here: the $\tilde{H}$ path in~\cref{fig:squark_decays} leads to a $b$-jet instead of $t$ if it proceeds via $\tilde{H}^{\pm}$. However, since we assume $\tilde{H}^{\pm}$ and $\tilde{H}^0$ are mass-degenerate, the corresponding path via $\tilde{H}^0$ is always equally likely.} 
This is a general feature, independent of the path it actually takes. Thus, any 
model with a $\tilde{q}$ LSP and a dominant $\lam''_{312}$ operator has a 
characteristic $t+3j_l$ signature, irrespective of any other spectrum 
details. We can target all such scenarios with a single search 
-- this observation is the most crucial aspect of this work. 

Using the above approach, we can compile the characteristic signatures 
arising from each LSP and dominant RPV coupling combination, in 
order to arrive at a minimal set of searches that would provide complete 
coverage for the RPV-MSSM, in a model-independent way. We present this 
set in the form of tables below. We will also compare it to what has 
been covered by the vast program of BSM searches by the \texttt{ATLAS} 
and \texttt{CMS} collaborations. Although only a small subset of these 
searches provides an explicit interpretation in terms of RPV-SUSY models, 
the wide range of final states considered covers the majority of 
signatures expected from RPV decays. Thus, appropriately reinterpreted,
they could be used to restrict the RPV parameter space.

In order to facilitate a systematic exploration of the RPV-MSSM landscape with our approach, we have developed an
RPV Python library called \texttt{abc-rpv}. This library provides a powerful toolkit containing a range of features for analyzing the characteristic signatures arising from various RPV scenarios. The main functionalities include identifying signatures and decay chains for any LSP and RPV coupling combination, as well as going in the other direction: identifying potential RPV scenarios leading to a user-given final state. Using this library, one can reproduce all signature tables in this paper -- for instance, (Tables~\ref{tab:LLE1}-\ref{tab:LQD6}) shown below, as well as~\cref{tab:optimisesearch} in~\cref{sec:B}. The information in~\cref{fig:squark_decays} (possible decay chains for
a given LSP) can also be generated easily up to a fixed number
of vertices. An introduction to the \texttt{abc-rpv} library, including a quick user manual is provided in~\cref{sec:C}.

We note that one downside of our approach is that only final state objects 
arising in the LSP decay are targeted, and all objects arising in the 
cascade decays are neglected. In specific models -- for instance, one with 
squark pair-production and a neutralino LSP -- one could certainly optimize by targeting the additional jets arising in the cascade decays of 
the parent squarks, thus improving the search sensitivity. However, 
in order to analyze the status of complete coverage 
while being model-independent, our approach is necessary. For completeness, 
we compile a list of additional objects that can arise in cascade decays 
for various production channels in~\cref{tab:optimisesearch} in~\cref{sec:B}. That table may be used to optimize the searches compiled below for particular scenarios when the model details are known. Further, it can help understand the loss in sensitivity for searches that veto additional objects to help with background suppression. 

\subsection{LLE Tables}
We depict the signatures corresponding to the decay of a pair of LSPs for the 
$LL\Bar{E}$ operators of~\cref{eq:TFeq1a} in Tables~\ref{tab:LLE1} and~\ref{tab:LLE2}. 
The tables have been written assuming that LSPs are gauge eigenstates, and the pair decays via the same coupling. However, if one is interested in scenarios where the mass eigenstates have significant mixing, or where several dominant RPV couplings contribute, the results can be generalized by considering linear combinations of the table entries.

\begin{table}[t]
\caption{Characteristic signatures arising from LSP decays for $L_iL_j\Bar{E}_k$ 
operators. The first column depicts the LSPs. The second and third columns represent
the signatures from \textit{pair-production} of LSPs for the cases where the indices $i,j,k\in\{1,2\}$, and where the indices 
$i,k\in\{1,2\}$ and $j=3$, respectively. For cases involving degenerate LSPs, \textit{e.g.}, $\tilde{\ell}\left(\tilde{\nu}\right)$, all pair combinations are considered. Further, only the relevant signatures are retained and we have introduced color-coding to improve the readability of the table; the details are in the main text.}
\centering
\scalebox{0.75}{
\renewcommand{\arraystretch}{1.3}
\begin{adjustbox}{width=1.3\textwidth}
\begin{tabular}{ccc}
\hline \hline
{\bf LSP} & $\mathbf{LLE}$ & $\mathbf{LL_3E}$\\

\hline \hline

$\tilde{\ell}\left(\tilde{\nu}\right)$ & \textcolor{chromeyellow}{$3\ell + \ETmiss$}/\textcolor{coral}{$4\ell$} & \textcolor{chromeyellow}{$2\ell + \tau + \ETmiss$}/\textcolor{coral}{$2\ell + 2\tau$} \\

$\tilde{e}$ & \textcolor{cadmiumred}{$2\ell + \ETmiss$} & \textcolor{cadmiumred}{$2\ell +\ETmiss$}/\textcolor{cadmiumred}{$\ell + \tau + \ETmiss$} \\

$\tilde{\tau}_L\left(\tilde{\nu}_\tau\right)$ & \textcolor{cambridgeblue}{$4\ell + 2\tau + \ETmiss$}/\textcolor{naplesyellow}{$4\ell + \tau + \ETmiss$} & \textcolor{chromeyellow}{$3\ell + \ETmiss$}/\textcolor{coral}{$4\ell$} \\

$\tilde{\tau}_R$ &\textcolor{cambridgeblue}{$4\ell+ 2\tau+\ETmiss$} & \textcolor{cambridgeblue}{$4\ell+ 2\tau+\ETmiss$}/\textcolor{cambridgeblue}{$3\ell + 3\tau + \ETmiss$} \\

$\tilde{g}$ & \textcolor{cadetblue}{$4\ell + 4j + \ETmiss$} & \textcolor{cadetblue}{$4\ell + 4j + \ETmiss$}/\textcolor{cadetblue}{$3\ell + \tau + 4j + \ETmiss$} \\

$\tilde{q},\; \tilde{u},\; \tilde{d}$ & \textcolor{cadetblue}{$4\ell + 2j_l + \ETmiss$} & \textcolor{cadetblue}{$4\ell + 2j_l + \ETmiss$}/\textcolor{cadetblue}{$3\ell + \tau + 2j_l + \ETmiss$} \\

$\tilde{t}_L(\tilde{b}_L)$ & \textcolor{cadetblue}{$4\ell + 2j_3 + \ETmiss$} & \textcolor{cadetblue}{$4\ell + 2j_3 + \ETmiss$}/\textcolor{cadetblue}{$3\ell + \tau +2j_3 + \ETmiss$} \\

$\tilde{t}_R$ & \textcolor{cadetblue}{$4\ell + 2t + \ETmiss$} & \textcolor{cadetblue}{$4\ell + 2t + \ETmiss$}/\textcolor{cadetblue}{$3\ell + \tau +2t + \ETmiss$} \\

$\tilde{b}_R$ & \textcolor{cadetblue}{$4\ell + 2b + \ETmiss$} & \textcolor{cadetblue}{$4\ell + 2b + \ETmiss$}/\textcolor{cadetblue}{$3\ell + \tau + 2b +\ETmiss$} \\

$\tilde{B}, \; \tilde{W}, \;\tilde{H}$ & \textcolor{cadetblue}{$4\ell + \ETmiss$} & \textcolor{cadetblue}{$4\ell + \ETmiss$}/\textcolor{cadetblue}{$3\ell + \tau + \ETmiss$}\\
\hline \hline
\end{tabular}
\end{adjustbox}}

\label{tab:LLE1}
\end{table}

\begin{table}[htbp]
\caption{Same as~\cref{tab:LLE1} but for $L_iL_j\Bar{E}_k$ operators with $i,j\in\{1,2\}$ and $k=3$ (second column), and $j,k=3$ and $i\in\{1,2\}$ (third column).}
\centering
\scalebox{0.75}{
\renewcommand{\arraystretch}{1.3}
\begin{adjustbox}{width=1.3\textwidth}
\begin{tabular}{ccc}
\hline \hline
{\bf LSP} &  $\mathbf{LLE_3} $ & $\mathbf{LL_3E_3}  $\\
\hline \hline

$\tilde{\ell}\left(\tilde{\nu}\right)$ & \textcolor{chromeyellow}{$\ell+2\tau+\ETmiss$}/\textcolor{coral}{$2\ell+2\tau$} & \textcolor{chromeyellow}{$3\tau + \ETmiss$}/\textcolor{coral}{$4\tau$}\\

$\tilde{e}$ &  \textcolor{cambridgeblue}{$4\ell + 2\tau + \ETmiss$} & \textcolor{cambridgeblue}{$4\ell + 2\tau + \ETmiss$}/\textcolor{cambridgeblue}{$3\ell + 3\tau +\ETmiss$} \\

$\tilde{\tau}_L\left(\tilde{\nu}_\tau\right)$  & \textcolor{cambridgeblue}{$2\ell + 4\tau + \ETmiss$}/\textcolor{naplesyellow}{$2\ell + 3\tau + \ETmiss$} & \textcolor{coral}{$2\ell+2\tau$}/\textcolor{chromeyellow}{$\ell + 2\tau + \ETmiss$}\\

$\tilde{\tau}_R$ &  \textcolor{cadmiumred}{$2\ell + \ETmiss$} & \textcolor{cadmiumred}{$2\ell + \ETmiss$}/\textcolor{cadmiumred}{$\ell + \tau + \ETmiss$}\\

$\tilde{g}$ & \textcolor{cadetblue}{$2\ell+2\tau+4j+\ETmiss$} & \textcolor{cadetblue}{$2\ell+2\tau+4j+\ETmiss$}/\textcolor{cadetblue}{$\ell + 3\tau+4j+\ETmiss$}\\

$\tilde{q}, \;\tilde{u}, \;\tilde{d}$ &  \textcolor{cadetblue}{$2\ell + 2\tau + 2j_l + \ETmiss$} & \textcolor{cadetblue}{$2\ell + 2\tau + 2j_l + \ETmiss$}/\textcolor{cadetblue}{$\ell + 3\tau + 2j_l + \ETmiss$} \\

$\tilde{t}_L(\tilde{b}_L), $ & \textcolor{cadetblue}{$2\ell + 2\tau + 2j_3 + \ETmiss$} & \textcolor{cadetblue}{$2\ell + 2\tau + 2j_3 + \ETmiss$}/\textcolor{cadetblue}{$\ell + 3\tau + 2j_3 + \ETmiss$}\\

$\tilde{t}_R$ & \textcolor{cadetblue}{$2\ell + 2\tau + 2t + \ETmiss$} & \textcolor{cadetblue}{$2\ell + 2\tau + 2t + \ETmiss$}/\textcolor{cadetblue}{$\ell + 3\tau + 2t + \ETmiss$}\\

$\tilde{b}_R$ & \textcolor{cadetblue}{$2\ell + 2\tau + 2b + \ETmiss$}  & \textcolor{cadetblue}{$2\ell + 2\tau + 2b + \ETmiss$}/\textcolor{cadetblue}{$\ell + 3\tau + 2b + \ETmiss$}\\

$\tilde{B}, \; \tilde{W}, \;\tilde{H}$ & \textcolor{cadetblue}{$2\ell+2\tau + \ETmiss$} & \textcolor{cadetblue}{$2\ell+2\tau + \ETmiss$}/\textcolor{cadetblue}{$\ell + 3\tau + \ETmiss$}\\
\hline \hline
\end{tabular}
\end{adjustbox}}

\label{tab:LLE2}
\end{table}

The tables show the LSP in the first column. The second and third columns depict the 
resulting signature depending on the generation structure of the $LL\bar E$
operator responsible for decay; we employ the compact notation of \cref{tab:notation}. Note that, to be concise, we assume all RPV operators within a given category are non-zero, \textit{e.g.}, both $L_1L_2\Bar{E}_1$ and $L_1L_2\Bar{E}_2$ are non-zero for the category $LL\Bar{E}$. Otherwise, more objects may arise, \textit{e.g.}, with a $\tilde{\mu}_R$ LSP and a non-zero $L_1L_2\Bar{E}_1$ operator, the smuon would first need to transition into $\tilde{\ell}(\tilde{\nu})$ or $\tilde{e}_R$ leading to two extra muons; the extension is straightforward. In some cases, there is more than one signature possible. If two signatures are equally likely, we have listed the one that contains more electrons or muons, since we expect it to be more readily observable. In cases where a signature with fewer $e/\mu$ can have a higher cross-section, we have retained both separated by a `/'. For instance, in the case of a $\tilde{\ell}(\tilde{\nu})$ LSP (we assume mass degeneracy of $\mathrm{SU}(2)_{L}$-doublets, \textit{cf.}~\cref{tab:notation}) decaying via $\lam_{121}$, the $\tilde{\ell}$ decays into one charged lepton and one neutrino, while the $\tilde{\nu}$ decays into two charged leptons. Thus, the possible signatures from pair production are: $4\ell,\;3\ell+\ETmiss, \; 2\ell + \ETmiss$. In the table, we retain the first and second signatures: the former because it has the highest number of charged leptons, and the latter because it has the highest cross-section. $2\ell + \ETmiss$ is not retained since it has both a lower cross-section compared to the $3\ell+\ETmiss$ signature, as well as fewer leptons and, hence, will never be the most relevant final state for searches.

From the tables, we see that the $LL\Bar{E}$ case can be completely covered through the following six searches:
\begin{enumerate}
    \item \textcolor{cadmiumred}{$2L + \ETmiss$}
    \item \textcolor{chromeyellow}{$3L+\ETmiss$}
    \item \textcolor{coral}{$4L$}
    \item \textcolor{cadetblue}{$4L + (0-4)j + \ETmiss$}
    \item  \textcolor{naplesyellow}{$5L + \ETmiss$}
    \item \textcolor{cambridgeblue}{$6L + \ETmiss$}
\end{enumerate}
 
To improve the readability of the table, we have introduced a color scheme based on the number of charged leptons in the search region: red (two), blue (three), yellow (four without missing energy), green (four with missing energy), purple (five), and gray (six).

Thus, indeed -- in spite of the large number of possibilities that RPV offers -- it is possible to organize experimental searches into a small, workable set. The identification of these minimal signatures and the corresponding experimental coverage is one of the main results of this paper. We stress that this is more than just a convenient notational scheme. As will be shown, all signatures that we will classify in our tables -- except for one -- are experimentally covered by \texttt{ATLAS} and \texttt{CMS} in one form or another, although in some cases strong improvements in sensitivity are required to reach the electroweak production cross-sections. In~\cref{sec:results}, we will further apply these to see how the same small set of searches provides exclusion limits across a broad class of RPV models. 

One point to note is that, in the above, we have only classified the total number of leptons in each search. However, often it may be useful to know the flavor/sign combinations of these leptons. While we do not employ them in our numerical studies, we provide tables in~\cref{sec:B} that explicitly show these configurations. These may be useful in developing more sensitive search regions, in case one wishes to target specific scenarios.

We now discuss the experimental coverage of the above signatures. The six final states identified include multiple leptons, 
may include additional jets, and may come with or without \ETmiss. Searches for 
R-parity Conserving SUSY (RPC-SUSY) typically have good coverage for signatures 
with \ETmiss or with at least three leptons (or two with the same charge), with 
several of these searches providing some interpretations in RPV-SUSY models as 
well. Other searches sensitive to the $LL\Bar{E}$ case include analyses targeting heavy leptons or 
additional Higgs bosons. LHC searches relevant for the $LL\Bar{E}$ coupling broadly span the final states of
(1.) ${2\ell+\ETmiss}$~\cite{CMS:2018eqb,CMS:2018kag,CMS:2021edw,ATLAS:2019lff,ATLAS:2019gti,CMS:2020bfa,CMS:2022rqk,CMS:2019eln,CMS:2020cpy}, 
(2.) ${3\ell+\ETmiss}$~\cite{CMS:2013pkf,CMS:2019lwf,CMS:2020cpy,CMS:2021cox,ATLAS:2021moa,CMS:2021edw}, 
(3.) ${4\ell}$~\cite{ATLAS:2021wob,ATLAS:2018mrn,ATLAS:2022yhd,CMS:2021cox,ATLAS:2020zms,CMS:2016zgb}, and (4., 5., 6.) ${\geq4\ell+\ETmiss}$~\cite{ATLAS:2021yyr,CMS:2021cox,CMS:2020cpy,CMS:2019lwf}.
Searches with four leptons are typically inclusive and include events with more than four leptons, therefore covering also the $5L$ and $6L$ categories. 

\subsection{UDD Tables}
Next, we show analogous results for the $\Bar{U}\Bar{D}\Bar{D}$ case in 
Tables~\ref{tab:UDD1} and~\ref{tab:UDD2}. The comments from before apply here too. These scenarios can be completely covered through the following five searches:
\begin{enumerate}
    \item \textcolor{buff}{$4j$}   
    \item \textcolor{carrotorange}{$2j_l+4j$}
    \item \textcolor{coolgrey}{$2j_l+6j$} 
    \item \textcolor{junglegreen}{$1L+2j_l+4j+\ETmiss$} 
    \item \textcolor{charcoal}{$2L+2j_l+ 4j$}
\end{enumerate}

The color scheme is based on the number of jets and charged leptons: red (four jets), blue (six jets, no leptons), yellow (eight jets), green (six jets, one lepton), and purple (six jets, two leptons).

\begin{table}[t]
\caption{Characteristic signatures arising from LSP decays for $\Bar{U}_i\Bar{D}_j\Bar{D}_k$ operators. The first column depicts the LSPs. The second and third columns represent
the signatures from \textit{pair-production} of LSPs for the cases where the indices $i,j,k\in\{1,2\}$, and where the indices $i,k\in\{1,2\}$ and $j=3$, respectively. For cases involving degenerate LSPs, \textit{e.g.}, $\tilde{\ell}\left(\tilde{\nu}\right)$, all pair combinations are considered. Further, only the relevant signatures are retained and we have introduced color-coding to improve the readability of the table; the details are in the main text.}
\label{tab:UDD1}
\centering
\scalebox{0.75}{
\renewcommand{\arraystretch}{1.3}
\begin{adjustbox}{width=1.3\textwidth}
\begin{tabular}{ccc}
\hline \hline
{\bf LSP} & $\mathbf{UDD} $ & $\mathbf{UD_3D} $\\

\hline \hline

$\tilde{\ell}\left(\tilde{\nu}\right)$ & \textcolor{charcoal}{$2\ell+ 6 j_l$}/\textcolor{junglegreen}{$\ell+ 6 j_l+ \ETmiss$} & \textcolor{charcoal}{$2\ell+ 2b + 4j_l$}/\textcolor{junglegreen}{$\ell+2b+4j_l+\ETmiss$} \\

$\tilde{e}$ & \textcolor{charcoal}{$2\ell+ 6 j_l$} & \textcolor{charcoal}{$2\ell+ 2b + 4j_l$} \\

$\tilde{\tau}_L\left(\tilde{\nu}_\tau\right)$ & \textcolor{charcoal}{$2\tau+ 6 j_l$}/\textcolor{junglegreen}{$\tau+ 6 j_l+ \ETmiss$} & \textcolor{charcoal}{$2\tau+ 2b + 4j_l$}/\textcolor{junglegreen}{$\tau+2b+4j_l+\ETmiss$} \\

$\tilde{\tau}_R$ & \textcolor{charcoal}{$2\tau+ 6 j_l$} & \textcolor{charcoal}{$2\tau+ 2b + 4j_l$} \\

$\tilde{g}$ & \textcolor{carrotorange}{$6 j_l$} & \textcolor{carrotorange}{$2b + 4j_l$} \\

$\tilde{q}$ & \textcolor{coolgrey}{$8 j_l$} & \textcolor{coolgrey}{$2b + 6j_l$} \\

$\tilde{u}$ & \textcolor{buff}{$4 j_l$} &  \textcolor{buff}{$2b + 2j_l$} \\

$\tilde{d}$ & \textcolor{buff}{$4 j_l$} &  \textcolor{buff}{$2b + 2j_l$} \\

$\tilde{t}_L(\tilde{b}_L)$ & \textcolor{coolgrey}{$6 j_l+2j_3$} & \textcolor{coolgrey}{$2b+4j_l+2j_3$}\\

$\tilde{t}_R$ & \textcolor{coolgrey}{$2t+6 j_l$} & \textcolor{coolgrey}{$2t+2b+4j_l$}\\

$\tilde{b}_R$ & \textcolor{coolgrey}{$2b+6 j_l$} & \textcolor{buff}{$4j_l$} \\

$\tilde{B}, \; \tilde{W}, \;\tilde{H}$ & \textcolor{carrotorange}{$6 j_l$} & \textcolor{carrotorange}{$2b + 4j_l$}\\
\hline \hline
\end{tabular}
\end{adjustbox}}
\end{table}

\begin{table}[htbp]
\caption{Same as~\cref{tab:UDD1} but for $\Bar{U}_i\Bar{D}_j\Bar{D}_k$ operators with $j,k\in\{1,2\}$ and $i=3$ (second column), and $i,j=3$ and $k\in\{1,2\}$ (third column).}
\centering
\scalebox{0.75}{
\renewcommand{\arraystretch}{1.3}
\begin{adjustbox}{width=1.3\textwidth}
\begin{tabular}{ccc}
\hline \hline
{\bf LSP} &  $\mathbf{U_3DD} $ & $\mathbf{U_3D_3D}  $\\
\hline \hline

$\tilde{\ell}\left(\tilde{\nu}\right)$ &  \textcolor{charcoal}{$2\ell + 4j_l + 2j_3$}/\textcolor{junglegreen}{$\ell + 4j_l +2j_3+ \ETmiss$} & \textcolor{charcoal}{$2\ell+2b+2j_l+2j_3$}/\textcolor{junglegreen}{$\ell+2b +2j_l+2j_3 + \ETmiss$}\\

$\tilde{e}$ &  \textcolor{charcoal}{$2\ell + 4j_l+ 2j_3$} & \textcolor{charcoal}{$2\ell+2b+2j_l+2j_3$} \\

$\tilde{\tau}_L\left(\tilde{\nu}_\tau\right)$  & \textcolor{charcoal}{$2\tau + 4j_l+ 2j_3$}/\textcolor{junglegreen}{$\tau + 4j_l+2j_3 + \ETmiss$} & \textcolor{charcoal}{$2\tau+2b+2j_l+2j_3$}/\textcolor{junglegreen}{$\tau+2b +2j_l+2j_3 + \ETmiss$}\\

$\tilde{\tau}_R$ &  \textcolor{charcoal}{$2\tau + 4j_l+ 2j_3$} & \textcolor{charcoal}{$2\tau+2b+2j_l+2j_3$} \\

$\tilde{g}$ & \textcolor{carrotorange}{$2t + 4j_l$} & \textcolor{carrotorange}{$2t+2b + 2j_l$}\\

$\tilde{q}$ &  \textcolor{coolgrey}{$2t + 6j_l$} &  \textcolor{coolgrey}{$2t+2b + 4j_l$} \\

$\tilde{u}$ &  \textcolor{coolgrey}{$2t + 6j_l$} &  \textcolor{coolgrey}{$2t+2b + 4j_l$} \\

$\tilde{d}$ &  \textcolor{buff}{$2t + 2j_l$} & \textcolor{buff}{$2t+2b$} \\

$\tilde{t}_L(\tilde{b}_L)$ &  \textcolor{coolgrey}{$4j_l+4j_3$} &  \textcolor{coolgrey}{$2b+2j_l+4j_3$}\\

$\tilde{t}_R$  & \textcolor{buff}{$4j_l$} & \textcolor{buff}{$2b+2j_l$}\\

$\tilde{b}_R$ & \textcolor{coolgrey}{$2t+2b+4j_l$}  & \textcolor{buff}{$2t+2j_l$}\\

$\tilde{B}, \; \tilde{W}, \;\tilde{H}$ & \textcolor{carrotorange}{$4j_l+2j_3$} & \textcolor{carrotorange}{$2b + 2j_l+2j_3$}\\
\hline \hline
\end{tabular}
\end{adjustbox}}
\label{tab:UDD2}
\end{table}
One interesting point worth noting is that we write $j_3$ and not $t$ in~\cref{tab:UDD2} for the  non-colored LSPs ($j_3$ indicates that the jet could be $t/b$, \textit{cf.}~\cref{tab:notation}). This is to account for the possibility that kinematic suppression may lead to the decay into a $b$ (via a virtual chargino) to be preferred over the decay into a $t$ (via a neutralino). Generally, in all tables to follow, we will take this consideration into account for all the non-colored LSPs.

Three of the five $U\Bar{D}\Bar{D}$ final states listed above contain only jets and correspond to the largest fraction of the possible LSP decays. However, up to two of the jets listed 
could be top quarks in certain configurations. This would result in additional 
final state jets or leptons which can be used as experimental handles to improve 
sensitivity. The last two signatures listed arise from slepton LSPs and always 
include leptons and/or \ETmiss in the final state. \texttt{ATLAS} and 
\texttt{CMS} have covered the signatures of 
(1.) 4~jets~\cite{CMS:2018mts,CMS:2022usq,ATLAS:2017jnp}, (2.) 6~jets~\cite{ATLAS:2018umm,CMS:2018ikp}, 
(3.) 8~jets~\cite{ATLAS:2020wgq,CMS:2018pdq,CMS:2016vfw,CMS:2016zgb,CMS:2021knz}, (4.) 1 lepton plus at least 6 jets~\cite{ATLAS:2021fbt,CMS:2017szl,CMS:2021knz,CMS:2020cur,CMS:2022cpe}, 
and (5.) 2 leptons plus 6 jets~\cite{CMS:2016ooq,CMS:2020cpy,CMS:2022cpe}.

Some of these searches explicitly require a minimum number of $b$-tagged jets, whereas others are more inclusive. The searches considering leptons typically only consider electrons or muons, which reduces the sensitivity to scenarios featuring tau leptons. 
The searches for signatures (1.), (2.), and (3.) reduce the potentially overwhelming multijet background by requiring the presence of two same-mass resonances in each event. Even so, while some final states are nominally covered, the large difference in production cross-sections leads to exclusion limits being available for some production modes (\textit{e.g.}, $\tilde{g} \rightarrow 3j_l$) but still requiring orders of magnitude of improvement to reach others (\textit{e.g.}, $\tilde{H} \rightarrow 3j_l$).

\subsection{LQD Tables}
Lastly, we show the results for the $LQ\Bar{D}$ case in 
Tables~\ref{tab:LQD1}-\ref{tab:LQD6}. The comments from before apply here. Analyzing 
the tables, we see that the $LQ\Bar{D}$ scenarios can be completely covered through 
the following six searches:
\begin{enumerate}
    \item \textcolor{darkolivegreen}{$4j$}   
    \item \textcolor{cardinal}{$2b+2j+\ETmiss$} 
    \item \textcolor{darkpastelblue}{$1L+(2-6)j+\ETmiss$}
    \item \textcolor{taupegray}{$2L+(2-6)j + (\ETmiss)$} 
    \item \textcolor{atomictangerine}{$3L+4j+\ETmiss$}
    \item \textcolor{vegasgold}{$4L+4j$}
\end{enumerate}

The color scheme is based on the number of charged leptons and jets: red (no charged leptons, four jets, without missing energy), blue (no charged leptons, four jets, with missing energy), yellow (one charged lepton), green (two charged leptons), purple (three charged leptons), and gray (four charged leptons).

As can be seen from the tables, $LQ\Bar{D}$ operators result in a wide range of possible 
final states, typically including at least one lepton and several jets. Therefore, 
searches targeting a wide range of BSM models beyond RPV-SUSY can be sensitive, 
\textit{e.g.}, searches for RPC-SUSY, leptoquarks, etc. It is important 
to consider whether one of the generation indices of the $L_iQ_j\bar{D}_k$ 
operator is 3 since this changes the experimental signature
significantly. For example, searches explicitly requiring $b$-tagged jets 
typically are the most sensitive for $j,k=3$. An operator with $i=3$ requires 
searches exploiting final states with $\tau$ leptons. The relevant existing searches
for the $LQ\Bar{D}$ coupling cover the final states of 
(1.) 4~jets~\cite{CMS:2018mts,CMS:2022usq,ATLAS:2017jnp},
(2.) $\geq 4$~jets (including $b$-tags) plus $\ETmiss$~\cite{CMS:2019ybf,CMS:2019zmd}, (3.) $1\ell$ plus 2 jets~\cite{ATLAS-CONF-2022-059,CMS:2020wzx} or $1\ell$ plus 6 jets~\cite{ATLAS:2021fbt,CMS:2017szl,CMS:2021knz,CMS:2022idi,CMS:2021eha,CMS:2022cpe}, 
(4.) $2\ell$-same-sign plus 2 jets~\cite{ATLAS-CONF-2022-057}, or $2\ell$-same-sign plus 6 jets~\cite{CMS:2020cpy,ATLAS:2019fag}, or $2\ell$-opposite-sign plus 2 or more  jets~\cite{ATLAS:2017jvy,ATLAS:2021yij,ATLAS:2020dsf,ATLAS:2022wcu,ATLAS:2020xov,CMS:2020bfa,CMS:2019lrh, CMS:2022cpe,CMS:2020cay}, 
(5.) $3\ell$ plus 4 jets~\cite{ATLAS-CONF-2023-017,CMS:2021cox, CMS:2020cpy}, and (6.) $4\ell$ plus 4 jets~\cite{ATLAS:2021yyr,CMS:2021cox}.

It is important to note that for signatures 1.~($4j$) and 2.~($2b+2j+\ETmiss$), the relevant searches target strong production cross-sections. As seen from Tables~\ref{tab:LQD2} to \ref{tab:LQD6}, these signatures arise from the decays $\tilde{\ell} \rightarrow j j$ and $\tilde{\chi}_1^0 \rightarrow \nu j b$ respectively. While the latter is experimentally less sensitive than the competing $\tilde{\chi}_1^0 \rightarrow \ell j t$ decay, phase-space effects due to the top-quark mass can lead to a strong suppression of channels involving $t$. In both cases, the existing analyses target strong production, via $\tilde{q} \rightarrow j j$ and $\tilde{g} \rightarrow b j \tilde{\chi}^0_1$, respectively,\footnote{The scenario with an almost massless neutralino matches the $LQ\bar{D}$ signature of $\tilde{g} \rightarrow b j \nu$.} and have no sensitivity to low masses and electroweak cross sections. We do note the special case of $\tilde{\chi}_1^0 \rightarrow \nu b b$, leading to the 4$b$+\ETmiss final state which has already been explored for Higgsino production~\cite{ATLAS:2018tti,CMS:2017nin}. However, crucially, the searches require an intermediate Higgs resonance which is not present in the RPV case.

\begin{table}[hp]
\caption{Characteristic signatures arising from LSP decays for $L_iQ_j\Bar{D}_k$ operators. The first column depicts the LSPs. The second and third columns represent
the signatures from \textit{pair-production} of LSPs for the cases where the indices $i,j,k\in\{1,2\}$, and where the indices $i,j\in\{1,2\}$ and $k=3$, respectively. For cases involving degenerate LSPs, \textit{e.g.}, $\tilde{\ell}\left(\tilde{\nu}\right)$, all pair combinations are considered. Further, only the relevant signatures are retained and we have introduced color-coding to improve the readability of the table; the details are in the main text.}
\centering
\scalebox{0.69}{
\renewcommand{\arraystretch}{1.33}
\begin{adjustbox}{width=1.33\textwidth}
\begin{tabular}{ccc}
\hline \hline
{\bf LSP} & $\mathbf{LQD} $ & $\mathbf{LQD_3} $\\

\hline \hline

$\tilde{\ell}\left(\tilde{\nu}\right)$ & \textcolor{darkolivegreen}{$4j_l$} & \textcolor{darkolivegreen}{$2b + 2j_l$} \\

$\tilde{e}$ & \textcolor{vegasgold}{$4\ell+4j_l$}/\textcolor{atomictangerine}{$3\ell+4j_l+\ETmiss$} & \textcolor{vegasgold}{$4\ell+2b+2j_l$}/\textcolor{atomictangerine}{$3\ell+2b+2j_l+\ETmiss$} \\

\multirow{2}{*}{$\tilde{\tau}_L\left(\tilde{\nu}_\tau\right)$} & \textcolor{vegasgold}{$2\ell+2\tau+4j_l$}/\textcolor{atomictangerine}{$\ell+2\tau+4j_l+\ETmiss$}/ & \textcolor{vegasgold}{$2\ell+2\tau+2b+2j_l$}/\textcolor{atomictangerine}{$\ell+2\tau+2b+2j_l+\ETmiss$}/ \\

& \textcolor{atomictangerine}{$2\ell+\tau+4j_l+\ETmiss$}/\textcolor{taupegray}{$\ell+\tau+4j_l+\ETmiss$} & \textcolor{atomictangerine}{$2\ell+\tau+2b+2j_l+\ETmiss$}/\textcolor{taupegray}{$\ell+\tau+2b+2j_l+\ETmiss$} \\

$\tilde{\tau}_R$ & \textcolor{vegasgold}{$2\ell+2\tau+4j_l$}/\textcolor{atomictangerine}{$\ell+2\tau+4j_l+ \ETmiss $} & \textcolor{vegasgold}{$2\ell+2\tau+2b+2j_l$}/\textcolor{atomictangerine}{$\ell+2\tau+2b+2j_l+\ETmiss$} \\

$\tilde{g}$ & \textcolor{taupegray}{$2\ell+4j_l$}/\textcolor{darkpastelblue}{$\ell+4j_l+\ETmiss$} & \textcolor{taupegray}{$2\ell+2b+2j_l$}/\textcolor{darkpastelblue}{$\ell+2b+2j_l + \ETmiss$} \\

$\tilde{q}$ & \textcolor{taupegray}{$2\ell + 2j_l$} & \textcolor{taupegray}{$2\ell+2b$}\\

$\tilde{u}$ & \textcolor{taupegray}{$2\ell+6j_l$}/\textcolor{darkpastelblue}{$\ell+6j_l+ \ETmiss$} & \textcolor{taupegray}{$2\ell+2b+4j_l$}/\textcolor{darkpastelblue}{$\ell+2b+4j_l+\ETmiss$}\\ 

$\tilde{d}$ &\textcolor{taupegray}{$2\ell + 2j_l$}/\textcolor{darkpastelblue}{$\ell+2j_l+\ETmiss$} & \textcolor{taupegray}{$2\ell+2b+4j_l$}/\textcolor{darkpastelblue}{$\ell+2b+4j_l+\ETmiss$} \\

$\tilde{t}_L(\tilde{b}_L)$ & \textcolor{taupegray}{$2\ell+4j_l+2j_3$}/\textcolor{darkpastelblue}{$\ell+4j_l+2j_3+\ETmiss$} & \textcolor{taupegray}{$2\ell+2b+2j_l+2j_3$}/\textcolor{darkpastelblue}{$\ell+2b+2j_l+2j_3+\ETmiss$} \\

$\tilde{t}_R$ & \textcolor{taupegray}{$2\ell+2t+4j_l$}/\textcolor{darkpastelblue}{$\ell+2t+4j_l+\ETmiss$} & \textcolor{taupegray}{$2\ell+2t+2b+2j_l$}/\textcolor{darkpastelblue}{$\ell+2t+2b+2j_l+\ETmiss$} \\

$\tilde{b}_R$ & \textcolor{taupegray}{$2\ell+2b+4j_l$}/\textcolor{darkpastelblue}{$\ell+2b+4j_l+\ETmiss$} & \textcolor{taupegray}{$2\ell+2j_l$}/\textcolor{darkpastelblue}{$\ell+2j_l+\ETmiss$}\\

$\tilde{B}, \; \tilde{W}, \;\tilde{H}$ & \textcolor{taupegray}{$2\ell+4j_l$}/\textcolor{darkpastelblue}{$\ell+4j_l+\ETmiss$} & \textcolor{taupegray}{$2\ell+2b+2j_l$}/\textcolor{darkpastelblue}{$\ell+2b+2j_l+\ETmiss$}\\
\hline \hline
\end{tabular}
\end{adjustbox}}
\label{tab:LQD1}
\end{table}

\begin{table}[hp]
\caption{Same as~\cref{tab:LQD1} but for $L_iQ_j\Bar{D}_k$ operators with $i,k\in\{1,2\}$ and $j=3$.}
\centering
\scalebox{0.75}{
\renewcommand{\arraystretch}{1.3}
\begin{adjustbox}{width=1.3\textwidth}
\begin{tabular}{cc}
\hline \hline
{\bf LSP} &  $\mathbf{LQ_3D} $ \\
\hline \hline

$\tilde{\ell}\left(\tilde{\nu}\right)$ & \textcolor{darkolivegreen}{$2j_l + 2j_3$}\\

$\tilde{e}$ &  \textcolor{vegasgold}{$4\ell+2t+2j_l$}/\textcolor{atomictangerine}{$3\ell+t+b+2j_l+\ETmiss$}/\textcolor{taupegray}{$2\ell+2b+2j_l+\ETmiss$} \\

\multirow{2}{*}{$\tilde{\tau}_L\left(\tilde{\nu}_\tau\right)$} &  \textcolor{vegasgold}{$2\ell+2\tau+2t+2j_l$}/\textcolor{atomictangerine}{$\ell+2\tau+t+b+2j_l+\ETmiss$}/\textcolor{taupegray}{$2\tau+2b+2j_l+\ETmiss$}/\textcolor{atomictangerine}{$2\ell+\tau+2t+2j_l+\ETmiss$}/\\
& \textcolor{taupegray}{$\ell+\tau+t+b+2j_l+\ETmiss$}/\textcolor{darkpastelblue}{$\tau+2b+2j_l+\ETmiss$}\\

$\tilde{\tau}_R$ &  \textcolor{vegasgold}{$2\ell+2\tau+2t+2j_l$}/\textcolor{atomictangerine}{$\ell+2\tau+t+b+2j_l+\ETmiss$}/\textcolor{taupegray}{$2\tau+2b+2j_l+\ETmiss$}\\

$\tilde{g}$ & \textcolor{taupegray}{$2\ell+2t+2j_l$}/\textcolor{darkpastelblue}{$\ell+t+b+2j_l+\ETmiss$}\\

$\tilde{q}$ & \textcolor{taupegray}{$2\ell+2t+4j_l$}/\textcolor{darkpastelblue}{$\ell+t+b+4j_l+\ETmiss$} \\

$\tilde{u}$ & \textcolor{taupegray}{$2\ell+2t+4j_l$}/\textcolor{darkpastelblue}{$\ell+t+b+4j_l+\ETmiss$} \\

$\tilde{d}$ &  \textcolor{taupegray}{$2\ell+2t$}/\textcolor{darkpastelblue}{$\ell+t+b+\ETmiss$} \\

$\tilde{t}_L(\tilde{b}_L)$ & \textcolor{taupegray}{$2\ell+2j_l$}\\

$\tilde{t}_R$ & \textcolor{taupegray}{$2\ell+4t+2j_l$}/\textcolor{darkpastelblue}{$\ell+3t+b+2j_l+\ETmiss$}\\

$\tilde{b}_R$ & \textcolor{taupegray}{$2\ell+2t+2b+2j_l$}/\textcolor{darkpastelblue}{$\ell+t+3b+2j_l+\ETmiss$}\\

$\tilde{B}, \; \tilde{W}, \;\tilde{H}$ & \textcolor{taupegray}{$2\ell+2j_l+2j_3$}/\textcolor{darkpastelblue}{$\ell+2j_l+2j_3+\ETmiss$}/\textcolor{cardinal}{$2b+2j_l+\ETmiss$}\\
\hline \hline
\end{tabular}
\end{adjustbox}}
\label{tab:LQD2}
\end{table}

\begin{table}[hp]
\caption{Same as~\cref{tab:LQD1} but for $L_iQ_j\Bar{D}_k$ operators with $j,k=3$ and $i\in\{1,2\}$.}
\centering
\scalebox{0.75}{
\renewcommand{\arraystretch}{1.3}
\begin{adjustbox}{width=1.3\textwidth}
\begin{tabular}{cc}
\hline \hline
{\bf LSP} &   $\mathbf{LQ_3D_3}  $\\
\hline \hline

$\tilde{\ell}\left(\tilde{\nu}\right)$  & \textcolor{darkolivegreen}{$2b+2j_3$}\\

$\tilde{e}$  & \textcolor{vegasgold}{$4\ell+2t+2b$}/\textcolor{atomictangerine}{$3\ell+t+3b+\ETmiss$}/\textcolor{taupegray}{$2\ell+4b+\ETmiss$} \\

\multirow{2}{*}{$\tilde{\tau}_L\left(\tilde{\nu}_\tau\right)$} &  \textcolor{vegasgold}{$2\ell+2\tau+2t+2b$}/\textcolor{atomictangerine}{$\ell+2\tau+t+3b+\ETmiss$}/\textcolor{taupegray}{$2\tau+4b+\ETmiss$}/\textcolor{atomictangerine}{$2\ell+\tau+2t+2b+\ETmiss$}/\\
& \textcolor{taupegray}{$\ell+\tau+t+3b+\ETmiss$}/\textcolor{darkpastelblue}{$\tau+4b+\ETmiss$}\\

$\tilde{\tau}_R$ & \textcolor{vegasgold}{$2\ell+2\tau+2t+2b$}/\textcolor{atomictangerine}{$\ell+2\tau+t+3b+\ETmiss$}/\textcolor{taupegray}{$2\tau+4b+\ETmiss$}\\

$\tilde{g}$ & \textcolor{taupegray}{$2\ell+2t+2b$}/\textcolor{darkpastelblue}{$\ell+t+3b+\ETmiss$} \\

$\tilde{q}$ & \textcolor{taupegray}{$2\ell+2t+2b+2j_l$}/\textcolor{darkpastelblue}{$\ell+t+3b+2j_l+\ETmiss$} \\

$\tilde{u}$ & \textcolor{taupegray}{$2\ell+2t+2b+2j_l$}/\textcolor{darkpastelblue}{$\ell+t+3b+2j_l+\ETmiss$} \\

$\tilde{d}$ & \textcolor{taupegray}{$2\ell+2t+2b+2j_l$}/\textcolor{darkpastelblue}{$\ell+t+3b+2j_l+\ETmiss$} \\

$\tilde{t}_L(\tilde{b}_L)$ & \textcolor{taupegray}{$2\ell+2b$} \\

$\tilde{t}_R$ & \textcolor{taupegray}{$2\ell+4t+2b$}/\textcolor{darkpastelblue}{$\ell+3t+3b+\ETmiss$}\\

$\tilde{b}_R$ & \textcolor{taupegray}{$2\ell+2t$}/\textcolor{darkpastelblue}{$\ell+t+b+\ETmiss$}\\

$\tilde{B}, \; \tilde{W}, \;\tilde{H}$ & \textcolor{taupegray}{$2\ell+2b+2j_3$}/\textcolor{darkpastelblue}{$\ell+2b+2j_3+\ETmiss$}/\textcolor{cardinal}{$4b+\ETmiss$}\\
\hline \hline
\end{tabular}
\end{adjustbox}}
\label{tab:LQD3}
\end{table}

\begin{table}[h]
\caption{Same as~\cref{tab:LQD1} but for $L_iQ_j\Bar{D}_k$ operators with $j,k\in\{1,2\}$ and $i=3$ (second column), and $i,k=3$ and $j\in\{1,2\}$ (third column).}
\centering
\scalebox{0.75}{
\renewcommand{\arraystretch}{1.3}
\begin{adjustbox}{width=1.3\textwidth}
\begin{tabular}{ccc}
\hline \hline
{\bf LSP} & $\mathbf{L_3QD} $ & $\mathbf{L_3QD_3} $\\

\hline \hline

\multirow{2}{*}{$\tilde{\ell}\left(\tilde{\nu}\right)$} & \textcolor{vegasgold}{$2\ell+2\tau+4j_l$}/\textcolor{atomictangerine}{$2\ell+\tau+4j_l+\ETmiss$}/ & \textcolor{vegasgold}{$2\ell+2\tau+2b+2j_l$}/\textcolor{atomictangerine}{$2\ell+\tau+2b+2j_l+\ETmiss$}/ \\

& \textcolor{atomictangerine}{$\ell+2\tau+4j_l+\ETmiss$}/\textcolor{taupegray}{$\ell+\tau+4j_l+\ETmiss$} & \textcolor{atomictangerine}{$\ell+2\tau+2b+2j_l+\ETmiss$}/\textcolor{taupegray}{$\ell+\tau+2b+2j_l+\ETmiss$} \\

$\tilde{e}$ & \textcolor{vegasgold}{$2\ell+2\tau+4j_l$}/\textcolor{atomictangerine}{$2\ell+\tau+4j_l+\ETmiss$} & \textcolor{vegasgold}{$2\ell+2\tau+2b+2j_l$}/\textcolor{atomictangerine}{$2\ell+\tau+2b+2j_l+\ETmiss$} \\

$\tilde{\tau}_L\left(\tilde{\nu}_\tau\right)$ & \textcolor{darkolivegreen}{$4j_l$} & \textcolor{darkolivegreen}{$2b + 2j_l$} \\

$\tilde{\tau}_R$ & \textcolor{vegasgold}{$4\tau+4j_l$}/\textcolor{atomictangerine}{$3\tau+4j_l+\ETmiss$} & \textcolor{vegasgold}{$4\tau+2b+2j_l$}/\textcolor{atomictangerine}{$3\tau+2b+2j_l+\ETmiss$} \\

$\tilde{g}$ & \textcolor{taupegray}{$2\tau+4j_l$}/\textcolor{darkpastelblue}{$\tau+4j_l+\ETmiss$} & \textcolor{taupegray}{$2\tau+2b+2j_l$}/\textcolor{darkpastelblue}{$\tau+2b+2j_l + \ETmiss$} \\

$\tilde{q}$ & \textcolor{taupegray}{$2\tau + 2j_l$} & \textcolor{taupegray}{$2\tau+2b$} \\

$\tilde{u}$ & \textcolor{taupegray}{$2\tau+6j_l$}/\textcolor{darkpastelblue}{$\tau+6j_l+ \ETmiss$} & \textcolor{taupegray}{$2\tau+2b+4j_l$}/\textcolor{darkpastelblue}{$\tau+2b+4j_l+\ETmiss$}\\ 

$\tilde{d}$ & \textcolor{taupegray}{$2\tau + 2j_l$}/\textcolor{darkpastelblue}{$\tau+2j_l+\ETmiss$} & \textcolor{taupegray}{$2\tau+2b+4j_l$}/\textcolor{darkpastelblue}{$\tau+2b+4j_l+\ETmiss$} \\

$\tilde{t}_L(\tilde{b}_L)$ &\textcolor{taupegray}{ $2\tau+4j_l+2j_3$}/\textcolor{darkpastelblue}{$\tau+4j_l+2j_3+\ETmiss$} & \textcolor{taupegray}{$2\tau+2b+2j_l+2j_3$}/\textcolor{darkpastelblue}{$\tau+2b+2j_l+2j_3+\ETmiss$} \\

$\tilde{t}_R$ & \textcolor{taupegray}{$2\tau+2t+4j_l$}/\textcolor{darkpastelblue}{$\tau+2t+4j_l+\ETmiss$} & \textcolor{taupegray}{$2\tau+2t+2b+2j_l$}/\textcolor{darkpastelblue}{$\tau+2t+2b+2j_l+\ETmiss$} \\

$\tilde{b}_R$ & \textcolor{taupegray}{$2\tau+2b+4j_l$}/\textcolor{darkpastelblue}{$\tau+2b+4j_l+\ETmiss$} & \textcolor{taupegray}{$2\tau+2j_l$}/\textcolor{darkpastelblue}{$\tau+2j_l+\ETmiss$}\\

$\tilde{B}, \; \tilde{W}, \;\tilde{H}$ & \textcolor{taupegray}{$2\tau+4j_l$}/\textcolor{darkpastelblue}{$\tau+4j_l+\ETmiss$} & \textcolor{taupegray}{$2\tau+2b+2j_l$}/\textcolor{darkpastelblue}{$\tau+2b+2j_l+\ETmiss$}\\
\hline \hline
\end{tabular}
\end{adjustbox}}

\label{tab:LQD4}
\end{table}

\begin{table}[h]
\caption{Same as~\cref{tab:LQD1} but for $L_iQ_j\Bar{D}_k$ operators with $i,j=3$ and $k\in\{1,2\}$.}
\centering
\scalebox{0.75}{
\renewcommand{\arraystretch}{1.3}
\begin{adjustbox}{width=1.3\textwidth}
\begin{tabular}{cc}
\hline \hline
{\bf LSP} &  $\mathbf{L_3Q_3D} $ \\
\hline \hline

\multirow{2}{*}{$\tilde{\ell}\left(\tilde{\nu}\right)$} & \textcolor{vegasgold}{$2\ell+2\tau+2t+2j_l$}/\textcolor{atomictangerine}{$2\ell+\tau+t+b+2j_l+\ETmiss$}/\textcolor{taupegray}{$2\ell+2b+2j_l+\ETmiss$}/\textcolor{atomictangerine}{$\ell+2\tau+2t+2j_l+\ETmiss$}/\\
& \textcolor{taupegray}{$\ell+\tau+t+b+2j_l+\ETmiss$}/\textcolor{darkpastelblue}{$\ell+2b+2j_l+\ETmiss$}\\

$\tilde{e}$ & \textcolor{vegasgold}{$2\ell+2\tau+2t+2j_l$}/\textcolor{atomictangerine}{$2\ell+\tau+t+b+2j_l+\ETmiss$}/\textcolor{taupegray}{$2\ell+2b+2j_l+\ETmiss$} \\

$\tilde{\tau}_L\left(\tilde{\nu}_\tau\right)$  & \textcolor{darkolivegreen}{$2j_l+2j_3 $}\\

$\tilde{\tau}_R$ &  \textcolor{vegasgold}{$4\tau+2t+2j_l$}/\textcolor{atomictangerine}{$3\tau+t+b+2j_l+\ETmiss$}/\textcolor{taupegray}{$2\tau+2b+2j_l+\ETmiss$}\\

$\tilde{g}$ & \textcolor{taupegray}{$2\tau+2t+2j_l$}/\textcolor{darkpastelblue}{$\tau+t+b+2j_l+\ETmiss$}\\

$\tilde{q}$ & \textcolor{taupegray}{$2\tau+2t+4j_l$}/\textcolor{darkpastelblue}{$\tau+t+b+4j_l+\ETmiss$} \\

$\tilde{u}$ & \textcolor{taupegray}{$2\tau+2t+4j_l$}/\textcolor{darkpastelblue}{$\tau+t+b+4j_l+\ETmiss$} \\

$\tilde{d}$ &  \textcolor{taupegray}{$2\tau+2t$}/\textcolor{darkpastelblue}{$\tau+t+b+\ETmiss$} \\

$\tilde{t}_L(\tilde{b}_L)$ & \textcolor{taupegray}{$2\tau+2j_l$} \\

$\tilde{t}_R$ & \textcolor{taupegray}{$2\tau+4t+2j_l$}/\textcolor{darkpastelblue}{$\tau+3t+b+2j_l+\ETmiss$}\\

$\tilde{b}_R$ & \textcolor{taupegray}{$2\tau+2t+2b+2j_l$}/\textcolor{darkpastelblue}{$\tau+t+3b+2j_l+\ETmiss$}\\

$\tilde{B}, \; \tilde{W}, \;\tilde{H}$ & \textcolor{taupegray}{$2\tau+2j_l+2j_3$}/\textcolor{darkpastelblue}{$\tau+2j_l+2j_3+\ETmiss$}/\textcolor{cardinal}{$2b+2j_l+\ETmiss$}\\
\hline \hline
\end{tabular}
\end{adjustbox}}

\label{tab:LQD5}
\end{table}

\begin{table}[h]
\caption{Same as~\cref{tab:LQD1} but for $L_iQ_j\Bar{D}_k$ operators with $i,j,k=3$.}
\centering
\scalebox{0.75}{
\renewcommand{\arraystretch}{1.3}
\begin{adjustbox}{width=1.3\textwidth}
\begin{tabular}{cc}
\hline \hline
{\bf LSP} &   $\mathbf{L_3Q_3D_3}  $\\
\hline \hline

\multirow{2}{*}{$\tilde{\ell}\left(\tilde{\nu}\right)$} & \textcolor{vegasgold}{$2\ell+2\tau+2t+2b$}/\textcolor{atomictangerine}{$2\ell+\tau+t+3b+\ETmiss$}/\textcolor{taupegray}{$2\ell+4b+\ETmiss$}/\textcolor{atomictangerine}{$\ell+2\tau+2t+2b+\ETmiss$}/\\
& \textcolor{taupegray}{$\ell+\tau+t+3b+\ETmiss$}/\textcolor{darkpastelblue}{$\ell+4b+\ETmiss$}\\

$\tilde{e}$ & \textcolor{vegasgold}{$2\ell+2\tau+2t+2b$}/\textcolor{atomictangerine}{$2\ell+\tau+t+3b+\ETmiss$}/\textcolor{taupegray}{$2\ell+4b+\ETmiss$} \\

$\tilde{\tau}_L\left(\tilde{\nu}_\tau\right)$  & \textcolor{darkolivegreen}{$2b+2j_3$}\\

$\tilde{\tau}_R$ & \textcolor{vegasgold}{$4\tau+2t+2b$}/\textcolor{atomictangerine}{$3\tau+t+3b+\ETmiss$}/\textcolor{taupegray}{$2\tau+4b+\ETmiss$}\\

$\tilde{g}$ & \textcolor{taupegray}{$2\tau+2t+2b$}/\textcolor{darkpastelblue}{$\tau+t+3b+\ETmiss$} \\

$\tilde{q}$ & \textcolor{taupegray}{$2\tau+2t+2b+2j_l$}/\textcolor{darkpastelblue}{$\tau+t+3b+2j_l+\ETmiss$} \\

$\tilde{u}$ & \textcolor{taupegray}{$2\tau+2t+2b+2j_l$}/\textcolor{darkpastelblue}{$\tau+t+3b+2j_l+\ETmiss$} \\

$\tilde{d}$ & \textcolor{taupegray}{$2\tau+2t+2b+2j_l$}/\textcolor{darkpastelblue}{$\tau+t+3b+2j_l+\ETmiss$} \\

$\tilde{t}_L(\tilde{b}_L)$ & \textcolor{taupegray}{$2\tau+2b$} \\

$\tilde{t}_R$ & \textcolor{taupegray}{$2\tau+4t+2b$}/\textcolor{darkpastelblue}{$\tau+3t+3b+\ETmiss$}\\

$\tilde{b}_R$ & \textcolor{taupegray}{$2\tau+2t$}/\textcolor{darkpastelblue}{$\tau+t+b+\ETmiss$}\\

$\tilde{B}, \; \tilde{W}, \;\tilde{H}$ & \textcolor{taupegray}{$2\tau+2b+2j_3$}/\textcolor{darkpastelblue}{$\tau+2b+2j_3+\ETmiss$}/\textcolor{cardinal}{$4b+\ETmiss$}\\
\hline \hline
\end{tabular}
\end{adjustbox}}

\label{tab:LQD6}
\end{table}


\section{Sample Application of the Framework: LLE Couplings}\label{sec:results}
\subsection{Benchmark Scenarios}
We now demonstrate the practical application of our framework by using it to calculate 
mass bounds on SUSY particles in a wide range of RPV scenarios. Throughout this section, for simplicity, we assume that the only non-zero RPV coupling corresponds to a single $LL\Bar{E}$ operator, although -- as mentioned -- generalization to several non-zero RPV couplings is possible by combining the different rows of our signature tables. Further, we assume all mass eigenstates are aligned 
with the gauge eigenstates, except for the neutral Higgsinos which are assumed to be 
maximally mixed.

As discussed in~\cref{sec:classification}, the signatures in our `RPV Dictionary' 
have significant coverage through existing \texttt{ATLAS} and \texttt{CMS} searches, even if only indirectly. We can, therefore, 
reinterpret these searches in the context of RPV scenarios to set limits on the latter. In the $LL\bar E$ scenario, these can be comparable to or even more constraining than the MSSM limits.

In order to calculate the mass limits, we have simulated SUSY processes at leading order using the program \texttt{MadGraph5\_aMC@NLO}~\cite{Alwall:2014hca} linked to 
\texttt{PYTHIA\;8.2}~\cite{Sjostrand:2014zea}. We have employed the \texttt{UFO} 
\texttt{RPV-MSSM} model file available at Ref.~\cite{FR:RPVMSSM}. The decays are 
computed under the narrow-width approximation. The branching ratios for two-body decays
are computed by \texttt{MadGraph5\_aMC@NLO}, while for higher-multiplicity decays -- to 
save computational time -- we set them by hand; the details of how we do this are given in~\cref{sec:A}. The width is always set by hand to a small arbitrary value (smaller than the experimental resolution) such that the decay of the LSP remains prompt; under the narrow-width approximation, the results are independent of the number. \texttt{PYTHIA\;8.2} then produces the final decayed and showered event samples. These are passed through \texttt{CheckMATE\;2}~\cite{Dercks:2016npn,Cacciari:2011ma,Cacciari:2005hq,Cacciari:2008gp,Read:2002hq} which uses a database of several existing \texttt{ATLAS} and \texttt{CMS} analyses in order to determine whether the RPV-MSSM parameter point used to generate the event sample can be excluded or not.\footnote{We note that we limit ourselves to the analyses already implemented in \texttt{CheckMATE\;2} as of December 2022; the list of implemented analyses can be found at Ref.~\cite{CM:WS}. Some analyses explicitly targeting $LL\Bar{E}$ models such as the most relevant SRs from Ref.~\cite{ATLAS:2021yyr} are not implemented. Despite this, we observe excellent coverage.} Detector effects have been accounted for through the \texttt{DELPHES\;3}~\cite{deFavereau:2013fsa} detector simulation module linked with \texttt{CheckMATE\;2}. 

We now describe the various benchmark scenarios we study. These have been designed to 
cover what, we believe, should be all relevant possibilities for the $LL\Bar{E}$ case, 
subject to our minimal assumptions.

\paragraph{Gluino LSP:} The first set of scenarios we consider is with a gluino LSP. It is sufficient to consider only direct gluino-pair 
production since the cross-section is higher than any channel where the gluino LSP is produced in cascade decays (see discussion in~\cref{sec:B}). Thus, in our simulation, we consider the rest of 
the spectrum to be decoupled; this gives us the most conservative, model-independent exclusion limits. However, we assume that, despite this decoupling, the gluino LSP still decays promptly; see~\cref{sec:A} for details on the specific decay modes chosen in the simulation. In the first three scenarios, we consider $\lambda_{121}$ to be the only non-zero RPV operator. The characteristic signature 
for the gluino decay is $2\ell+2j+\ETmiss$, \textit{cf.}~\cref{tab:LLE1}. Here $j$ can be a light, top, or bottom jet depending on the nature of the virtual squark involved in the decay; the 
three scenarios target the possible dependence of the coverage on this choice. Next, to study how the results are affected if the RPV operator leads to more muons or taus instead of 
electrons, we consider three more scenarios corresponding to $\lambda_{122}, \lambda_{311}$, and $\lambda_{313}$, respectively, being the sole non-zero RPV couplings. The details of all gluino benchmarks have been summarized in~\cref{tab:gluinoBMs}.  

\begin{table}[htpb]
\caption{ Details of our benchmarks: the first two columns depict the LSP and the production mode considered, respectively; the RPV coupling assumed to be non-zero is shown in the third column; the fourth column represents the possible decays of the LSP (these are split into two columns for better readability); the last column shows the notation we use for labeling the scenario.}
\centering
\begin{tabular}{ccccccc}
\hline \hline
{\bf LSP}& {\bf Production} & {\bf Coupling} & \multicolumn{2}{c}{\bf LSP Decay} &{\bf Label} \\
\hline \hline
\multirow{8}{*}{$\tilde{g}$} & Direct & $\lambda_{121}$ & $2e + 2 j_l + \nu_\mu$ & $e + \mu + 2j_l + \nu_e $ & $\text{D}^{e\mu e}_{\tilde{g}}$\\[0.18 cm]
& Direct & $\lambda_{121}$ & $2e + 2b + \nu_\mu$ & $e + \mu + 2b + \nu_e $ &$\text{D}^{e\mu e-b}_{\tilde{g}}$\\[0.18 cm]
& Direct &$\lambda_{121}$ & $2e + 2t + \nu_\mu$ & $e + \mu + 2t + \nu_e $ & $\text{D}^{e\mu e-t}_{\tilde{g}}$\\[0.18 cm]
& Direct & $\lambda_{122}$ & $2\mu + 2j_l + \nu_e$ & $e + \mu  + 2j_l + \nu_\mu $ & $\text{D}^{e\mu \mu}_{\tilde{g}}$\\[0.18 cm]
& Direct & $\lambda_{311}$ & $2e + 2j_l + \nu_\tau$ & $e + \tau + 2j_l + \nu_e $ & $\text{D}^{\tau e e}_{\tilde{g}}$\\[0.18 cm]
& Direct & $\lambda_{313}$ & $2\tau + 2j_l + \nu_e$ & $e + \tau + 2j_l + \nu_\tau $ & $\text{D}^{\tau e \tau}_{\tilde{g}}$ \\[0.18 cm]
\hline \hline
\end{tabular}
\label{tab:gluinoBMs}
\end{table}

\paragraph{Squark LSPs:} Similar to above, for squark LSP scenarios, we first consider only direct pair production with the other sparticles decoupled. Thus, we have selected two scenarios each for the light-flavor squarks ($\tilde{q},\; \tilde{u},\; \tilde{d}$), and the heavy-flavor ones ($\tilde{q}_3,\;\tilde{t},\;\tilde{b}$), corresponding to the couplings $\lambda_{121}$ and $\lambda_{313}$.\footnote{These couplings correspond to the 
two extreme cases: maximum and minimum number of light leptons in the final state. For the remaining scenarios we will only consider these cases; the results for the other coupling configurations 
can be interpolated from the gluino LSP results.} All squarks within a particular scenario are considered mass-degenerate for simplicity. 
    
For the squarks, cascade decays involving gluino production channels (pair as well as associated) can also be relevant since these can have a higher cross-section than the direct production channels, \textit{cf.} the discussion in~\cref{sec:B}. Thus, we include four more scenarios -- covering the two couplings for each of the two squark groups -- 
where the gluino and squarks are both kinematically accessible, while the rest of the spectrum is decoupled (again, in a way that the squarks still decay promptly according to branching ratios described in~\cref{sec:A}). The corresponding results are presented as two-dimensional plots in the gluino mass vs. squark mass plane. The details of all the squark benchmarks have been summarized in~\cref{tab:qBMs}. 

\begin{table}[htpb]
\caption{As in \cref{tab:gluinoBMs} but for the squark LSP benchmarks.}
\centering
\begin{tabular}{ccccccc}
\hline \hline
{\bf LSP}& {\bf Production} & {\bf Coupling} & \multicolumn{2}{c}{\bf LSP Decay} &{\bf Label} \\
\hline \hline
\multirow{5.75}{*}{$\tilde{q}/\tilde{u}/\tilde{d}$} 
& Direct & \multirow{2.5}{*}{$\lambda_{121}$} & \multirow{2.5}{*}{$2 e + j_l + \nu_\mu$} & \multirow{2.5}{*}{$ e + \mu + j_l + \nu_e$} & $\text{D}^{{e\mu e}}_{\tilde{q}}$\\[0.18 cm]
& $\tilde{g}$ &  &  &  & $\text{I}^{e\mu e}_{\tilde{g}\shortto\tilde{q}}$\\[0.2 cm] \cmidrule{2-6}
& Direct & \multirow{2.5}{*}{$\lambda_{313}$} & \multirow{2.5}{*}{$ 2 \tau + j_l + \nu_e$} & \multirow{2.5}{*}{$ e + \tau + j_l + \nu_\tau$} & $\text{D}^{{\tau e \tau}}_{\tilde{q}}$\\[0.18 cm]
& $\tilde{g}$ & & & & $\text{I}^{\tau e \tau}_{\tilde{g}\shortto\tilde{q}}$\\[0.3 cm]
\hline \\[-0.28 cm]
\multirow{5.75}{*}{$\tilde{q}_3/\tilde{t}/\tilde{b}$} & Direct & \multirow{2.5}{*}{$\lambda_{121}$} & \multirow{2.5}{*}{$2e  + j_3 + \nu_\mu$} & \multirow{2.5}{*}{$ e + \mu  + j_3 + \nu_e$} & $\text{D}^{{e\mu e}}_{\tilde{q}_3}$\\[0.18 cm]
& $\tilde{g}$ &  &  &  & $\text{I}^{e\mu e}_{\tilde{g}\shortto\tilde{q}_3}$\\[0.2 cm] \cmidrule{2-6}
& Direct & \multirow{2.5}{*}{$\lambda_{313}$} & \multirow{2.5}{*}{$ 2 \tau  + j_3 + \nu_e$} & \multirow{2.5}{*}{$ e + \tau  + j_3 + \nu_\tau$} & $\text{D}^{{\tau e \tau}}_{\tilde{q}_3}$  \\[0.18 cm]
& $\tilde{g}$ &  &  &  & $\text{I}^{\tau e \tau}_{\tilde{g}\shortto\tilde{q}_3}$\\[0.18 cm]
\hline \hline
\end{tabular}

\label{tab:qBMs}
\end{table}

\paragraph{Electroweakino LSPs:} For the electroweakinos, we study three sets of scenarios corresponding to the winos ($\tilde{W}$), the Higgsinos ($\tilde{H}$), or the bino ($\tilde{B}$) being the LSP(s), respectively. 
    
For the winos and the higgsinos, as before, we look at scenarios focusing on the direct modes, as well as the relevant indirect modes mentioned in~\cref{sec:B}. For winos, the latter includes production of gluinos, light-flavor squarks, or heavy-flavor squarks. However, the latter two scenarios have similar features, so we only focus on the light-flavor squarks. For the higgsinos, we include only production of gluinos and the heavy-flavor squarks since their coupling to the light-flavor squarks is suppressed. 
    
For the bino, direct production is not relevant due to the small cross-section, and thus we only study indirect modes. This time, we need to consider the possibility of each of the other SUSY particles being the parent: this includes the colored sector, the winos, the Higgsinos, and the sleptons.
 
As before, apart from the LSP(s) and the relevant parent sparticle(s), all other SUSY fields are considered decoupled, in a way that the LSP decay remains prompt. We study scenarios corresponding to both $\lambda_{121}$ and $\lambda_{313}$. The details for all benchmarks corresponding to electroweakino LSPs have been summarized in~\cref{tab:ewBMs}. 

\paragraph{Slepton LSPs:} Finally, we have the slepton LSP scenarios. For each case -- light-flavor sleptons ($\tilde{\ell}/\tilde{\nu}/\tilde{e}$), and heavy-flavor sleptons ($\tilde{\tau}_L/\tilde{\nu}_{\tau}/\tilde{\tau}_R$) -- we study direct and indirect 
production, once again for the couplings $\lambda_{121}$ 
and $\lam_{313}$. The relevant indirect modes include every sparticle except the Bino, \textit{cf.} discussion in~\cref{sec:B}. We only study scenarios with $\tilde{g}$ or $\tilde{W}$ parents; results for other colored sparticles or electroweakinos can be interpolated.

Unlike sparticles considered so far, sleptons can couple directly to the $LL\Bar{E}$ operators, depending on the flavor configuration. This can significantly affect the decay modes for a given slepton. To study this effect, we also include scenarios with non-zero $\lambda_{122}$ and $\lambda_{311}$ for both slepton classes. The details of the slepton benchmarks are summarized in~\cref{tab:lBMs}.

\begin{table}[htpb]
\caption{As in \cref{tab:gluinoBMs} but for the electroweakino LSP benchmarks.}
\centering
\begin{tabular}{ccccccc}
\hline \hline
{\bf LSP}& {\bf Production} & {\bf Coupling} & \multicolumn{2}{c}{\bf LSP Decay} &{\bf Label} \\
\hline \hline
\multirow{8.35}{*}{$\tilde{W}$} & Direct & \multirow{3.75}{*}{$\lambda_{121}$} & \multirow{3.75}{*}{$2 e + \nu_\mu/2e + \mu$} & \multirow{3.75}{*}{$  e + \mu + \nu_{e}/e+\nu_e+\nu_\mu $} & $\text{D}^{{e\mu e}}_{\tilde{W}}$\\[0.18 cm]
& $\tilde{g}$ &  &  &  & $\text{I}^{e \mu e}_{\tilde{g}\shortto\tilde{W}}$\\[0.18 cm]
& $\tilde{q}/\tilde{u}/\tilde{d}$ &  &  &  & $\text{I}^{e \mu e}_{\tilde{q}\shortto\tilde{W}}$\\[0.18 cm]
\cmidrule{2-6}
& Direct & \multirow{3.75}{*}{$\lambda_{313}$} & \multirow{3.75}{*}{$2 \tau + \nu_e/ e+2\tau$} & \multirow{3.75}{*}{$ e + \tau + \nu_\tau/\tau+\nu_e+\nu_\tau$} & $\text{D}^{{\tau e \tau}}_{\tilde{W}}$\\[0.18 cm]
& $\tilde{g}$ &  &  &  & $\text{I}^{\tau e \tau}_{\tilde{g}\shortto\tilde{W}}$\\[0.18 cm]
& $\tilde{q}/\tilde{u}/\tilde{d}$ & &  & & $\text{I}^{\tau e \tau}_{\tilde{q}\shortto\tilde{W}}$\\[0.18 cm]
\hline \\[-0.28 cm]
\multirow{8.35}{*}{$\tilde{H}$} & Direct & \multirow{3.75}{*}{$\lambda_{121}$} & \multirow{3.75}{*}{$2e + V + \nu_\mu$} & \multirow{3.75}{*}{$e + \mu + V + \nu_{e} $} & $\text{D}^{{e\mu e}}_{\tilde{H}}$\\[0.18 cm]
& $\tilde{g}$ &  &  &  & $\text{I}^{e \mu e}_{\tilde{g}\shortto\tilde{H}}$\\[0.18 cm]
& $\tilde{q}_3/\tilde{t}/\tilde{b}$ &  & &  & $\text{I}^{e \mu e}_{\tilde{q}_3\shortto\tilde{H}}$\\[0.2 cm] \cmidrule{2-6}
& Direct & \multirow{3.75}{*}{$\lambda_{313}$} & \multirow{3.75}{*}{$2\tau + V + \nu_e$} & \multirow{3.75}{*}{$e + \tau + V + \nu_\tau $} & $\text{D}^{{\tau e \tau}}_{\tilde{H}}$\\[0.18 cm]
& $\tilde{g}$ & &  &  & $\text{I}^{\tau e \tau}_{\tilde{g}\shortto\tilde{H}}$\\[0.18 cm]
& $\tilde{q}_3/\tilde{t}/\tilde{b}$ &  &  &  & $\text{I}^{\tau e \tau}_{\tilde{q}_3\shortto\tilde{H}}$\\[0.3 cm]
\hline \\[-0.28 cm]
\multirow{19.25}{*}{$\tilde{B}$} & $\tilde{g}$ & \multirow{9}{*}{$\lambda_{121}$} & \multirow{9}{*}{$2 e + \nu_\mu$} & \multirow{9}{*}{$  e + \mu + \nu_{e} $} & $\text{I}^{e \mu e}_{\tilde{g}\shortto\tilde{B}}$\\[0.18 cm]
& $\tilde{q}/\tilde{u}/\tilde{d}$ &  &  &  & $\text{I}^{e \mu e}_{\tilde{q}\shortto\tilde{B}}$\\[0.18 cm]
&$\tilde{q}_3/\tilde{t}/\tilde{b}$&   &  &  & $\text{I}^{e \mu e}_{\tilde{q}_3\shortto\tilde{B}}$\\[0.18 cm]
&$\tilde{\ell}/\tilde{\nu}/\tilde{e}$ & &  &  & $\text{I}^{e \mu e}_{\tilde{\ell}\shortto\tilde{B}}$\\[0.18 cm]
&$\tilde{\tau}_L/\tilde{\nu}_{\tau}/\tilde{\tau}_R$ & &  &  & $\text{I}^{e \mu e}_{\tilde{\tau}\shortto\tilde{B}}$\\[0.18 cm]
& $\tilde{W}$ &  &  &  & $\text{I}^{e \mu e}_{\tilde{W}\shortto\tilde{B}}$\\[0.18 cm]
& $\tilde{H}$ &  &  &  & $\text{I}^{e \mu e}_{\tilde{H}\shortto\tilde{B}}$\\[0.2 cm] \cmidrule{2-6}
& $\tilde{g}$ & \multirow{9}{*}{$\lambda_{313}$} & \multirow{9}{*}{$2\tau + \nu_e$} & \multirow{9}{*}{$  e + \tau + \nu_{\tau} $} & $\text{I}^{\tau e \tau}_{\tilde{g}\shortto\tilde{B}}$\\[0.18 cm]
& $\tilde{q}/\tilde{u}/\tilde{d}$ &  &  &  & $\text{I}^{\tau e \tau}_{\tilde{q}\shortto\tilde{B}}$\\[0.18 cm]
&$\tilde{q}_3/\tilde{t}/\tilde{b}$&   &  &  & $\text{I}^{\tau e \tau}_{\tilde{q}_3\shortto\tilde{B}}$\\[0.18 cm]
&$\tilde{\ell}/\tilde{\nu}/\tilde{e}$ &&  &  & $\text{I}^{\tau e \tau}_{\tilde{\ell}\shortto\tilde{B}}$\\[0.18 cm]
&$\tilde{\tau}_L/\tilde{\nu}_{\tau}/\tilde{\tau}_R$ & &  &  & $\text{I}^{\tau e \tau}_{\tilde{\tau}\shortto\tilde{B}}$\\[0.18 cm]
& $\tilde{W}$ &  &  &  & $\text{I}^{\tau e \tau}_{\tilde{W}\shortto\tilde{B}}$\\[0.18 cm]
& $\tilde{H}$ &  &  &  & $\text{I}^{\tau e \tau}_{\tilde{H}\shortto\tilde{B}}$\\[0.18 cm]
\hline \hline
\end{tabular}
\label{tab:ewBMs}
\end{table}

\begin{table}[htpb]
\caption{As in \cref{tab:gluinoBMs} but for the slepton LSP benchmarks. For brevity, we skip showing decay modes explicitly (indicated by $*$) for some sleptons that do not couple directly to the relevant RPV operator (\textit{e.g.}, $\tilde{\mu}_R$ LSP with $\lambda_{121}$). However, the details of how we include these modes in our simulations can be found in~\cref{sec:A}.}
\centering
\begin{adjustbox}{width=1\textwidth}
\begin{tabular}{ccccccc}
\hline \hline
{\bf LSP}& {\bf Production} & {\bf Coupling} & \multicolumn{2}{c}{\bf LSP Decay} &{\bf Label} \\
\hline \hline
\multirow{16}{*}{$\tilde{\ell}/\tilde{\nu}/\tilde{e}$} & Direct & \multirow{6.35}{*}{$\lambda_{121}$} & &  & $\text{D}^{{e\mu e}}_{\tilde{\ell}}$\\[0.18 cm]
&  &  & $2e/e+\mu$  & $ e+\nu_e/e + \nu_\mu$  & \\[0.18 cm]
&$\tilde{g}$ &  &  &  & $\text{I}^{e \mu e}_{\tilde{g}\shortto\tilde{\ell}}$\\[0.18 cm]
& &   &  $\mu+\nu_e$ & $*$ &\\[0.18 cm]
& $\tilde{W}$ &  &  & & $\text{I}^{e \mu e}_{\tilde{W}\shortto\tilde{\ell}}$\\[0.18 cm]
\cmidrule{2-6}
& Direct & \multirow{6.35}{*}{$\lambda_{313}$} &  \multirow{6.35}{*}{$2\tau/\tau + \nu_\tau$} &  \multirow{6.35}{*}{$*$}  & $\text{D}^{{\tau e \tau}}_{\tilde{\ell}}$\\[0.18 cm]
&  &  &  &  & \\[0.18 cm]
& $\tilde{g}$ &  &  &  & $\text{I}^{\tau e \tau}_{\tilde{g}\shortto\tilde{\ell}}$\\[0.18 cm]
&&   &  &  & \\[0.18 cm]
& $\tilde{W}$ &  &  &  & $\text{I}^{\tau e \tau}_{\tilde{W}\shortto\tilde{\ell}}$\\[0.18 cm]
\cmidrule{2-6}
& Direct & $\lambda_{122}$ & $ 2\mu/e+\mu/e+\nu_\mu$ & $\mu + \nu_e/\mu+\nu_\mu/*$ & $\text{D}^{{e\mu \mu}}_{\tilde{\ell}}$\\[0.2 cm] \cmidrule{2-6}
& Direct & $\lambda_{311}$ & $e+\tau/e+\nu_\tau$ & $\tau+\nu_e/*$ & $\text{D}^{{\tau e e}}_{\tilde{\ell}}$\\[0.3 cm]
\hline \\[-0.28 cm]
\multirow{16}{*}{$\tilde{\tau}_L/\tilde{\nu}_{\tau}/\tilde{\tau}_R$} &Direct & \multirow{6.35}{*}{$\lambda_{121}$} &  &  & $\text{D}^{{e\mu e}}_{\tilde{\tau}}$\\[0.18 cm]
&  &  & $ 2e + \tau + \nu_\mu$  & $2e + \nu_\mu + \nu_\tau$  & \\[0.18 cm]
& $\tilde{g}$ &  &  &  & $\text{I}^{e \mu e}_{\tilde{g}\shortto\tilde{\tau}}$\\[0.18 cm]
&&   & $e + \mu + \tau + \nu_e$ & $e +\mu + \nu_e + \nu_\tau$  & \\[0.18 cm]
& $\tilde{W}$ &  &  &  & $\text{I}^{e \mu e}_{\tilde{W}\shortto\tilde{\tau}}$\\[0.18 cm]
\cmidrule{2-6}
& Direct &  \multirow{6.35}{*}{$\lambda_{313}$} &  \multirow{6.35}{*}{$e+\tau / e + \nu_\tau $} &  \multirow{6.35}{*}{$\tau +\nu_e$} & $\text{D}^{{\tau e \tau}}_{\tilde{\tau}}$\\[0.18 cm]
&  &  &  &  & \\[0.18 cm]
& $\tilde{g}$ &  &  &  & $\text{I}^{\tau e \tau}_{\tilde{g}\shortto\tilde{\tau}}$\\[0.18 cm]
&&   &  &  & \\[0.18 cm]
& $\tilde{W}$ &  &  &  & $\text{I}^{\tau e \tau}_{\tilde{W}\shortto\tilde{\tau}}$\\[0.18 cm]
\cmidrule{2-6}
& Direct & $\lambda_{122}$ & $2\mu+ \tau + \nu_e/e+\mu+\tau+\nu_\mu$ & $2\mu+\nu_e+ \nu_\tau/e+\mu+\nu_\mu+\nu_\tau$ & $\text{D}^{{e\mu \mu}}_{\tilde{\tau}}$\\[0.2 cm] \cmidrule{2-6}
& Direct & $\lambda_{311}$ & $2e/e+\nu_e$ & $2e+\tau+\nu_\tau/e+2\tau+\nu_e$ & $\text{D}^{{\tau e e}}_{\tilde{\tau}}$\\[0.18 cm]
\hline \hline
\end{tabular}
\end{adjustbox}

\label{tab:lBMs}
\end{table} 

\clearpage

\subsection{Results}
Before presenting the numerical results of our simulations, we stress one important detail: even though our benchmarks correspond to simple scenarios where all sparticles other than the LSP (and NLSP) are decoupled, we expect our results to be more general. Since the characteristic signature from the LSP decay -- which provides the exclusion, as we show below -- is independent of the spectrum details, the sensitivity should only be slightly modified for scenarios with arbitrary sparticle mass spectra, as long as the objects in the characteristic topology do not become too soft.

We now discuss our results. All relevant details for the \texttt{ATLAS} and \texttt{CMS} searches implemented in \texttt{CheckMATE\;2} that show sensitivity to our scenarios have been summarized in~\cref{tab:searches} for reference. This list is merely meant to illustrate the searches with the strongest sensitivity and is not exhaustive. When there are multiple overlapping searches offering comparable sensitivity, we have omitted some of them.

\begin{table}[h!]
\caption{Summary of the most sensitive searches in our numerical simulations. The first column lists existing \texttt{ATLAS} and \texttt{CMS} searches providing sensitivity and our shorthand notation for each; the second column summarizes the relevant cuts; and the last column refers to 
the scenario labels presented in Tables~\ref{tab:gluinoBMs}-\ref{tab:lBMs}. We have color-coded the labels according to the final state topologies of~\cref{sec:classification}: \textcolor{chromeyellow}{$3L+\ETmiss$}, \textcolor{cadetblue}{$4L + (0-4)j + \ETmiss$}, and \textcolor{naplesyellow}{$5L + \ETmiss$}. The same searches also constrain the  $\text{I}_{\tilde{x}\shortto\tilde{p}}$ scenarios (not shown here).}
\centering
\begin{adjustbox}{width=1\textwidth}
\def\arraystretch{1.2}
    \begin{tabular}{c|c|c}
\hline \hline
Reference and search region & Representative cuts & Most sensitive for \\
\hline
\CMSewkfourlep~\cite{CMS:2017moi} SR \textbf{G05}  & $\geq 4\ell$, 0$b$, $\ETmiss$ & \textcolor{cadetblue}{$D^{e\mu e}_{\tilde{g}}$}, \textcolor{cadetblue}{$D^{e\mu \mu}_{\tilde{g}}$}, \textcolor{cadetblue}{$D^{e\mu e}_{\tilde{q},\tilde{W},\tilde{H}}$}, \textcolor{naplesyellow}{$D^{\tau e e}_{\tilde{\ell}}$}, \textcolor{naplesyellow}{$D^{e \mu e}_{\tilde{\tau}}$}, \textcolor{naplesyellow}{$D^{e \mu \mu}_{\tilde{\tau}}$} \\

\ATLASstrongSS~\cite{ATLAS:2019fag} SR \textbf{Rpv2L} & $\geq 2\ell$, $\geq 6j$  & \textcolor{cadetblue}{$D^{e\mu e}_{\tilde{g}}$}, \textcolor{cadetblue}{$D^{e\mu \mu}_{\tilde{g}}$}, \textcolor{cadetblue}{$D^{e\mu e-b}_{\tilde{g}}$}, \textcolor{cadetblue}{$D^{e\mu e-t}_{\tilde{g}}$}, \textcolor{cadetblue}{$D^{\tau e e}_{\tilde{g}}$}, \textcolor{cadetblue}{$D^{e \mu e}_{\tilde{q}_3}$} \\

\ATLASrpv~\cite{ATLAS:2021fbt} SR \textbf{SS-6j100-0b} & $\geq 2\ell$, $\geq 6j$, $0b$  & \textcolor{cadetblue}{$D^{\tau e e}_{\tilde{g}}$}, \textcolor{cadetblue}{$D^{\tau e \tau}_{\tilde{g}}$}, \textcolor{cadetblue}{$D^{\tau e \tau}_{\tilde{q}}$} \\

\ATLASstrongSSb~\cite{ATLAS:2017tmw} SR \textbf{Rpc3L1bH} & $\geq 3\ell$, $\geq 4 j$, $\geq 1 b$, $\ETmiss$ & \textcolor{cadetblue}{$D^{e \mu e}_{\tilde{q}_3}$}, \textcolor{cadetblue}{$D^{\tau e \tau}_{\tilde{q}_3}$} \\

\CMSewktwotautwolep~\cite{CMS:2017moi} SR \textbf{K03} & $2\ell$, $2\tau$, $\ETmiss$ & \textcolor{cadetblue}{$D^{\tau e \tau}_{\tilde{W}}$}, \textcolor{cadetblue}{$D^{\tau e \tau}_{\tilde{H}}$} \\

\CMSewkthreelep~\cite{CMS:2017moi} SR \textbf{A44} & $3\ell$, $\ETmiss$ & \textcolor{chromeyellow}{$D^{e \mu e}_{\tilde{\ell}}$}, \textcolor{chromeyellow}{$D^{e \mu \mu}_{\tilde{\ell}}$}, \textcolor{chromeyellow}{$D^{\tau e e}_{\tilde{\tau}}$}\\

\CMSewkonetauthreelep~\cite{CMS:2017moi} SR \textbf{I04} & $3\ell$, $1\tau$, $\ETmiss$  & \textcolor{naplesyellow}{$D^{\tau e \tau}_{\tilde{\ell}}$}\\

\CMSewktwotauonelep~\cite{CMS:2017moi} SR \textbf{F12}  & $1\ell$, $2\tau$, $\ETmiss$ 
 & \textcolor{chromeyellow}{$D^{\tau e \tau}_{\tilde{\tau}}$} \\

\hline \hline
\end{tabular}
\end{adjustbox}
\label{tab:searches}
\end{table}

\subsubsection{Direct Production}
\begin{figure}[ht!]
	\centering
	\includegraphics[width=0.82\textwidth]{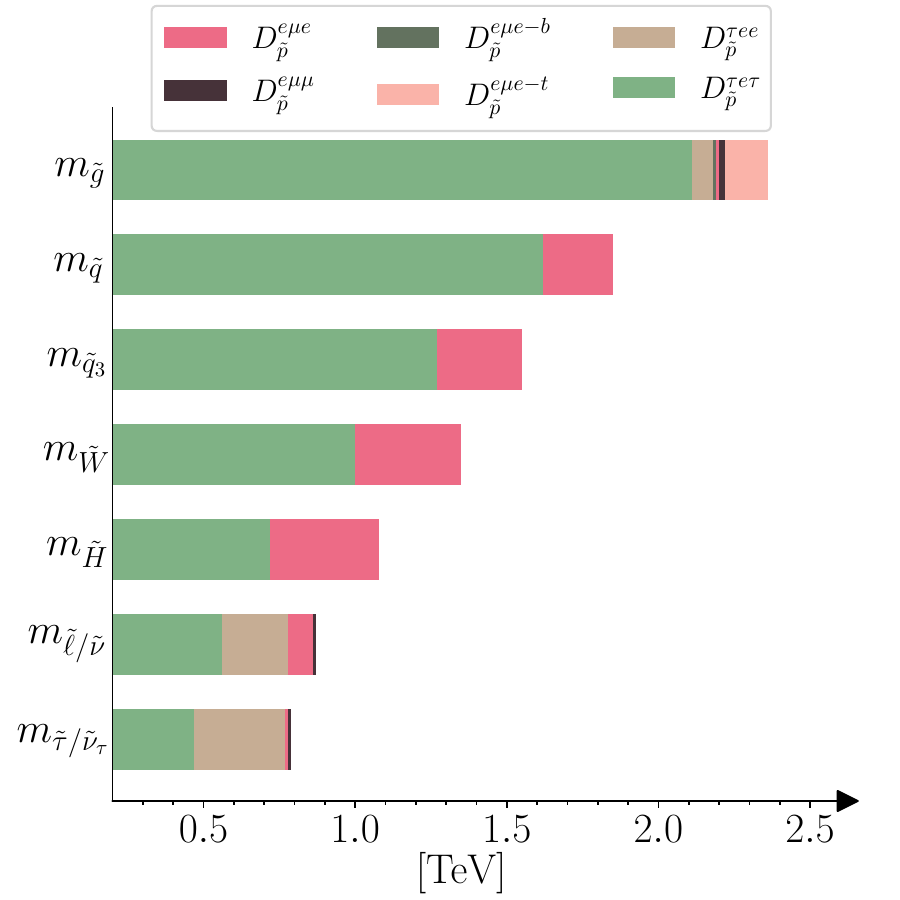}
	\caption{$95\%$ confidence-level mass-exclusion limits for various LSPs corresponding to direct pair production. In each scenario, all sparticles other than the LSP(s) ($\tilde{p}$) are assumed to be decoupled, while the LSP decays are still prompt.}
	\label{fig:direct_bounds}
\end{figure}

\cref{fig:direct_bounds} shows a summary of the mass 
limits corresponding to $95\%$ confidence level for the 
direct-production scenarios, \textit{i.e.}, all the $\text{D}_{\tilde{p}}$ scenarios from Tables~\ref{tab:gluinoBMs}-\ref{tab:lBMs}, where $\tilde{p}$ stands for the relevant LSP. The rest of the spectrum is assumed to be decoupled in these 
benchmarks, while the LSP decays remain prompt. We see that the exclusion limits are comparable to the current mass bounds corresponding to the regular MSSM (see, for instance, Ref.~\cite{ATLAS:2022rcw}). We now discuss the results in more detail in the following paragraphs.

The $\tilde{g}$-LSP scenarios can be ruled out up to about $m_{\tilde{g}}\sim \SIrange{2.1}{2.4}{\tera\electronvolt}$, with the weaker limits corresponding to cases where the $\lambda_{ijk}$ coupling 
involves third-generation indices. The strongest limit is achieved for scenarios involving couplings to light leptons and decay via off-shell top squarks. The signature from pair 
production for the gluino benchmarks is $4L + 4j + \ETmiss$. In general, the strongest sensitivity comes, as expected, from 
multilepton searches, especially \CMSewkfourlep\ and 
\ATLASstrongSS.\footnote{See \cref{tab:searches} for the notation we employ for searches.} In scenarios with heavy-flavor squarks,  $\text{D}^{{e\mu e - b}}_{\tilde{g}}$ and $\text{D}^{{e\mu e - t}}_{\tilde{g}}$, \CMSewkfourlep shows a weaker sensitivity due 
to the veto of $b$-jets. Finally, for couplings that involve $\tau$ leptons, $\text{D}^{{\tau e e}}_{\tilde{g}}$ and $\text{D}^{{\tau e \tau}}_{\tilde{g}}$, the most relevant 
analyses are \ATLASstrongSS and \ATLASrpv. Both searches offer 
sensitivity despite the fact that they focus only on light leptons. This is due to the fraction of gluino decays into electrons (\textit{cf.}~\cref{tab:gluinoBMs}), and the leptonic decay of taus. The hadronic decays of taus are reconstructed as additional jets which satisfy the jet multiplicity requirement of both searches. None of the searches are optimized for our signal but they still provide great sensitivity.

For the squarks and the electroweakinos, the final states are similar to the gluino case, albeit with fewer jets: $4L + (0-2)j + \ETmiss$. The most stringent limits for the $\text{D}^{{e\mu e}}_{\tilde{p}}$ scenarios are provided by \CMSewkfourlep: $m_{\tilde{q}}\gsim\SI{1.85}{\tera\electronvolt},\; m_{\tilde{W}}\gsim\SI{1.35}{\tera\electronvolt},\; m_{\tilde{H}}\gsim\SI{1.1}{\tera\electronvolt}$. The reduced jet multiplicity limits the sensitivity of \ATLASstrongSS and \ATLASrpv. The $\text{D}^{{\tau e \tau}}_{\tilde{W}}$ and $\text{D}^{{\tau e \tau}}_{\tilde{H}}$ scenarios are now instead covered by \CMSewktwotautwolep, a search explicitly targeting two hadronic taus, leading to the limits, $m_{\tilde{W}}\gsim\SI{1}{\tera\electronvolt}$ and $m_{\tilde{H}}\gsim\SI{720}{\GeV}$. However, similar to the gluino case, \ATLASrpv is the most sensitive for $\text{D}^{{\tau e \tau}}_{\tilde{q}}$ and rules out this scenario up to $m_{\tilde{q}}\approx\SI{1.6}{\tera\electronvolt}$.

The production of stops and sbottoms is special due to the presence of additional $b$ jets, which are vetoed by \CMSewkfourlep and \ATLASrpv. Thus, the best limits in this case come from \ATLASstrongSSb and \ATLASstrongSS for $\text{D}^{{e \mu e}}_{\tilde{q}_3}$ ($m_{\tilde{q}_3}\gsim\SI{1.55}{\tera\electronvolt}$) and $\text{D}^{{\tau e \tau}}_{\tilde{q}_3}$ ($m_{\tilde{q}_3}\gsim\SI{1.3}{\tera\electronvolt}$).

Finally, we have the slepton-LSP scenarios. For $\tilde{\ell}/\tilde{\nu}/\tilde{e}$, the exclusion limits lie in the broad range $m_{\tilde{\ell}}\sim \SIrange{560}{860}{\GeV}$. The most constraining search for scenarios $D^{e\mu e}_{\tilde{\ell}}$ and $D^{e\mu \mu}_{\tilde{\ell}}$ turns out to be \CMSewkthreelep. This search matches the $3L+\ETmiss$ topology from $\tilde{\ell}\tilde{\nu}$ production, as listed in~\cref{tab:LLE1}. For scenarios $D^{\tau e e}_{\tilde{\ell}}$ and $D^{\tau e \tau}_{\tilde{\ell}}$, the most relevant searches are \CMSewkfourlep and \CMSewkonetauthreelep, respectively. The latter needs at least three light leptons and at least one hadronic tau.

In the case of $\tilde{\tau}_L/\tilde{\nu}_{\tau}/\tilde{\tau}_R$, a large gap in sensitivity is observed between scenarios $D^{e \mu e}_{\tilde{\tau}}$, $D^{e \mu \mu}_{\tilde{\tau}}$, and $D^{\tau e e}_{\tilde{\tau}}$ which are excluded up to $m_{\tilde{\tau}}\sim780-\SI{790}{\GeV}$; and the $D^{\tau e \tau}_{\tilde{\tau}}$ scenario with a reach of just $m_{\tilde{\tau}}\gsim\SI{470}{\GeV}$. The former are covered by \CMSewkfourlep and \CMSewkthreelep, while the latter is targeted by \CMSewktwotauonelep.
The topologies targeted by all the above search regions match those in Tables~\ref{tab:LLE1}-\ref{tab:LLE2}.

\subsubsection{Cascade Decays}
We next look at the results for the indirect-production/cascade-decay scenarios, \textit{i.e.}, all the $\text{I}_{\tilde{x}\shortto\tilde{p}}$ benchmarks from Tables~\ref{tab:gluinoBMs}-\ref{tab:lBMs}, where $\tilde{p}$ is the LSP and $\tilde{x}$ denotes the directly produced parent particle decaying into the
LSP. Cascade decays are especially important for scenarios 
with a bino LSP, where direct production is irrelevant. For 
all other LSP types, the limits from direct LSP production (corresponding to $D_{\tilde{p}}$) are also taken into account.

In general, exclusion limits are mostly independent of the LSP mass (with a few exceptions) as the signal regions have high acceptance and the limit is 
driven by the production cross-section. A loss in sensitivity is observed in regions with small mass splittings only for models 
where the most sensitive signal region requires additional jets. 
In the bino scenarios, a loss in sensitivity is also observed for low LSP masses as its decay products carry energies that are too low to survive the search region cuts. This effect is not observed for other scenarios as the direct production of LSP becomes dominant for lower 
masses.

\begin{figure}[!ht]
\begin{subfigure}{0.45\textwidth}
\includegraphics[width=\linewidth]{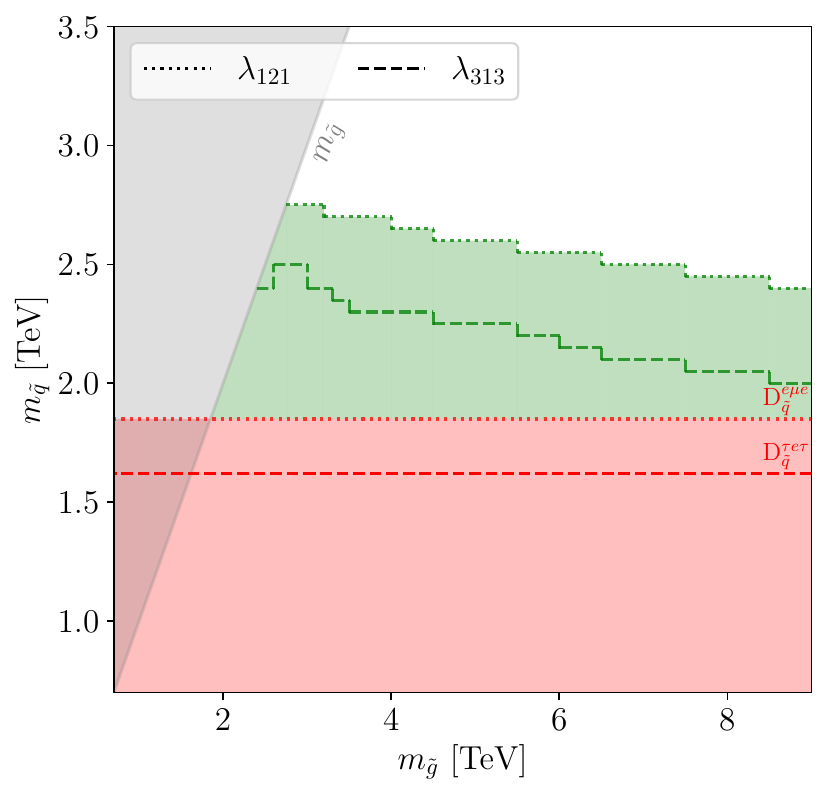}
  \caption{$\tilde{q}/\tilde{u}/\tilde{d}$-LSPs with non-decoupled $\tilde{g}$. \label{fig:I1a}}
  \end{subfigure}\hfill
\begin{subfigure}{0.45\textwidth}
   \includegraphics[width=\linewidth]{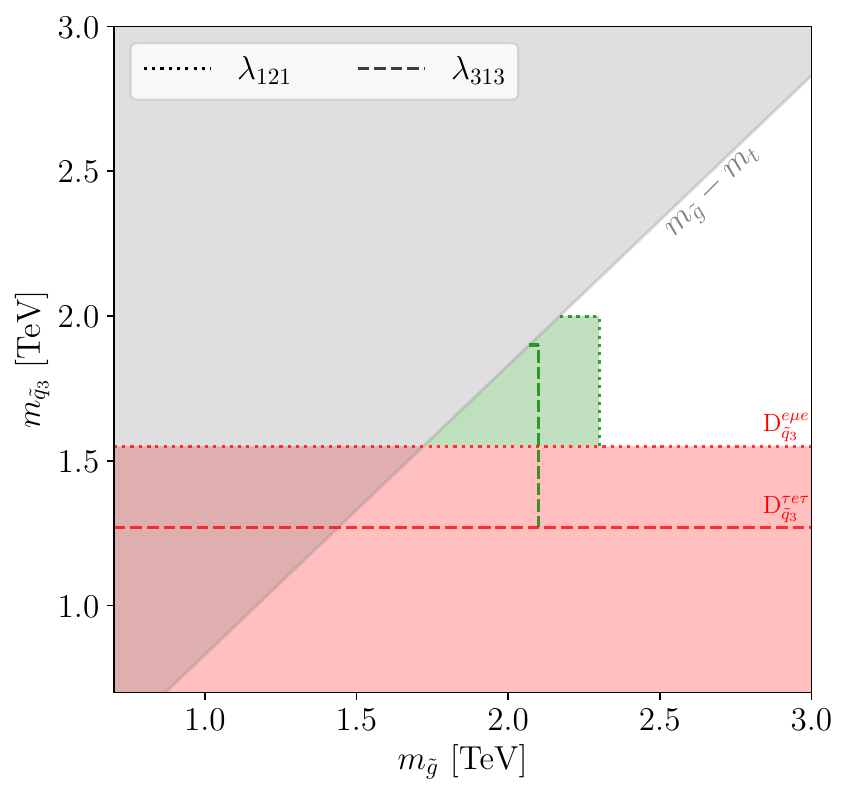}
  \caption{$\tilde{q}_3/\tilde{t}/\tilde{b}$-LSPs with non-decoupled $\tilde{g}$. \label{fig:I1b}}
\end{subfigure}
\caption{Exclusion regions (in green) corresponding to $95\%$ confidence level for the $I_{\tilde{g}\shortto\tilde{q}}$ (left) and $I_{\tilde{g}\shortto\tilde{q}_3}$ (right) scenarios. The bounds of~\cref{fig:direct_bounds} from direct squark production also apply to the scenario and are shown in red. The gray region is kinematically disallowed in the scenario. The dotted (dashed) contours correspond to coupling $\lambda_{121}\;(\lambda_{313})$.}
\label{fig:I1}
\end{figure}

\paragraph{Squark LSPs:}
In~\cref{fig:I1}, we show the exclusion limits for $\tilde{q}/\tilde{u}/\tilde{d}$-LSPs (\cref{fig:I1a}) and $\tilde{q}_3/\tilde{t}/\tilde{b}$-LSPs (\cref{fig:I1b}) for 
a non-decoupled gluino. The relevant production processes are gluino-gluino, squark-squark, and associated gluino-squark production, followed by 
the decay of the gluino into the squark LSP(s) and a jet, and finally the LSP decay via the RPV operator into $2L + j + \ETmiss$.\footnote{See~\cref{sec:A} for a detailed discussion on 
the specific decay modes we pick for each simulation.} The 
phase-space region $m_{\tilde{g}}<m_{\tilde{q}}$ ($m_{\tilde{g}}
<m_{\tilde{q}_3}+m_t$) is kinematically disallowed\footnote{Technically, for the heavy-flavor scenario, 
the region $m_{\tilde{q}_3} \leq m_{\tilde{g}}<m_{\tilde{q}_3}+m_t$ lets the gluino decay into a sbottom 
(ignoring the $b$-quark mass), and is allowed. However, for simplicity, we will ignore this here.} in the light-flavor 
(heavy-flavor) scenario, where we have neglected the masses of all SM fermions except the top quark. These regions are depicted in gray in the plot. 

From~\cref{fig:I1a}, we see that $I_{\tilde{g}\shortto\tilde{q}}$ can be excluded roughly up to the kinematic limit as long as we are below the threshold for $\tilde{g}\tilde{g}$ production, \textit{cf.}~\cref{fig:direct_bounds}. However, even above this threshold, we can exclude large regions of the parameter space that lie beyond the bounds from direct squark-pair production with a decoupled gluino. For instance, we see that even with $m_{\tilde{g}}\sim\SI{8}{\TeV}$, we get higher exclusion in the squark mass compared to the limit coming from $D_{\tilde{q}}$ (shown in red in the figure). This is due to two reasons. First, the associated-production channel (involving a single gluino) can stay kinematically accessible for longer. More importantly, a non-decoupled gluino significantly boosts direct squark-pair production cross-sections through its $t$-channel contributions~\cite{Borschensky:2014cia}. For very high masses, the gluino is essentially decoupled and the limits start converging, \textit{i.e.}, the scenarios reduce to the $D_{\tilde{q}}$ cases. 

For~\cref{fig:I1b}, the exclusion limits behave differently. For both couplings, roughly all kinematically viable regions can be excluded up to the corresponding $m_{\tilde{g}}$ limits of~\cref{fig:direct_bounds}. However, the limits reduce sharply to the $D_{\tilde{q}_3}$ bounds beyond this. For third-generation squarks, associated production as well as the boost in squark-squark cross-sections due to non-decoupled gluinos are suppressed by the small parton distribution functions (PDFs) for the heavy quarks inside the proton. Thus, as soon as gluino-pair production becomes kinematically inaccessible, the scenarios reduce to the $D_{\tilde{q}_3}$ cases. 

\begin{figure}[!ht]
\begin{subfigure}{0.43\textwidth}
  \includegraphics[width=\linewidth]{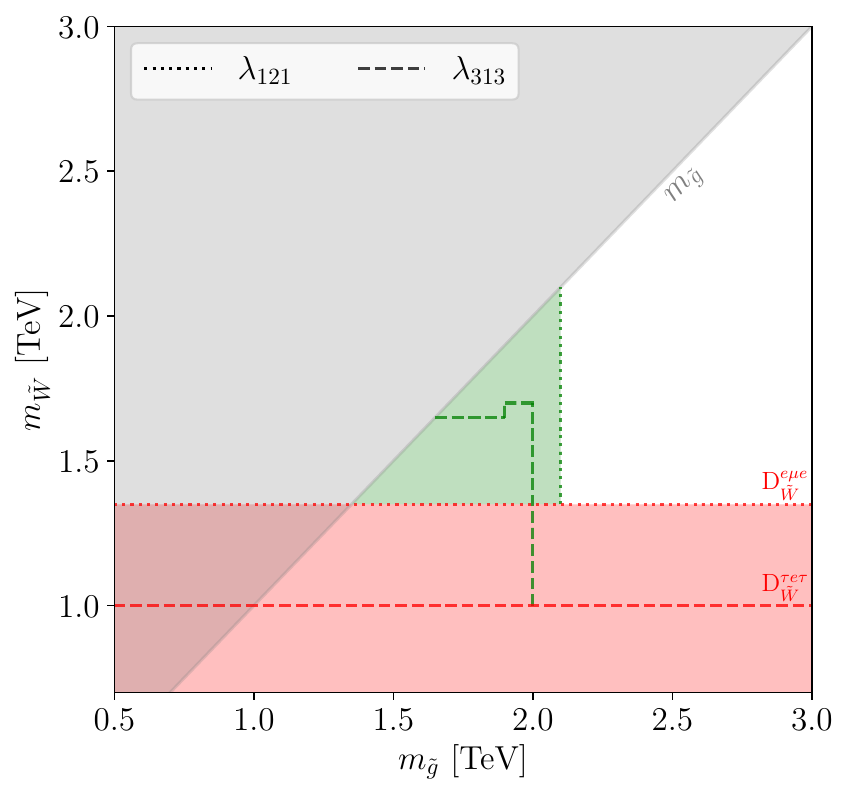}
  \caption{$\tilde{W}$ LSP with non-decoupled $\tilde{g}$. \label{fig:I2a}}
  \end{subfigure}\hfill
\begin{subfigure}{0.43\textwidth}
   \includegraphics[width=\linewidth]{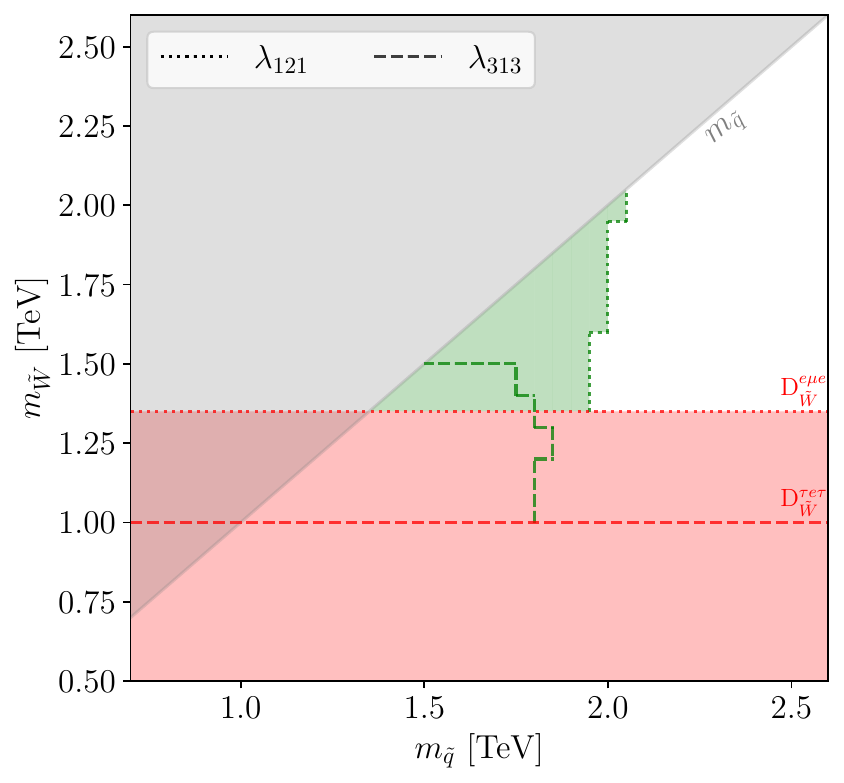}
  \caption{$\tilde{W}$ LSP with non-decoupled $\tilde{q}/\tilde{u}/\tilde{d}$. \label{fig:I2b}}
\end{subfigure}
\caption{As in~\cref{fig:I1} but for the $I_{\tilde{g}\shortto\tilde{W}}$ (left) and $I_{\tilde{q}\shortto\tilde{W}}$ (right) scenarios.}
\label{fig:I2}
\end{figure}

\paragraph{Electroweakino LSPs:} We next show the results for wino-LSP production with a non-decoupled gluino in~\cref{fig:I2a} and non-decoupled light-flavor squarks in~\cref{fig:I2b}.

\begin{figure}[!ht]
\begin{subfigure}{0.43\textwidth}
  \includegraphics[width=\linewidth]{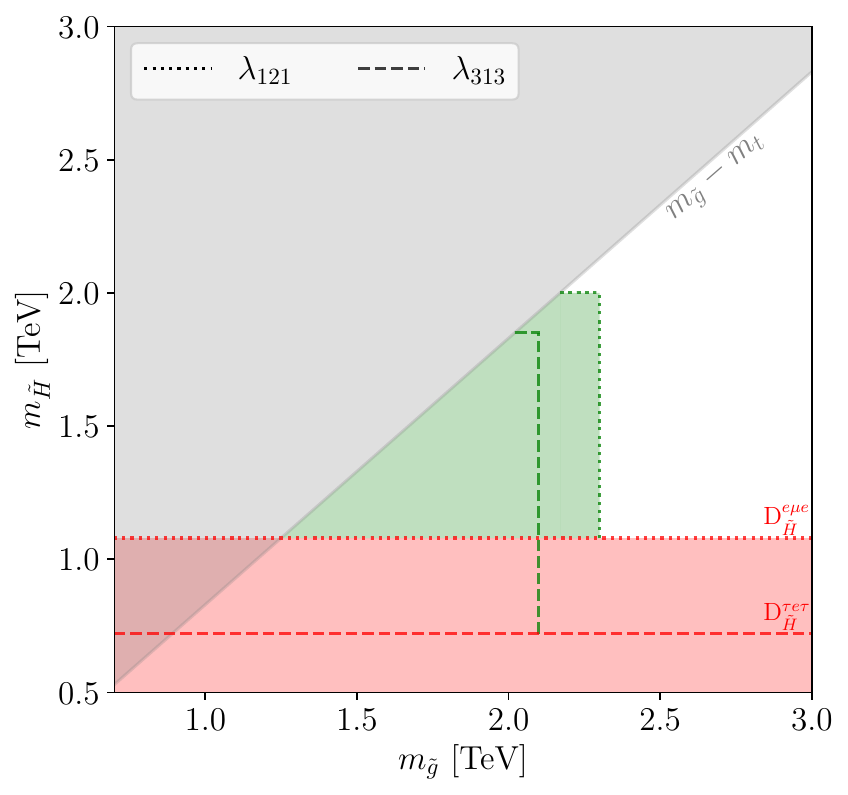}
  \caption{$\tilde{H}$ LSP with non-decoupled $\tilde{g}$. \label{fig:I3a}}
  \end{subfigure}\hfill
\begin{subfigure}{0.43\textwidth}
   \includegraphics[width=\linewidth]{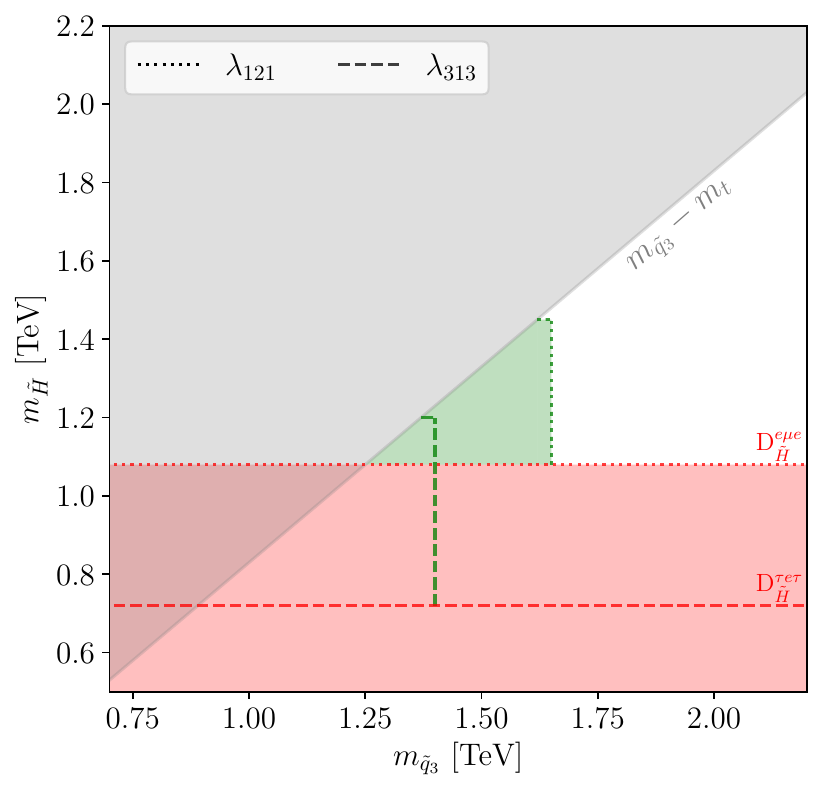}
  \caption{$\tilde{H}$ LSP with non-decoupled $\tilde{q}_3/\tilde{t}/\tilde{b}$. \label{fig:I3b}}
\end{subfigure}
\caption{As in~\cref{fig:I1} but for the $I_{\tilde{g}\shortto\tilde{H}}$ (left) and $I_{\tilde{q}_3\shortto\tilde{H}}$ (right) scenarios.}
\label{fig:I3}
\end{figure}

For the gluino case, we see features similar to~\cref{fig:I1b}. For both couplings, all phase-space regions almost up to the gluino-pair production threshold can be ruled out. Beyond this, the results from $D_{\tilde{W}}$ apply. One interesting feature is 
the flattening of the exclusion contour for $\lambda_{313}$ at $m_{\tilde{W}}\sim\SI{1.7}{\TeV}$ for gluino masses, $m_{\tilde{g}}\sim\SIrange{1.7}{2}{\TeV}$. This reduction in sensitivity occurs because the cuts in \ATLASrpv place a high 
demand on the transverse momentum of the six required jets, $p_\text{T} > \SI{100}{\GeV}$. If the wino and gluino are too close in mass, the jets produced in the gauge decay of the latter may not pass these requirements.

~\cref{fig:I2b} is more interesting. We again see that the parameter space roughly up to the squark-production thresholds can be ruled out and we observe the flattening effect mentioned above. However, we also see a 
new effect. The exclusion limit slightly weakens as we move lower in wino mass. This is clearly seen for the $\lambda_{121}$ 
case but the reduction in sensitivity occurs for both couplings throughout the phase space. 
This is because squark-pair production can also occur via $t$-channel wino exchange which can interfere negatively with the QCD contribution~\cite{Bornhauser:2007bf}; this interference term is bigger for lighter winos.

\begin{figure}[!ht]
\begin{subfigure}{0.31\textwidth}
  \includegraphics[width=\linewidth]{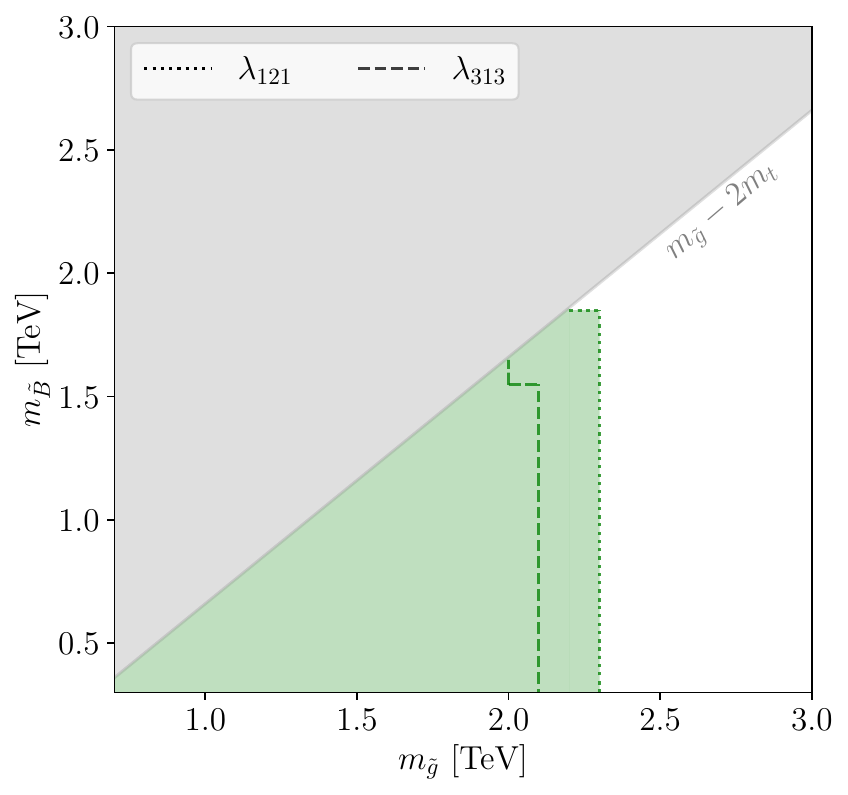}
  \caption{$\tilde{B}$ LSP with non-decoupled $\tilde{g}$. \label{fig:I4a}}
  \end{subfigure}\hfill
\begin{subfigure}{0.31\textwidth}
   \includegraphics[width=\linewidth]{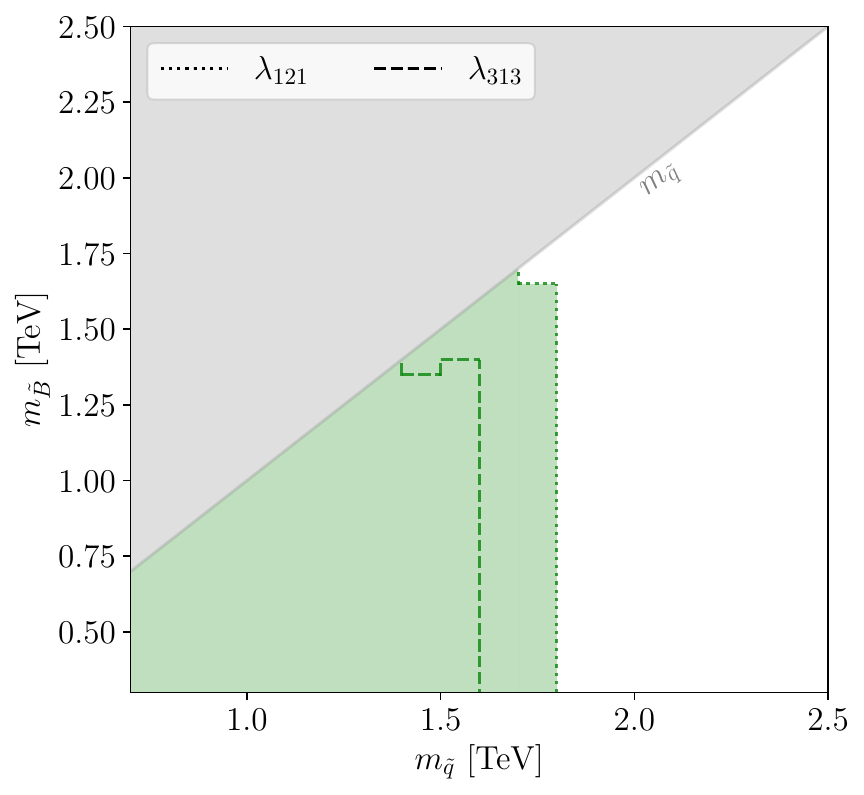}
  \caption{$\tilde{B}$ LSP with non-decoupled $\tilde{q}/\tilde{u}/\tilde{d}$. \label{fig:I4b}}
\end{subfigure}\hfill
\begin{subfigure}{0.31\textwidth}%
    \includegraphics[width=\linewidth]{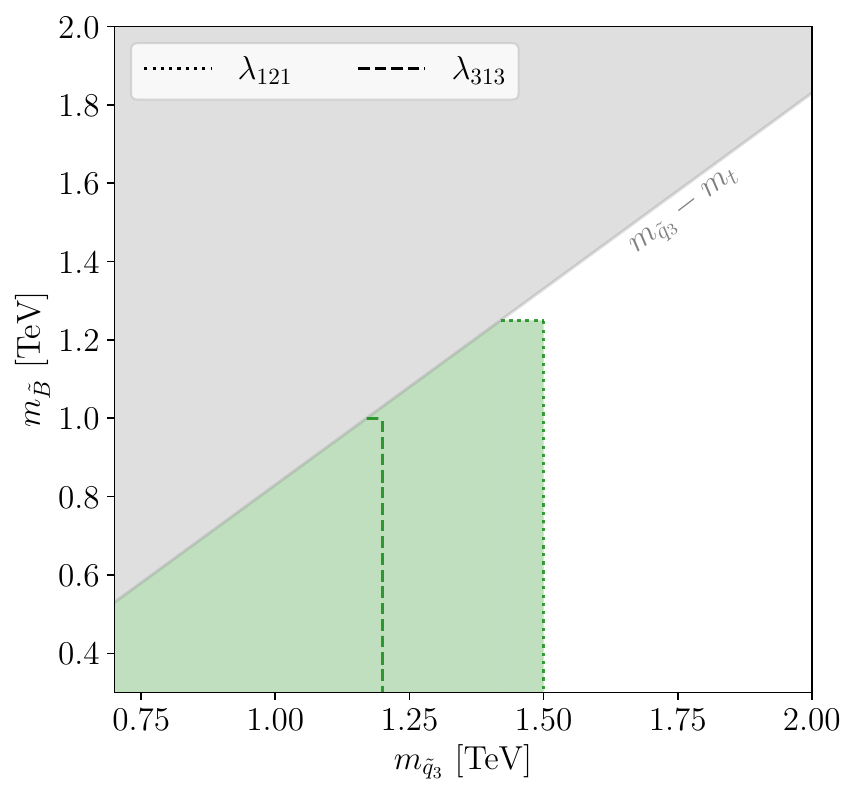}
  \caption{$\tilde{B}$ LSP with non-decoupled $\tilde{q}_3/\tilde{t}/\tilde{b}$. \label{fig:I4c}}
\end{subfigure}
\\
\begin{subfigure}{0.31\textwidth}
  \includegraphics[width=\linewidth]{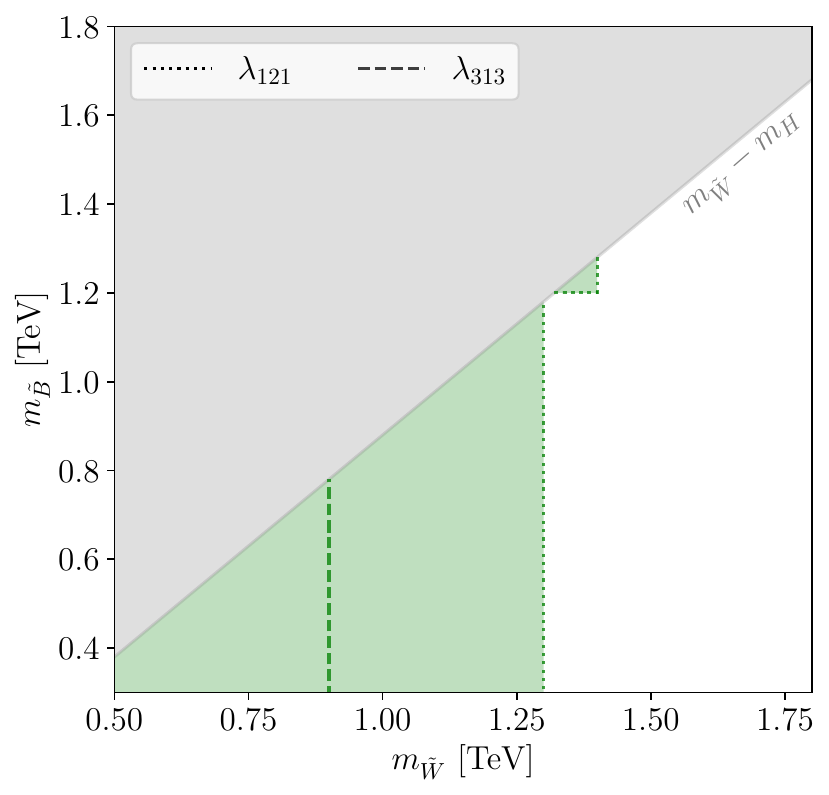}
  \caption{$\tilde{B}$ LSP with non-decoupled $\tilde{W}$. \label{fig:I4d}}
  \end{subfigure}\hfill
\begin{subfigure}{0.31\textwidth}
   \includegraphics[width=\linewidth]{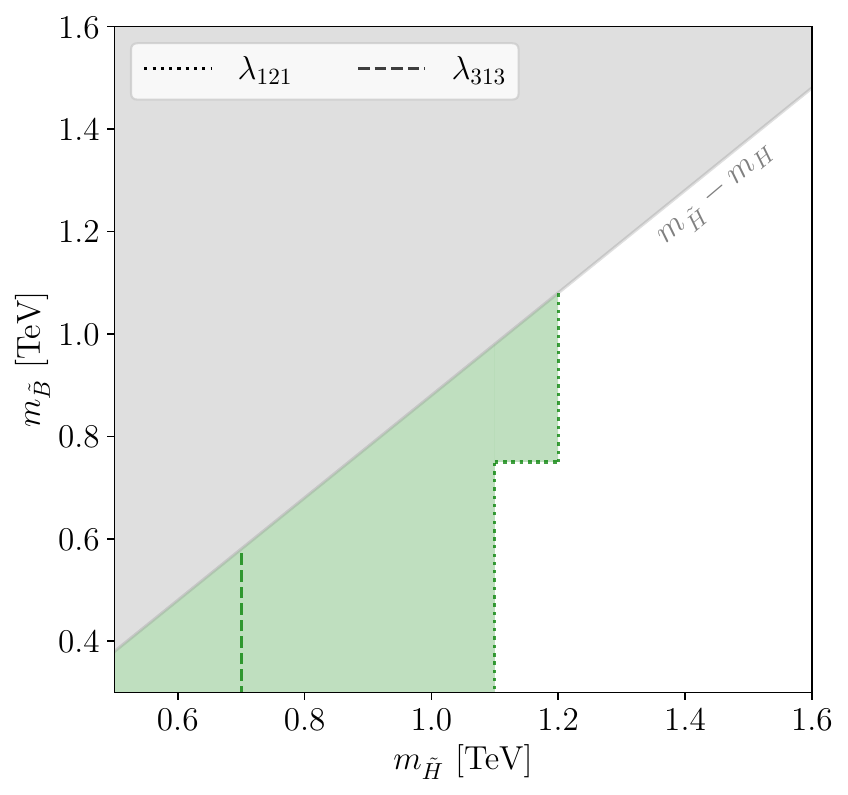}
  \caption{$\tilde{B}$ LSP with non-decoupled $\tilde{H}$. \label{fig:I4e}}
\end{subfigure}\hfill
\begin{subfigure}{0.31\textwidth}%
    \includegraphics[width=\linewidth]{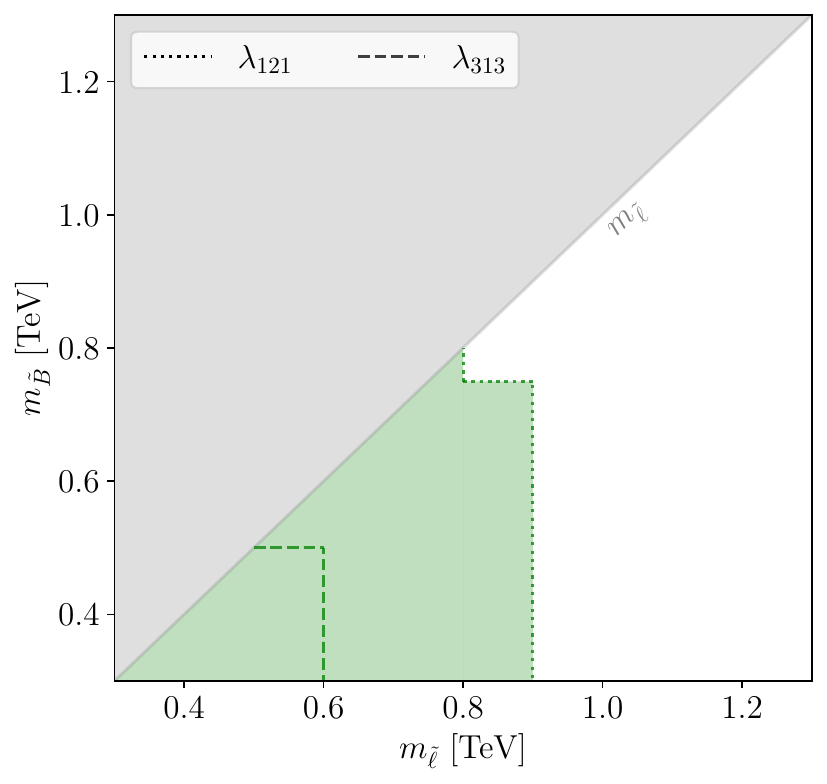}
  \caption{$\tilde{B}$ LSP with non-decoupled $\tilde{\ell}/\tilde{\nu}/\tilde{e}$. \label{fig:I4f}}
\end{subfigure}
\\
\begin{subfigure}{0.31\textwidth}
  \end{subfigure}\hfill
\begin{subfigure}{0.31\textwidth}
   \includegraphics[width=\linewidth]{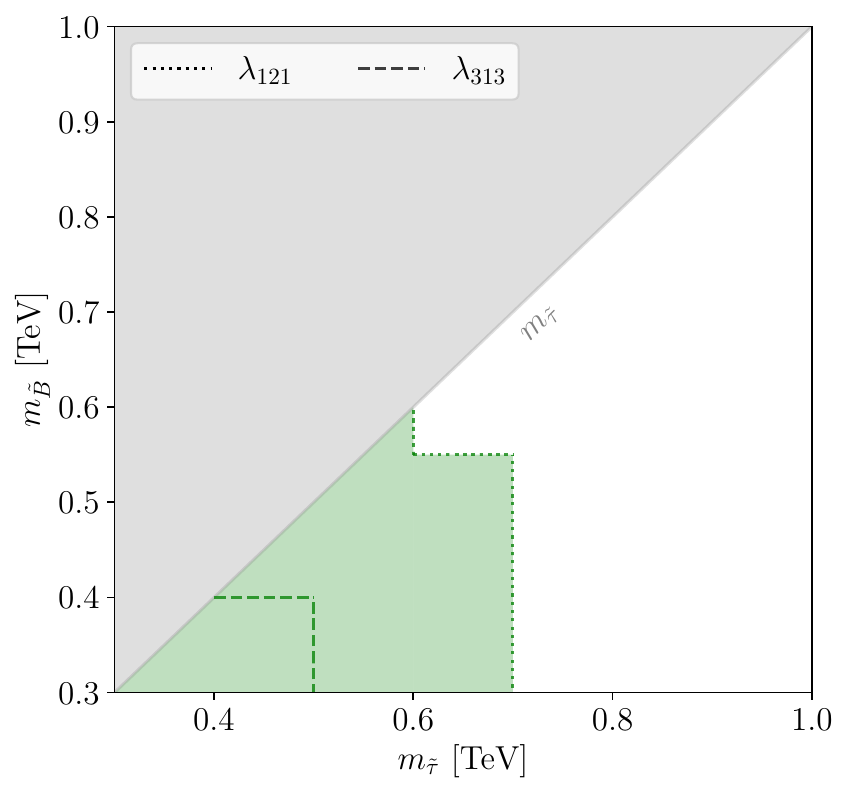}
  \caption{$\tilde{B}$ LSP with non-decoupled $\tilde{\tau}_L/\tilde{\nu}_{\tau}/\tilde{\tau}_R$. \label{fig:I4g}}
\end{subfigure}\hfill
\begin{subfigure}{0.31\textwidth}%
\end{subfigure}
\caption{As in~\cref{fig:I1} but for the $I_{\tilde{x}\shortto\tilde{B}}$ scenarios.}
\label{fig:I4}
\end{figure}

Next, we show the Higgsino-LSP results for non-decoupled gluinos and non-decoupled third-generation squarks in~\cref{fig:I3a} 
and~\cref{fig:I3b}, respectively. The exclusion limits show features similar to the earlier cases and are straightforward to interpret. Beyond the pair-production
thresholds for the parents, the benchmarks reduce to the respective $D_{\tilde{H}}$ scenarios.

The bino-LSP results are depicted in~\cref{fig:I4}: these correspond to scenarios with gluinos (\cref{fig:I4a}), light-flavor squarks (\cref{fig:I4b}), heavy-flavor squarks (\cref{fig:I4c}), winos (\cref{fig:I4d}), Higgsinos (\cref{fig:I4e}), light-flavor sleptons (\cref{fig:I4f}), and third-generation sleptons (\cref{fig:I4g}).
Generally, the exclusion limits cover almost the whole phase-space region up to the kinematic thresholds for the pair-production of the parents. However, there are a couple of features worth mentioning. First, we see the flattening effect, that we had described for~\cref{fig:I2}, in scenarios $I^{\tau e \tau}_{\tilde{x}\shortto\tilde{B}}$ with $\tilde{x}=\tilde{q}, \tilde{\ell}, \tilde{\tau}$. The other interesting effect is the slight increase in sensitivity as the $\tilde{B}$ mass increases from very low masses to higher values. This effect can be most clearly seen in~\cref{fig:I4e} but is a general feature in the other $\tilde{B}$ plots too. This happens due to the reason mentioned at the beginning of this subsection: for very low bino masses, the decay products are not energetic enough to pass the cuts of the analyses. We did not encounter it in the case of the other LSPs since the mass scales there were higher.

\paragraph{Slepton LSPs:} 
\FloatBarrier
\begin{figure}[H]
\begin{subfigure}{0.43\textwidth}
  \includegraphics[width=\linewidth]{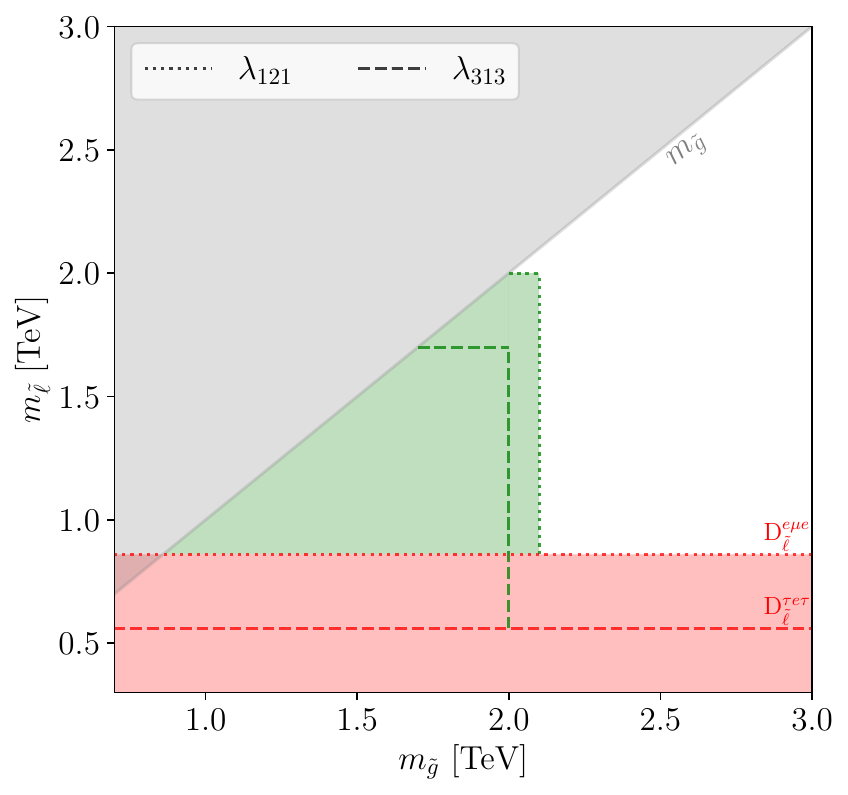}
  \caption{$\tilde{\ell}/\tilde{\nu}/\tilde{e}$ LSPs with non-decoupled $\tilde{g}$. \label{fig:I5a}}
  \end{subfigure}\hfill
\begin{subfigure}{0.43\textwidth}%
    \includegraphics[width=\linewidth]{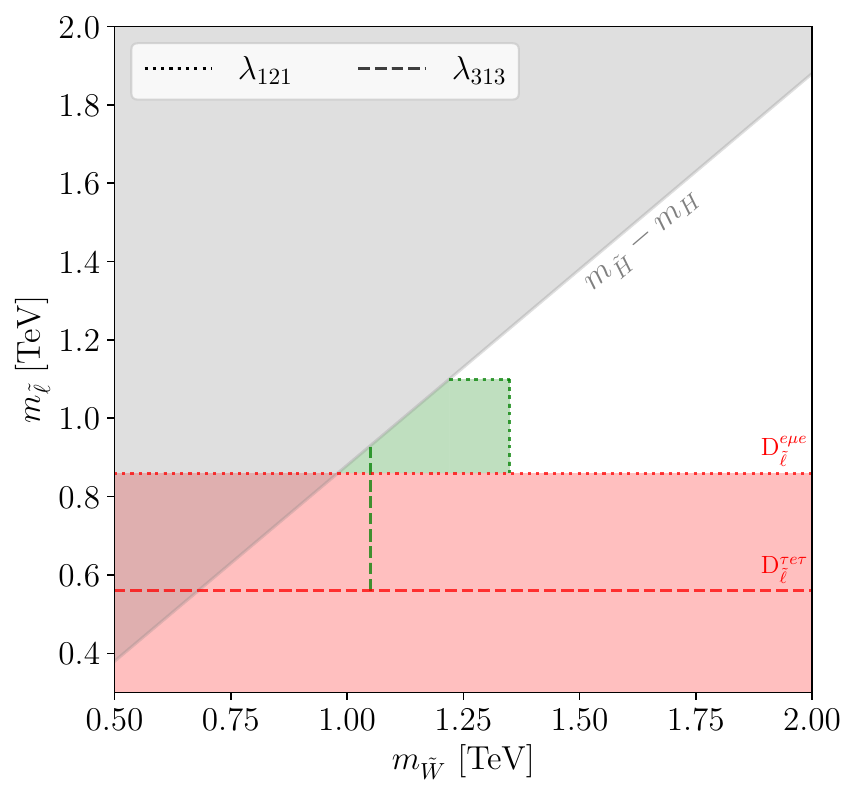}
  \caption{$\tilde{\ell}/\tilde{\nu}/\tilde{e}$ LSPs with non-decoupled $\tilde{W}$. \label{fig:I5b}}
\end{subfigure}
\\
\begin{subfigure}{0.43\textwidth}
  \includegraphics[width=\linewidth]{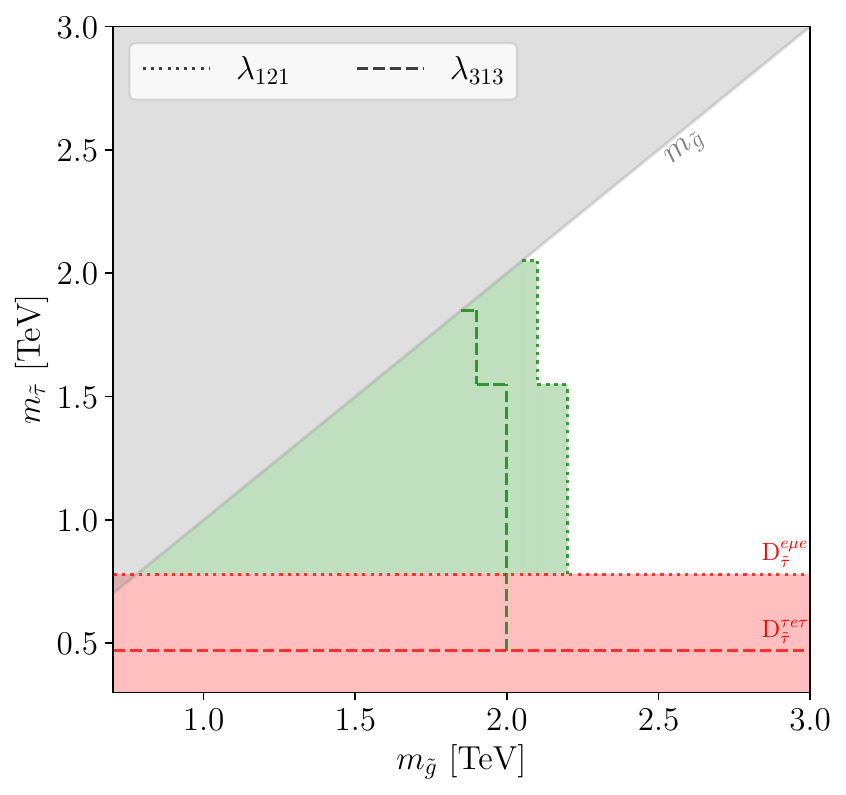}
  \caption{$\tilde{\tau}_L/\tilde{\nu}_{\tau}/\tilde{\tau}_R$ LSPs with non-decoupled $\tilde{g}$. \label{fig:I5c}}
  \end{subfigure}\hfill
\begin{subfigure}{0.43\textwidth}%
    \includegraphics[width=\linewidth]{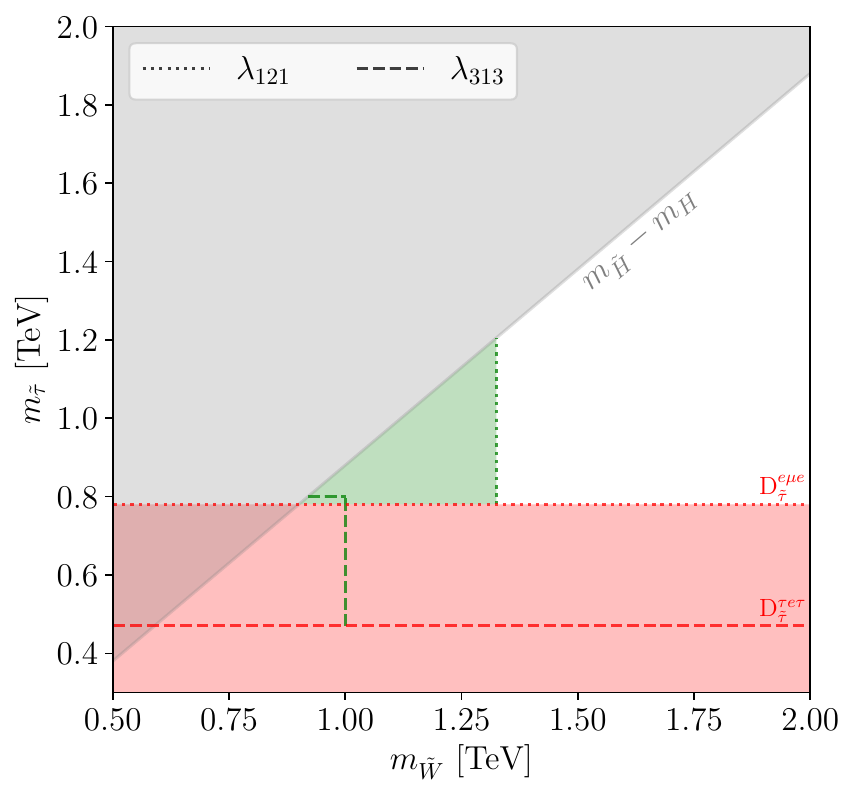}
  \caption{$\tilde{\tau}_L/\tilde{\nu}_{\tau}/\tilde{\tau}_R$ LSPs with non-decoupled $\tilde{W}$. \label{fig:I5d}}
\end{subfigure}
\caption{As in~\cref{fig:I1} but for the $I_{\tilde{x}\shortto\tilde{\ell}/\tilde{\tau}}$ scenarios.}
\label{fig:I5}
\end{figure}

\begin{figure}[!ht]
\begin{subfigure}{0.49\textwidth}
  \includegraphics[width=\linewidth]{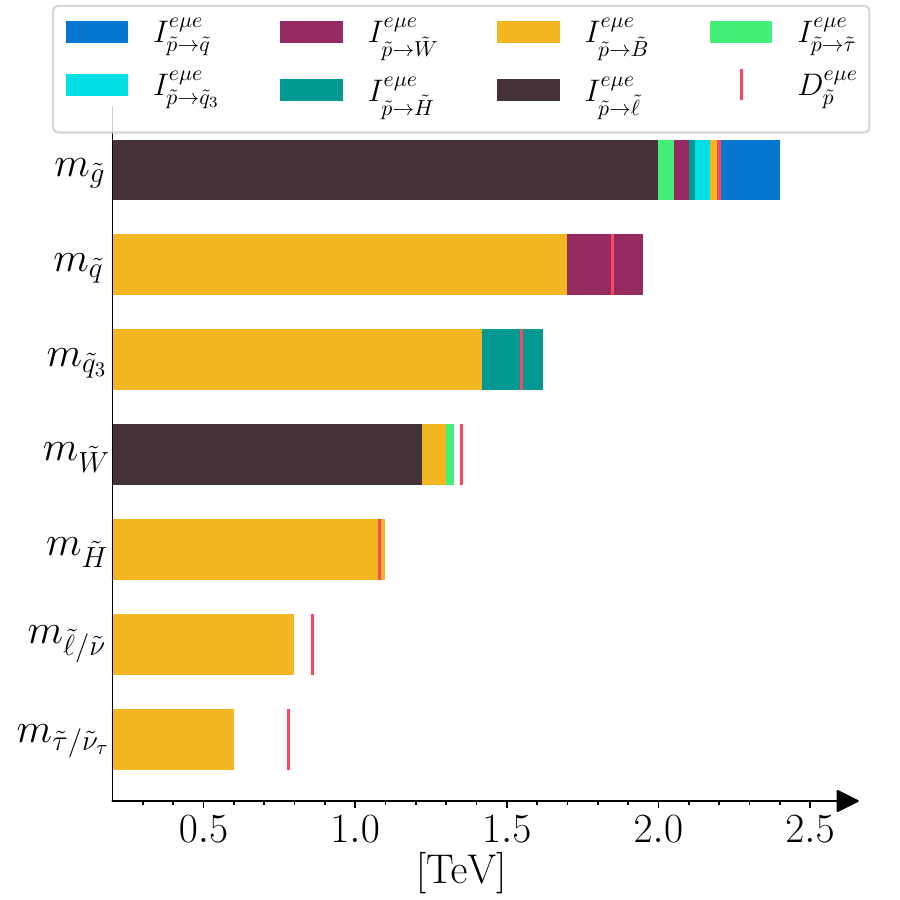}
  \caption{$\lambda_{121}$. \label{fig:Summarya}}
  \end{subfigure}\hfill
\begin{subfigure}{0.49\textwidth}%
    \includegraphics[width=\linewidth]{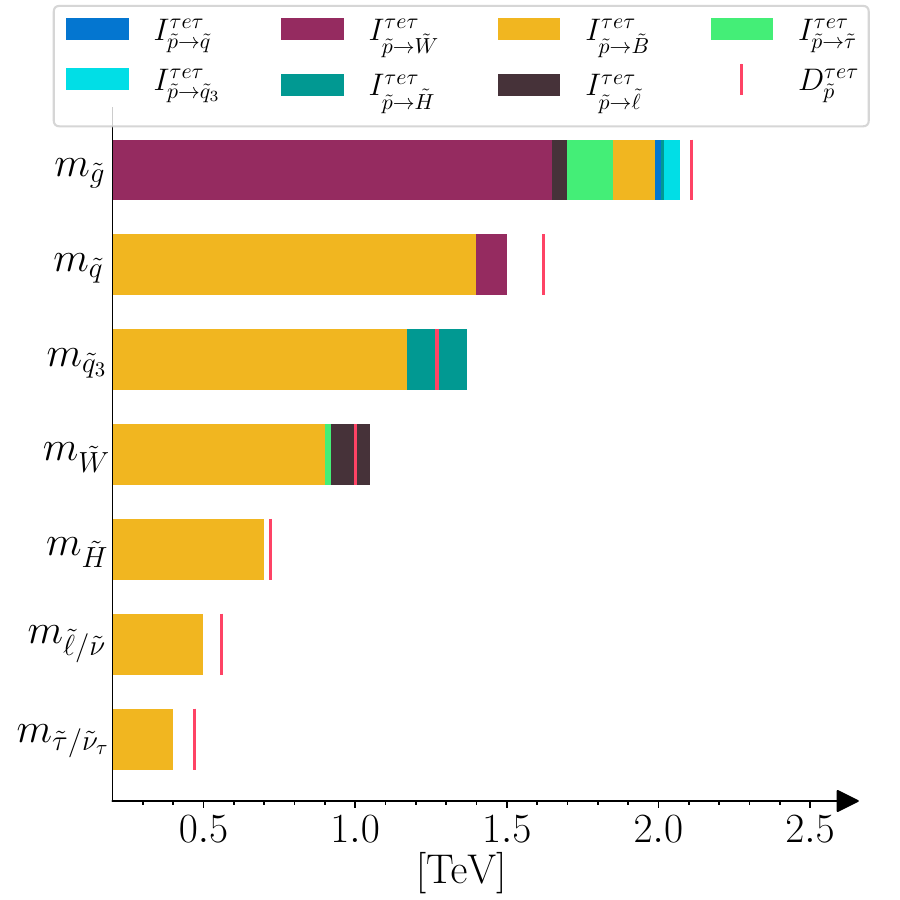}
  \caption{$\lambda_{313}$.\label{fig:Summaryb}}
\end{subfigure}
\caption{Summary of minimum mass bounds on sparticle $\tilde{p}$ across the various $\text{I}_{\tilde{p}\shortto\tilde{x}}$ benchmarks considered, where $\tilde{x}$ corresponds to the LSPs. The vertical red line represents the direct production mass bound when $\tilde{p}$ is the LSP, \textit{i.e.}, the limit corresponding to $\text{D}_{\tilde{p}}$.}
\label{fig:IndirectSummary}
\end{figure}

Finally, we show the slepton-LSP results in~\cref{fig:I5}. The exclusion limits can extend significantly in the cascade decay due to the much higher production cross-sections of other parent particles compared to direct slepton production.
\paragraph{LSP summary:} 
To summarize, we collect, in~\cref{fig:IndirectSummary}, the minimum excluded mass for each sparticle, $\tilde{p}$, undergoing a cascade decay (\textit{i.e.}, the minimum limits obtained for each of the $\text{I}_{\tilde{p}\shortto\tilde{x}}$ scenarios with $\tilde{x}$ the various LSPs), and compare it to the limit obtained from direct production of the sparticle when it is the LSP (\textit{i.e.}, the corresponding $\text{D}_{\tilde{p}}$ scenarios). It is interesting to note that, 
although cascade decays generally lead to final states with more visible objects, the sensitivity can be both degraded or improved. The reduction in $\ETmiss$ and the distribution of energy across more decay products can reduce the sensitivity. 
For example, the decay to a slepton or bino LSP yields in most cases the worst limits given that intermediate particles in the decay 
chain can become soft for compressed spectra, \textit{e.g.}, $\tilde{g}\rightarrow 2j+\ell+\tilde{\ell}(\rightarrow \ell\nu)$. However, changes in the decay modes due to the varying nature of the LSP can also lead to a higher number of leptons or third-generation quarks which leads to an improvement in the limits. It is worth highlighting that the degradation is around 20\% at maximum, and the exclusion limits remain for all sparticles under all variations of LSP hypotheses, LSP masses, and coupling choice.

\FloatBarrier
\section{Conclusions and Outlook}
\label{sec:conclusions}
In this paper, we have systematically analyzed the RPV-MSSM 
and classified the possible signatures at the LHC with the 
goal of compiling a minimal set of experimental searches 
that provides complete coverage. Our study provides, 
for the first time, a completely 
general and model-independent 
treatment of the phenomenology, for the case of small RPV
couplings. We briefly summarize the central results of the paper:

\begin{itemize}

\item As demonstrated, the phenomenological space for the 
    most general RPV-MSSM setup is vast and complicated. Nevertheless, we have shown that just $17$ final state topologies (six for $LL\Bar{E}$, five for $\Bar{U}\Bar{D}\Bar{D}$, and six for $LQ\Bar{D}$) are 
    sufficient to provide complete coverage for the RPV-MSSM at the LHC; we call this the `RPV Dictionary'. Our signature tables can be generated by using the accompanying \texttt{abc-rpv} Python library, described in~\cref{sec:C}.
   
\item Using the `RPV Dictionary', we have analyzed the current coverage of the RPV-MSSM at the LHC. In general, we find that even though most RPV scenarios have not been 
searched for directly, the vast landscape of searches implemented by \texttt{ATLAS} and \texttt{CMS} provides full coverage of the possible RPV-MSSM signatures.
    
\item However, we do point out the need for strong experimental improvements in some of the final states in order to achieve sensitivity to electroweak production cross-sections. Some examples are found for $LQ\bar{D}$ and $U\bar{D}\bar{D}$ decays, such as $\tilde{\ell} \rightarrow j j$, $\tilde{\chi}^0_1 \rightarrow j b \nu$, and $\tilde{\chi}^0_1 \rightarrow j j j$.

\item As an application of our framework, and in order to 
demonstrate the second point above, we have performed numerical 
simulations specifically for the case of a dominant $LL\bar E$ 
operator (single non-zero coupling at a time), in order to 
quantitively assess the coverage. We have derived mass bounds on SUSY particles within several 
RPV benchmark models corresponding to all relevant LSPs. We find that strong exclusion limits comparable to, or even better than, 
the RPC-MSSM are obtained, and these are robust across the wide range of models. Apart from clarifying the current status of 
several of these scenarios for which there are no explicit exclusion limits in the literature, our numerical examples 
demonstrate that our approach of using just a few characteristic topologies to cover the most general RPV-MSSM setting is not merely 
a reductionist fantasy, but can indeed offer a viable, model-independent search strategy. We have left the detailed 
analyses of the $LQ\bar D$ and $\bar U\bar D\bar D$ cases for future work.
\end{itemize}

We stress that there are a couple of limitations of our framework. As mentioned in the main body, we require that 
all decays in the cascade chain are prompt, including that of the LSP. Furthermore, we require that the final state decay
products of the LSP are not too soft to be detected. This 
assumption is crucial and restricts us to scenarios with $m_{\text{LSP}} > \mathcal{O}
\left(\SI{200}{\giga\electronvolt}\right)$. 
Further, while we require the RPV coupling to be large enough to 
cause the LSP to decay promptly, it cannot be too large, as 
that would modify the pattern of the cascade decays. Similarly, adding exotic particles to the MSSM spectrum that can modify the sparticle decay chains also affects our analysis. In such cases, our classification may not apply anymore. The most important restriction is that our approach -- in 
prioritizing model independence -- compromises on
search sensitivity for certain scenarios. 
For example, if the colored sector is always kinematically accessible at the LHC, stricter bounds can be obtained by including
the cascade decay products in the search signature, whereas, in our approach, we only target decay products from the LSP. The former approach is usually adopted by \texttt{ATLAS} and \texttt{CMS} in their searches for 
specific RPV-SUSY scenarios. 

On the other hand, we believe our unbiased approach is highly 
relevant, given that no supersymmetry has yet been discovered at 
the LHC, and with the HL-LHC era just around 
the corner. Nevertheless, we have compiled 
auxiliary tables in~\cref{sec:B} that can help in designing optimized search strategies in exchange for some 
model-independence.

As a continuation of this work, we shall pursue a detailed 
numerical treatment of the $LQ\Bar{D}$ and $\Bar{U}\Bar{D}
\Bar{D}$ scenarios analogous to the $LL\Bar{E}$ case considered 
in this paper. In those cases, present coverage is less comprehensive 
and it is important to identify potential gaps. Furthermore, we
would like to extend the present work to a systematic study of 
the large RPV coupling case, affecting both production and 
decay.

\section*{Acknowledgments}
 VML is funded by grant Mar\'ia Zambrano UP2021-044 (ZA2021-081) funded by Ministerio de Universidades and ``European Union-NextGenerationEU/PRTR'' and thanks support from the Spanish grants PID2020-113775GB-I00 (AEI/10.13039/501100011033) and CIPROM/2021/054 (Generalitat Valenciana). JMB is funded by grant RYC2021-030944-I funded by MCIN/AEI/10.13039/501100011033 and by the European Union NextGenerationEU/PRTR. NS is supported by the U.S. Department of Energy, Office of Science, Office of High Energy Physics under Award Number DE-SC0011845. 
\bigskip
\appendix
\section{Decay Modes for Numerical Simulations}
\label{sec:A}
We discuss the details of the decay modes used in the 
numerical simulations here. As mentioned 
in~\cref{sec:results}, all two-body decays are computed 
using \texttt{MadGraph5\_aMC@NLO}; we only 
discuss the higher-body decays that we set by hand here.

\paragraph{Gluino LSP Benchmarks:}
For a given coupling, $\lambda_{iki}$ with $i,k\in\{1,2,3\}$, we assume 
the chain: $\tilde{g}\to (\tilde{q}/\tilde{q}_3)^* + j \to \tilde{B}^* + 2j \to 2L + \ETmiss + 2j$, where $2L=\{L^+_i+ 
L^-_i\}\,, \{L^+_i+L^-_k\}\,, \{L^+_k+L^-_i\}$, and we set the corresponding branching ratios (BRs) to be $0.5$, 
$0.25$, and $0.25$, respectively. In the above, the asterisk denotes off-shell particles, and the nature of the (s)quark is fixed by the scenario considered. We stress that the chosen decay chain and BRs represent a non-trivial choice to facilitate numerical computation; alternate choices are possible -- \textit{e.g.}, the bino can be replaced by a wino or one flavor of slepton can be decoupled, thus, affecting the BRs. That is, we take the perspective where the branching ratios are taken to be the free variables, rather than the sparticle masses. Apart from being a simpler approach, this also saves computational time since we no longer need to calculate complicated high-multiplicity decays. Even with alternate BR choices, we expect the general implications of our results to hold.

\paragraph{Squark LSP Benchmarks:} This is very similar
to the above. For a coupling $\lambda_{iki}$, we assume the decay chain for the squarks to be: $(\tilde{q}/\tilde 
{q}_3)^* \to \tilde{B}^* + j \to 2L + \ETmiss + j$, where $2L 
= \{L^+_i+L^-_i\}\,, \{L^+_i+L^-_k\}\,, \{L^+_k+L^-_i\}$, and the BRs are set to $0.5$, $0.25$, and 
$0.25$, respectively.

\paragraph{Electroweakino LSP Benchmarks:}
\begin{itemize}
    \item $\text{D}_{\tilde{W}}$: The neutral wino decays 
    as: $\tilde{W}^{0}\to 2L + \ETmiss$, for a coupling 
    $\lambda_{iki}$, where $2L = \{L^+_i+L^-_i\}\,, \{L^+_i+L^-
    _k\}\,, \{L^+_k+L^-_i\}$. We set the corresponding 
    BRs to $0.5$, $0.25$, and $0.25$, respectively. For the charged wino, we have the decay modes: $\tilde{W}^{+}\to \{L^+_i+L^-_i+L^+_k\}\,, \{L^+_i+L^+
    _i+L^-_k\}\,, \{L^+_i + \ETmiss\}\,, \{L^+_k+\ETmiss\}$ (analogous for $\tilde{W}^-$). We set the BR to $0.25$ for each mode.
    \item $\text{I}_{\tilde{g} \shortto \tilde{W}}$: The 
    gluino decays as: $\tilde{g} \to \tilde{q}^* + j_l \to \tilde{W} + 2j_l$.
    \item $\text{I}_{\tilde{q} \shortto \tilde{W}}$: The 
    (singlet) squarks decay as: $\tilde{u}/\tilde{d} \to 
    \tilde{g}^* + j_l \to \tilde{q}^* + 2j_l \to \tilde{W} + 
    3j_l$. 
    \item $\text{D}_{\tilde{H}}$: The neutral Higgsinos 
    decay as: $\tilde{H}_{1(2)}^{0}\to \tilde{B}^* + Z(h) 
    \to 2L + \ETmiss + Z(h)$, for a coupling $\lambda_{iki}$, 
    where $2L = \{L^+_i+L^-_i\}\,, \{L^+_i+L^-_k\}\,, 
    \{L^+_k+L^-_i\}$, with the BRs fixed to $0.5$, 
    $0.25$, and $0.25$, respectively. The case for 
    $\tilde{H}^{\pm}$ is analogous, with $Z(h)$ replaced by $W^{\pm}$.
    \item $\text{I}_{\tilde{g} \shortto \tilde{H}}$: Here, the gluino is assumed to decay as:
    $\tilde{g} \to \tilde{q_3}^* + j_3 \to \tilde{H} + 2j_3$.
    \item In all $\tilde{B}$ LSP scenarios, the bino decays as:
    $\tilde{B} \to 2L + \ETmiss$, 
    for a coupling $\lambda_{iki}$, where $2L = \{L^+_i+L^-
    _i\}\,, \{L^+_i+L^-_k\}\,, \{L^+_k+L^-_i\}$, with the BRs set to
    $0.5$, $0.25$, and $0.25$, respectively.
    \item $\text{I}_{\tilde{g} \shortto \tilde{B}}$: 
    The gluino is assumed to decay as:
    $\tilde{g} \to \tilde{q_3}^* + j_3 \to \tilde{B} + 2j_3$.
\end{itemize}

\paragraph{Slepton LSP Benchmarks:}
\begin{itemize}
    \item For all $\text{D}_{\tilde{L}}$ scenarios, if a 
    particular lepton does not couple directly to the 
    considered operator, the decay is assumed to proceed via an off-shell bino, \textit{e.g.}, 
    $\tilde{\mu}_R \to \tilde{B}^* + \mu \to 2L  + \mu + 
    \ETmiss$, where $2L = \{L^+_i+L^-_i\}\,, \{L^+_i+L^-
    _k\}\,, \{L^+_k+L^-_i\}$ (for a coupling $\lambda_{iki}$). The corresponding BRs are set to 
    $0.5$, $0.25$, and $0.25$, respectively; and so on.
     \item $\text{I}_{\tilde{g} \shortto {\tilde{L}}}$: 
     The gluino is assumed to decay as: $\tilde{g} \to \tilde{q}^* + j_l \to \tilde{B}^* + 2j_l \to \tilde{L} + 2j_l + L$. Here, $\tilde{L}$ refers to all the sleptons.
     \item $\text{I}_{\tilde{W} \shortto \tilde{{L}}}$: We assume that the only contributing decays are the two-body decay modes of the wino into the 
     left-handed sleptons; we set the decay widths of the 
     modes into the right-handed sleptons to be zero. This can occur if, for instance, any virtual mediators that can lead to such a decay are completely decoupled. 
\end{itemize}

\section{Auxiliary Tables}
\label{sec:B}
\subsection{Production Table}
While compiling the `RPV Dictionary' in~\cref{sec:classification}, we have taken a model-independent approach. In order to completely cover the RPV-MSSM landscape (within our framework assumptions), it is necessary to perform the searches compiled in Tables~\ref{tab:LLE1}-\ref{tab:LQD6}. Often, however, one is not interested in being completely general but may have a bias for certain classes of models. For instance, it is usual in the literature to focus on scenarios where a given LSP is produced at the LHC in cascade decays of the colored sparticles; scenarios where all sparticles other than the LSP are completely decoupled are less common. Given such a bias, one can optimize the `RPV Dictionary' by adding the objects that would arise from such cascades. 

In order to facilitate the inclusion of the above, ~\cref{tab:optimisesearch} provides a list of the objects that arise in cascade-decays for each relevant production mode for each LSP. For instance, with a gluino LSP, the only relevant mode is gluino-pair production since every other channel will have a lower cross-section. For squark LSP(s), however, squark-pair, gluino-pair, and associated production are all relevant since the latter two contribute with high cross-sections when a gluino is not decoupled. \cref{tab:optimisesearch} shows that the cascade to the LSP from gluino-pair (associated-pair) production leads to 2 extra jets (1 extra jet). These can then be used to optimize searches for models with squark LSPs and non-decoupled gluinos by adding the extra jet(s) to the relevant squark LSP signatures in Tables~\ref{tab:LLE1}-\ref{tab:LQD6}. We note that each value in the table represents the maximal set of objects that is \textit{guaranteed to arise} in the cascade without knowing the details of the spectrum; however, more objects can always be present in specific scenarios. Further, to be economical, we have grouped left-handed and right-handed sparticles into one category but it is straightforward to expand them out.

\begin{table}[H]
\caption{Objects arising in the cascade decays of various pairs of parent sparticles (columns) down to the LSP(s) (rows). These can be added to the corresponding LSP signatures given in Tables~\ref{tab:LLE1}-\ref{tab:LQD6} to optimize searches. $+ -$ indicates an empty set while $\times$ indicates that the corresponding production channel is not relevant for the given LSP because the cross-section is either lower than or comparable to the cross-section for direct pair production of the LSP.}
\begin{center}
\scalebox{0.4}{
\renewcommand{\arraystretch}{1.9}
\begin{adjustbox}{width=1\textwidth}
\rotatebox{90}{
\begin{tabular}{c c c c c c c c c c}
\hline \hline
{\bf LSP} & $\tilde{g}\tilde{g}$  & $\tilde{g}\tilde{q}/\tilde{g}\tilde{u}/\tilde{g}\tilde{d}$ & Squark Pair (1st,2nd gen.) & Squark Pair (3rd gen.) &$\tilde{W}\tilde{W}$ & $\tilde{H}\tilde{H}$ & Slepton Pair (1st, 2nd gen.) & Slepton Pair (3rd gen.) & $\tilde{B}\tilde{B}$\\

\hline \hline
$\tilde{g}$ & $+ -$ & $\times$ & $\times$ & $\times$ & $\times$ & $\times$ & $\times$ & $\times$ & $\times$\\

$\tilde{q}/\tilde{u}/\tilde{d}$ & $+2j_l $ & $+j_l$ & $+ -$ & $\times$ & $\times$ & $\times$ & $\times$ & $\times$ & $\times$\\

$\tilde{q}_3/\tilde{t}/\tilde{b}$ & $+2j_3$ & $+j_l+2j_3$ & $+2j_l+2j_3$ & $+-$ & $\times$ & $\times$ & $\times$ & $\times$ & $\times$\\

{$\tilde{W}$} & $+4j$ & $+j_l+2j$ & $+2j_l$ & $+2j_3$ & $+-$ & $\times$ & $\times$ & $\times$ & $\times$\\

$\tilde{H}$ & $+4j$ & $+j_l+2j$ & $+2j_l$ & $+2j_3$ & $\times$ & $+ -$ & $\times$ & $\times$ & $\times$\\

\multirow{2}{*}{$\tilde{\ell}(\tilde{\nu})/\tilde{e}$} & $\tilde{\ell}\tilde{\ell}/\tilde{e}\tilde{e}:\;+2\ell+4j$ & $+2\ell+j_l+2j$ & $+2\ell+2j_l$ & $+2\ell+2j_3$ & $+\ell+\ETmiss$ & $+2\ell$ & $+ -$ & $\times$ & $\times$ \\ 

& $\tilde{\ell}\tilde{\nu}:\;+\ell+4j+\ETmiss$ & $+\ell+j_l+2j+\ETmiss$ & $+\ell+2j_l+\ETmiss$ & $+\ell+ 2j_3+\ETmiss$ & $+2\ell$ & $+\ell+\ETmiss$ & $+ -$ & $\times$ & $\times$ \\ 

\multirow{2}{*}{$\tilde{\tau}_L(\tilde{\nu}_{\tau})/\tilde{\tau}_R$} & $\tilde{\tau}_L\tilde{\tau}_L/\tilde{\tau}_R\tilde{\tau}_R:\;+2\tau+4j$ & $+2\tau+j_l+2j$ & $+2\tau+2j_l$ & $+2\tau+2j_3$ & $+\tau+\ETmiss$ & $+2\tau$ & $+ -$ & $\times$ & $\times$ \\ 

& $\tilde{\tau}_L\tilde{\nu}_{\tau}:\;+\tau+4j+\ETmiss$ & $+\tau+j_l+2j+\ETmiss$ & $+\tau+2j_l+\ETmiss$ & $+\tau+2j_3+\ETmiss$ & $+2\tau$ & $+\tau+\ETmiss$ & $+ -$ & $\times$ & $\times$ \\  

$\tilde{B}$ & $+4j$ & $+j_l+2j$ & $+2j_l$ & $+2j_3$ & $+3\ell+\ETmiss/2\ell+\ETmiss/4j/2v$ & $+4j_3/2v$ & $+2\ell/\ell+\ETmiss$ & $+2\tau/\tau+\ETmiss$ & $+ -$\\
\hline \hline
\end{tabular}
}
\end{adjustbox}}
\end{center}
\label{tab:optimisesearch}
\end{table}

\subsection{Flavor, Sign Configurations of Leptons}
Here, we compile tables that show the possible flavor and sign 
combinations of the leptons in the signatures of 
Tables~\ref{tab:LLE1}-\ref{tab:LQD6}. In all the tables below, the indices $i,j,k \in \{1,2,3\}$, while the indices $a,b \in \{1,2\}$; $\tilde{\ell}_3$ denotes a $\tau$. For each listed configuration, the charge conjugated state (if different) is also possible but we omit listing it explicitly.
\subsubsection*{LLE Signatures}
The flavor and sign combinations of the leptons corresponding to the various $LL\Bar{E}$ topologies are shown in~\cref{tab:B1} ($2L+\ETmiss$);~\cref{tab:B2} ($3L+\ETmiss$);~\cref{tab:B3} ($4L$);~\cref{tab:B4} ($4L + (0-4)j + \ETmiss$);~\cref{tab:B5} ($5L+\ETmiss$); and~\cref{tab:B6} ($6L+\ETmiss$). The combinations corresponding to $\tilde{B}$ also apply to the Higgsino and all colored-sector LSPs.
\begin{table}[h]
\fontsize{12pt}{12pt}\selectfont
\begin{minipage}{.5\linewidth}
\caption{$2L+\ETmiss$.}
\begin{center}
\fontsize{11pt}{11pt}\selectfont
\begin{tabular}{ccc}
\hline \hline
{\bf LSP} &  {\bf Coupling} &  {\bf Signature}\\
\hline \hline

$\tilde{e}_a$ & $\lambda_{aba}\ a\neq b$ & $\ell_a^+\ell_a^-/\ell_b^+\ell_b^-/\ell_a^+\ell_b^-$\\ 

$\tilde{e}_b$ & $\lambda_{a3b}$ & $\ell_a^+\ell_a^-/\ell_a^+\tau^-$ \\ 

$\tilde{\tau}_R$ & $\lambda_{ab3}\ a\neq b$ & $\ell_a^+\ell_a^-/\ell_b^+\ell_b^-/\ell_a^+\ell_b^-$\\ 

$\tilde{\tau}_R$ & $\lambda_{a33}$ & $\ell_a^+\ell_a^-/\ell_a^+\tau^-$\\ 

\hline \hline
\label{tab:B1}
\end{tabular}
\end{center}
\end{minipage}
\begin{minipage}{.5\linewidth}
\caption{$3L + \ETmiss$.}
\begin{center}
\fontsize{11pt}{11pt}\selectfont
\begin{tabular}{ccc}
\hline \hline
{\bf LSP} &  {\bf Coupling} &  {\bf Signature}\\
\hline \hline

$\tilde{\ell}_a \left(\tilde{\nu}_a\right)$ & $\lambda_{abc}\ a\neq b$ & $\ell_b^+\ell_c^+\ell_c^-/\ell_b^-\ell_c^+\ell_c^+$\\ 

$\tilde{\ell}_a \left(\tilde{\nu}_a\right)$ & $\lambda_{a3b}$ & $\ell_b^+\ell_b^+\tau^-/\ell_b^+\ell_b^-\tau^+$ \\ 

$\tilde{\ell}_a \left(\tilde{\nu}_a\right)$ & $\lambda_{ab3}\ a\neq b$ & $\ell^+_b \tau^+ \tau^-/\ell^-_b \tau^+ \tau^+$\\ 

$\tilde{\ell}_a \left(\tilde{\nu}_a\right)$ & $\lambda_{a33}$ & $\tau^+\tau^+\tau^-$\\ 

$\tilde{\tau}_L\left(\tilde{\nu}_\tau\right)$ & $\lambda_{a3b}$ & $\ell_a^-\ell_b^+\ell_b^+/\ell_a^+\ell_b^+\ell_b^-$ \\

$\tilde{\tau}_L\left(\tilde{\nu}_\tau\right)$ & $\lambda_{a33}$ & $\ell_a^+\tau^+\tau^-/\ell_a^-\tau^+\tau^+$ \\

\hline \hline
\label{tab:B2}
\end{tabular}
\end{center}
\end{minipage}
\end{table}

\begin{table}[h]
\fontsize{11pt}{11pt}\selectfont
\caption{$4L$.}
\begin{center}
\begin{tabular}{ccc}
\hline \hline
{\bf LSP} &  {\bf Coupling} &  {\bf Signature}\\
\hline \hline

$\tilde{\ell}_a \left(\tilde{\nu}_a\right)$ & $\lambda_{abc}\ a\neq b$ & $\ell_b^+\ell_b^+\ell_c^-\ell_c^-/\ell_b^+\ell_b^-\ell_c^+\ell_c^-$\\ 

$\tilde{\ell}_a \left(\tilde{\nu}_a\right)$ & $\lambda_{a3b}$ & $\ell_b^-\ell_b^-\tau^+\tau^+/\ell_b^+\ell_b^-\tau^+\tau^-$ \\ 

$\tilde{\ell}_a \left(\tilde{\nu}_a\right)$ & $\lambda_{ab3}\ a \neq b$ & $\ell_b^-\ell_b^-\tau^+\tau^+/\ell_b^+\ell_b^-\tau^+\tau^-$ \\

$\tilde{\ell}_a \left(\tilde{\nu}_a\right)$ & $\lambda_{a33}$ & $\tau^+\tau^+ \tau^- \tau^-$\\ 

$\tilde{\tau}_L\left(\tilde{\nu}_\tau\right)$ & $\lambda_{a3b}$ & $\ell_a^+\ell_a^+\ell_b^-\ell_b^-/\ell_a^+\ell_a^-\ell_b^+\ell_b^-$ \\

$\tilde{\tau}_L\left(\tilde{\nu}_\tau\right)$ & $\lambda_{a33}$ & $\ell_a^-\ell_a^-\tau^+\tau^+/\ell_a^+\ell_a^-\tau^+\tau^-$ \\

\hline \hline
\label{tab:B3}
\end{tabular}
\end{center}
\end{table}

\begin{table}[h]
\fontsize{11pt}{11pt}\selectfont
\caption{$4L + (0-4)j + \ETmiss$.}
\begin{center}
\begin{tabular}{ccc}
\hline \hline
{\bf LSP} &  {\bf Coupling} &  {\bf Signature}\\
\hline \hline

$\tilde{B}$ & $\lambda_{ijk}\ i\neq j$ & $\ell_i^+\ell_i^+\ell_k^-\ell_k^-/\ell_i^+\ell_i^-\ell_k^+\ell_k^-/\ell_j^+\ell_j^+\ell_k^-\ell_k^-/\ell_j^+\ell_j^-\ell_k^+\ell_k^-/\ell_i^+\ell_j^+\ell_k^-\ell_k^-/\ell_i^+\ell_j^-\ell_k^+\ell_k^-$\\ 

$\tilde{W}$ & $\lambda_{ijk}\ i\neq j$ & $\ell_i^+\ell_j^+\ell_k^-\ell_k^-/\ell_i^+\ell_j^-\ell_k^+\ell_k^-$\\ 
\hline \hline
\label{tab:B4}
\end{tabular}
\end{center}
\end{table}

\begin{table}[h]
\fontsize{11pt}{11pt}\selectfont
\caption{$5L+\ETmiss$.}
\begin{center}
\scalebox{0.75}{
\renewcommand{\arraystretch}{1.3}
\begin{adjustbox}{width=1.3\textwidth}
\begin{tabular}{ccc}
\hline \hline
{\bf LSP} &  {\bf Coupling} &  {\bf Signature}\\
\hline \hline

$\tilde{\tau}_L\left(\tilde{\nu}\right)$ & $\lambda_{aba}\ a\neq b$ & $\ell_a^+\ell_a^+\ell_a^-\ell_a^- \tau^+ / \ell_a^+\ell_a^-\ell_b^+\ell_b^-\tau^+/\ell_a^+\ell_a^+\ell_b^-\ell_b^- \tau^+/\ell_a^+\ell_a^+\ell_a^-\ell_b^+ \tau^+$\\ 

$\tilde{\tau}_L\left(\tilde{\nu}\right)$ & $\lambda_{ab3}\ a\neq b$ & $\ell^-_a \ell^-_a \tau^+ \tau^+\tau^+/\ell^-_b \ell^-_b \tau^+ \tau^+\tau^+/\ell^+_a \ell^-_a \tau^+\tau^+ \tau^-/\ell^+_b \ell^-_b \tau^+ \tau^+ \tau^-/\ell^-_a \ell^-_b \tau^+ \tau^+ \tau^+/\ell^+_a \ell^-_b \tau^+ \tau^+ \tau^-$\\

\hline \hline
\label{tab:B5}
\end{tabular}
\end{adjustbox}}
\end{center}
\end{table}

\begin{table}[h]
\begin{center}
\caption{$6L+\ETmiss$.}
\scalebox{0.75}{
\renewcommand{\arraystretch}{1.3}
\begin{adjustbox}{width=1.3\textwidth}
\begin{tabular}{ccc}
\hline \hline
{\bf LSP} &  {\bf Coupling} &  {\bf Signature}\\
\hline \hline

$\tilde{e}_a$ & $\lambda_{ab3}\ a\neq b$ & $\ell^+_a \ell^-_a\ell^-_a \ell^-_a \tau^+ \tau^+/\ell^+_a \ell^-_a\ell^-_b \ell^-_b \tau^+ \tau^+/\ell^+_a \ell^+_a \ell^-_a \ell^-_a \tau^+ \tau^-/\ell^+_a \ell^-_a\ell^+_b \ell^-_b \tau^+ \tau^-/\ell^+_a \ell^-_a\ell^-_a \ell^-_b \tau^+ \tau^+/\ell^+_a \ell^+_a \ell^-_a \ell^-_b \tau^+ \tau^-$\\ 

$\tilde{e}_b$ & $\lambda_{a33}$ & $\ell^+_a \ell^-_a\ell_a^-\tau^+\tau^+\tau^-/\ell^+_a \ell^-_a\ell^-_a \ell^-_a \tau^+ \tau^+/\ell^+_a \ell^+_a \ell^-_a \ell^-_a \tau^+ \tau^-$\\ 

$\tilde{\tau}_L\left(\tilde{\nu}\right)$ & $\lambda_{aba}\ a\neq b$ & $\ell_a^+\ell_a^+\ell_a^-\ell_a^- \tau^+\tau^- /\ell_a^+\ell_a^-\ell_b^+\ell_b^-\tau^+\tau^-/\ell_a^+\ell_a^+\ell_b^-\ell_b^- \tau^+\tau^- /\ell_a^+\ell_a^+\ell_a^-\ell_b^+ \tau^+\tau^-$\\ 

$\tilde{\tau}_L\left(\tilde{\nu}\right)$ & $\lambda_{ab3}\ a\neq b$ & $\ell^-_a \ell^-_a \tau^+ \tau^+ \tau^+\tau^- /\ell^-_b \ell^-_b \tau^+ \tau^+\tau^+\tau^-/\ell^+_a \ell^-_a \tau^+ \tau^+\tau^- \tau^-/\ell^+_b \ell^-_b \tau^+ \tau^+\tau^- \tau^-/\ell^-_a \ell^-_b \tau^+ \tau^+\tau^+\tau^-/\ell^+_a \ell^-_b \tau^+ \tau^+\tau^- \tau^-$\\

$\tilde{\tau}_R$ & $\lambda_{aba}\ a\neq b$ & $\ell_a^+\ell_a^+\ell_a^-\ell_a^-\tau^+\tau^-/\ell_a^+\ell_a^-\ell_b^+\ell_b^-\tau^+\tau^-/\ell_a^+\ell_a^+\ell_b^-\ell_b^-\tau^+\tau^-/\ell_a^+\ell_a^-\ell_a^-\ell_b^+\tau^+\tau^-$\\ 

$\tilde{\tau}_R$ & $\lambda_{a3b}$ & $\ell_a^+\ell_a^+\ell_b^-\ell_b^-\tau^+\tau^-/\ell_a^+\ell_a^-\ell_b^+\ell_b^-\tau^+\tau^-/\ell_a^+\ell_b^-\ell_b^-\tau^+\tau^+\tau^-/\ell_a^-\ell_b^+\ell_b^-\tau^+\tau^+\tau^-$ \\

\hline \hline
\label{tab:B6}
\end{tabular}
\end{adjustbox}}
\end{center}
\end{table}

\subsubsection*{UDD Tables}
For the $\Bar{U}\Bar{D}\Bar{D}$ topologies, the possible combinations are shown in~\cref{tab:B7} ($1L+2j_l+4j+\ETmiss$); and~\cref{tab:B8} ($2L+2j_l+4j$).
\begin{table}[h]
\fontsize{11pt}{11pt}\selectfont
\begin{minipage}{.5\linewidth}
\caption{$1L+2j_l+4j+\ETmiss$.}
\begin{center}
\begin{tabular}{ccc}
\hline \hline
{\bf LSP} &  {\bf Coupling} &  {\bf Signature}\\
\hline \hline

$\tilde{\ell}_a \left(\tilde{\nu}_a\right)$ & $\lambda''_{ijk}$ & $\ell_a^+$\\ 

$\tilde{\tau}_L\left(\tilde{\nu}\right)$ & $\lambda''_{ijk}$ & $\tau^+$\\ 

\hline \hline
\label{tab:B7}
\end{tabular}
\end{center}
\end{minipage}
\begin{minipage}{.5\linewidth}
\caption{$2L+2j_l+ 4j$.}
\begin{center}

\begin{tabular}{ccc}
\hline \hline
{\bf LSP} &  {\bf Coupling} &  {\bf Signature}\\
\hline \hline

$\tilde{\ell}_a \left(\tilde{\nu}_a\right)$ & $\lambda''_{ijk}$ & $\ell_a^+\ell_a^-$\\ 

$\tilde{e}_a$ & $\lambda''_{ijk}$ & $\ell_a^+\ell_a^-$ \\ 

$\tilde{\tau}_L\left(\tilde{\nu}\right)$ & $\lambda''_{ijk}$ & $\tau^+\tau^-$\\ 

$\tilde{\tau}_R$ & $\lambda''_{ijk}$ & $\tau^+\tau^-$\\ 

\hline \hline
\label{tab:B8}
\end{tabular}
\end{center}
\end{minipage}
\end{table}

\subsubsection*{LQD Tables}
Finally, for the $LQ\Bar{D}$ topologies, the possible configurations are shown in~\cref{tab:B9} ($1L+(2-6)j+\ETmiss$);~\cref{tab:B10} ($2L+(2-6)j+(\ETmiss)$);~\cref{tab:B11} ($3L+4j+\ETmiss$); and~\cref{tab:B12} ($4L+4j$). The $\tilde{B}$ configurations apply to the other electroweakinos, and the colored LSPs.

\begin{table}[h]
\fontsize{11pt}{11pt}\selectfont
\begin{minipage}{.5\linewidth}
\caption{$1L+(2-6)j+\ETmiss$.}
\begin{center}
\begin{tabular}{ccc}
\hline \hline
{\bf LSP} &  {\bf Coupling} &  {\bf Signature}\\
\hline \hline

$\tilde{B}$ & $\lambda'_{ijk}$ & $\ell_i^+$\\

$\tilde{\ell}_a (\tilde{\nu}_a)$ & $\lambda'_{33k}$ & $\ell_a^+$\\

$\tilde{\tau}_L\left(\tilde{\nu}_\tau\right)$ & $\lambda'_{a3k}$ & $\tau^+$\\

\hline \hline
\label{tab:B9}
\end{tabular}
\end{center}
\end{minipage}
\begin{minipage}{.5\linewidth}
\caption{$2L+(2-6)j + (\ETmiss)$.}
\begin{center}
\begin{tabular}{ccc}
\hline \hline
{\bf LSP} &  {\bf Coupling} &  {\bf Signature}\\
\hline \hline

$\tilde{B}$ & $\lambda'_{ijk}$ & $\ell_i^+\ell_i^-/\ell_i^+\ell_i^+$\\

$\tilde{\ell}_a (\tilde{\nu}_a)$ & $\lambda'_{3ak}$ & $\ell_a^+\tau^+/\ell_a^+\tau^-$\\

$\tilde{\ell}_a (\tilde{\nu}_a)$ & $\lambda'_{33k}$ & $\ell_a^+\ell_a^-/\ell_a^+\tau^+/\ell_a^+\tau^-$\\

$\tilde{\tau}_L\left(\tilde{\nu}_\tau\right)$ & $\lambda'_{abk}$ & $\ell_a^+\tau^+/\ell_a^+\tau^-$\\

$\tilde{\tau}_L\left(\tilde{\nu}_\tau\right)$ & $\lambda'_{a3k}$ & $\ell_a^+\tau^+/\ell_a^+\tau^-/\tau^+\tau^-$\\

$\tilde{e}_a$ & $\lambda'_{i3k}$ & $\ell_a^+\ell_a^-$\\

$\tilde{\tau}$ & $\lambda'_{i3k}$ & $\tau^+\tau^-$\\

\hline \hline
\label{tab:B10}
\end{tabular}
\end{center}
\end{minipage}
\end{table}

\begin{table}[h]
\caption{$3L+4j+\ETmiss$.}
\begin{center}
\begin{tabular}{ccc}
\hline \hline
{\bf LSP} &  {\bf Coupling} &  {\bf Signature}\\
\hline \hline
$\tilde{\ell}_a (\tilde{\nu}_a)$ & $\lambda'_{3jk}$ & $\ell_a^+\ell_a^-\tau^+/\ell_a^+\tau^+\tau^-/\ell_a^+\tau^+\tau^+/\ell_a^+\tau^-\tau^-$\\

$\tilde{\tau}_L\left(\tilde{\nu}_\tau\right)$ & $\lambda'_{ajk}$ & $\ell_a^+\ell_a^-\tau^+/\ell_a^+\ell_a^+\tau^+/\ell_a^-\ell_a^-\tau^+/\ell_a^+\tau^+\tau^-$\\

$\tilde{e}_a$ & $\lambda'_{ijk}$ & $\ell_a^+\ell_a^-\ell_i^+$\\

$\tilde{\tau}$ & $\lambda'_{i3k}$ & $\ell_i^+\tau^+\tau^-$\\

\hline \hline
\label{tab:B11}
\end{tabular}
\end{center}
\end{table}

\begin{table}[h]
\caption{$4L+4j$.}
\begin{center}
\begin{tabular}{ccc}
\hline \hline
{\bf LSP} &  {\bf Coupling} &  {\bf Signature}\\
\hline \hline

$\tilde{\ell}_a (\tilde{\nu}_a)$ & $\lambda'_{3jk}$ & $\ell_a^+\ell_a^-\tau^+\tau^-/\ell_a^+\ell_a^-\tau^+\tau^+$\\

$\tilde{\tau}_L\left(\tilde{\nu}_\tau\right)$ & $\lambda'_{ajk}$ & $\ell_a^+\ell_a^-\tau^+\tau^-/\ell_a^+\ell_a^+\tau^+\tau^-$\\

$\tilde{e}_a$ & $\lambda'_{ijk}$ & $\ell_a^+\ell_a^-\ell_i^+\ell_i^-/\ell_a^+\ell_a^-\ell_i^+\ell_i^+$\\

$\tilde{\tau}$ & $\lambda'_{i3k}$ & $\ell_i^+\ell_i^-\tau^+\tau^-/\ell_i^+\ell_i^+\tau^+\tau^-$\\

\hline \hline
\label{tab:B12}
\end{tabular}
\end{center}
\end{table}

\section{\texttt{abc-rpv}, the RPV Python Library}
\label{sec:C}
\texttt{abc-rpv}\footnote{\texttt{abc-rpv} Python library is available at: \href{https://github.com/kys-sheng/abctestrun.git}{\texttt{https://github.com/kys-sheng/abc-rpv.git}}} is a Python library that provides a framework for analyzing the collider signatures of the RPV-MSSM. Users are provided with various functionalities to explore the landscape of RPV-MSSM physics within the context of small RPV couplings. In this section, we provide a short introduction to the library. A complete manual will be provided as a separate document/paper in the future.  

\subsection{Introduction}
The code starts by generating all possible transitions from one sparticle to another, based on the vertices provided in the input table (\texttt{table\_notsup.csv}) stored in the input directory. Using this, it can obtain the resulting signature for a decay chain from any LSP to a sparticle directly coupled to an RPV operator; the latter, then, simply decays into purely Standard Model objects. Going through all combinations of LSP type, and RPV couplings (in terms of categories defined in Tables~\ref{tab:LLE1}-\ref{tab:LQD6}), all possible decay chains and signatures are compiled into tables. These tables are the output available to the user that can then be analyzed using the functions described below. By default, all output tables are already generated using the default input table, and are readily available in the data directory. The user does not need to generate the tables unless the input table is modified.
    
\subsection{Assumptions and Caveats}
In the implementation of our code, there are a few assumptions and caveats worth noting:
\begin{itemize}
    \item All possible transitions are constructed from vertices provided in \texttt{table\_notsup.csv} in the input directory. The vertices provided in this table need not be a 3-point vertex.
    \item The input table (\texttt{table\_notsup.csv}) contains vertices that allow transitions from one sparticle to another while producing standard model particles. By default, only non-suppressed transitions based on the MSSM interactions are included; we use modified versions of the tables compiled in Ref.~\cite{Dreiner:2012wm} for classifying vertices as suppressed or non-suppressed. Note that the input table can be modified by the user, as needed. This allows one to regenerate the output tables with custom vertices.
    \item While generating the decay chains for the LSPs, only the shortest chain is constructed by default. Users also have the option to generate all possible chains up to 3 transitions. 
    \item The decay chains do not contain repeating sparticles.
\end{itemize}

\subsection{Usage}
Please refer to \texttt{Tutorial.ipynb} available at \href{https://github.com/kys-sheng/abctestrun.git}{\texttt{https://github.com/kys-sheng/abc-rpv.git}} for a complete tutorial of the Python library. We only discuss basic functionality here.

\subsubsection*{Syntax}
Tables~\ref{tab:sparticles_syntax} and \ref{tab:final_state_objects_syntax} show the syntax used in the code. One can also refer to \texttt{rpv\_definitions.py} for more information.

\begin{table}
\begin{center}
\caption{Syntax for sparticles used in the code.}
    \begin{tabular}{cc}
    \hline \hline
    {\bf Code Syntax} & {\bf Sparticles} \\
    \hline \hline
    \vspace{-1em}&\\
$\texttt{B}$       &   Bino, B               \\
$\texttt{W\string^\string+}$    &   Charged Wino       \\
$\texttt{W\string^0}$    &   Neutral Wino       \\
$\texttt{G}$       &   Gluino             \\
$\texttt{H\string^+}$    &   Charged Higgsino   \\
$\texttt{H\string^0}$    &   Neutral Higgsino   \\
$\texttt{q}$       &   $\tilde{u}_L$, $\tilde{d}_L$, $\tilde{c}_L$, $\tilde{s}_L$ \\
$\texttt{d}$       &   $\tilde{d}_R$, $\tilde{s}_R$           \\
$\texttt{u}$       &   $\tilde{u}_R$, $\tilde{c}_R$           \\
$\texttt{l}$       &   $\tilde{e}_L$, $\tilde{\mu}_L$          \\
$\texttt{nu}$      &   $\tilde{\nu}_e$, $\tilde{\nu}_\mu$        \\
$\texttt{e}$       &   $\tilde{e}_R$, $\tilde{\mu}_R$          \\
$\texttt{t\_L}$    &   $\tilde{t}_L$                \\
$\texttt{b\_L}$    &   $\tilde{b}_L$                \\
$\texttt{t}$       &   $\tilde{t}_R$                \\
$\texttt{b}$       &   $\tilde{b}_R$                \\
$\texttt{tau\_L}$  &   $\tilde{\tau}_L$              \\
$\texttt{tau}$     &   $\tilde{\tau}_R$              \\
$\texttt{nu\_tau}$ &   $\tilde{\nu}_\tau$             \\
\hline \hline
\label{tab:sparticles_syntax}
\end{tabular}
\end{center}
\end{table}

\begin{table}
\begin{center}
\caption{One-character syntax for final state objects used in the code.}
    \begin{tabular}{cc}
    \hline \hline
    {\bf Symbol} & {\bf Particles (Final State Objects)} \\
    \hline \hline
    \vspace{-1em}&\\
    $\texttt{l}$ & $e/\mu$ \\
    $\texttt{T}$ & $\tau$\\
    $\texttt{L}$ & $e/\mu/\tau$\\
    $\texttt{j}$ & $u/d/c/s$ jets\\
    $\texttt{b}$ & $b$ jets \\
    $\texttt{t}$ & $t$ jets \\
    $\texttt{3}$ & $t/b$ jets \\
    $\texttt{J}$ & $u/d/c/s/t/b$ jets \\
    $\texttt{v}$ & $W/Z/h$\\
    $\texttt{X}$ & MET\\
    \hline \hline
    \label{tab:final_state_objects_syntax}
\end{tabular}
\end{center}
\end{table}

\subsubsection*{Dictionaries}
In the library, there are a few built-in dictionaries that contain the output tables generated from the code. 
\begin{itemize}
    \item \texttt{ONE\_LSP\_RPV\_DECAY\_DICT} : Contains details for all possible RPV decays of one LSP. Information regarding RPV coupling category, signature, decay chains, number of vertices is included.
    \item \texttt{TWO\_LSP\_RPV\_DECAY\_DICT} : Contains details for all possible RPV decays of a pair\footnote{We restrict to the case where both LSPs are the same, or belong to the same $\mathrm{SU}(2)_{L}$ doublet.} of LSPs (decay via same category of RPV coupling). Information regarding RPV coupling category, signature, decay chains, number of vertices is included.
    \item \texttt{TWO\_LSP\_MIXED\_RPV\_DECAY\_DICT} : Contains details for all possible RPV decays of a pair of LSPs (decay via different categories of RPV couplings). Information regarding RPV coupling categories, signature, decay chains, number of vertices is included.
    \item \texttt{ONE\_LSP\_SIG\_CAT\_DICT} : Contains final state signatures arising from decay of one LSP, categorized by RPV coupling; similar to Tables~\ref{tab:LLE1}-\ref{tab:LQD6}.
    \item \texttt{TWO\_LSP\_SIG\_CAT\_DICT} : Contains final state signatures arising from decay of pair of LSPs, categorized by RPV coupling; similar to Tables~\ref{tab:LLE1}-\ref{tab:LQD6}.
\end{itemize} 
Note that the above dictionaries are regenerated upon using different input transition tables, as well as different table generation choices (\textit{e.g.}, decay chain length). 

\subsubsection*{Main Functions}
Although the dictionaries by themselves contain all relevant information, it is more efficient and powerful to use the functions provided in the library to analyze the data. We describe the basic usage here; refer to \texttt{Tutorial.ipynb} for more details.

\paragraph{\small{One LSP Decay:}}
\begin{itemize}
\item \texttt{find\_one\_lsp\_from\_signature} \\
Using the signature as input, this function finds all LSPs with decay chains leading to the given final state. Alongside with the LSP, the relevant RPV couplings and decay chains are also returned.
\item \texttt{find\_one\_lsp\_from\_signature\_inclusive}\\
Similar to \texttt{find\_one\_lsp\_from\_signature}, but in the inclusive mode (\textit{e.g.}, one can choose $n_{jets}$ > 3 instead of $n_{jets}$ = 3).
\item \texttt{find\_signatures\_from\_one\_lsp} \\
Using the LSP as input, this function finds all possible signatures that can arise in the LSP decay. Alongside with the signatures, the relevant RPV couplings and decay chains are also returned.
\end{itemize}

\paragraph{\small{LSP Pair Decay; Same Coupling Category:}}
\begin{itemize}
\item \texttt{find\_two\_lsp\_from\_signature}: \\
Similar to \texttt{find\_one\_lsp\_from\_signature} but returns all pairs of LSPs leading to the input signature.
\item \texttt{find\_two\_lsp\_from\_signature\_inclusive}:\\
Inclusive mode of \texttt{find\_two\_lsp\_from\_signature}.
\item \texttt{find\_signatures\_from\_two\_lsp}:\\
Similar to \texttt{find\_signatures\_from\_one\_lsp} but for a pair of input LSPs.
\end{itemize}
In all of the above, the pair is assumed to decay via the same (category of) RPV coupling. 

\paragraph{\small{LSP Pair Decay; Different Coupling Categories:}}
\begin{itemize}
\item \texttt{find\_two\_lsp\_from\_signature\_mixed\_couplings}
\item \texttt{find\_two\_lsp\_from\_signature\_mixed\_couplings\_inclusive}
\item \texttt{find\_signatures\_from\_two\_lsp\_mixed\_couplings}
\end{itemize}
Analogous to the above but for LSP pair decaying via different (categories of) RPV couplings.

\subsection*{Advanced Usage}
By default, all dictionaries and tables are regenerated automatically from the input table if all the csv files in the data directory are deleted. Thus, users can generate all the tables based on their custom input table (\texttt{table\_notsup.csv}) by deleting the csv files in the data directory and reimporting the library. A step-by-step example demonstrating this will be provided in the complete manual.

\bibliographystyle{JHEP}
\bibliography{bibliography}
\end{document}